\documentclass[twocolumn,superscriptaddress,floatfix,nofootinbib]{revtex4-2}
\usepackage{microtype}
\usepackage{amsfonts}
\usepackage{dsfont} 
\usepackage{mathtools}
\usepackage{wrapfig}
\usepackage{graphicx}
\usepackage{subfig}
\usepackage{xcolor}
\usepackage{bbold}
\usepackage{bbm}
\usepackage{float}
\usepackage[font=small,
   justification=justified,
   format=plain]{caption}
   
\usepackage[colorlinks]{hyperref}
\DeclareMathOperator{\Tr}{Tr}
\DeclareMathOperator{\diag}{diag}
\DeclareMathOperator{\rk}{rk}

\begin{document}

\title{Cost and Routing of Continuous Variable Quantum Networks}

\author{Federico Centrone}
\email{fcentrone@icfo.net}
\affiliation{ICFO-Institut de Ciencies Fotoniques, The Barcelona Institute of Science and Technology,
08860 Castelldefels (Barcelona), Spain}

\author{Frederic Grosshans}
\email{frederic.grosshans@lip6.fr}
\affiliation{Sorbonne Université, CNRS, LIP6, 4 place Jussieu, F-75005 Paris, France}

\author{Valentina Parigi}
\email{valentina.parigi@lkb.upmc.fr}
\affiliation{Laboratoire Kastler Brossel, Sorbonne Universit\'{e}, CNRS, ENS-Universit\'{e} PSL, Coll\`{e}ge de France, 4 place Jussieu, F-75252 Paris, France}

\begin{abstract}
We study continuous-variable graph states with regular and complex network shapes and we report for their cost as a global measure of squeezing and number of squeezed modes that are necessary to build the network. We provide an analytical formula to compute the experimental resources required to implement the graph states and we use it to show  that  the scaling of the squeezing cost with the size of the network  strictly depends on its topology. We show that homodyne measurements along parallel paths between two nodes allow to increase the final entanglement in these nodes and we use this effect to boost the efficiency of an entanglement routing protocol. The devised routing protocol is particularly efficient in running-time for complex sparse networks.
\end{abstract}
\maketitle

\section{Introduction}

Networks science has been used to model the structures and properties of many biological, physical and technological systems, including internet and the world wide web. Photonics quantum networks are essential resources for quantum information processing and notably for quantum internet applications, where quantum states of light will allow for the efficient distribution and manipulation of information \cite{Wehner18,pirandola2019end,Pant19,Guonet20}. In order to develop large scale quantum communications and build a quantum internet it is compulsory to grasp the potentialities of quantum networks and exploit all their exceptional features. We can expect that complex networks theory can be used, like in the case of classical networks, to study and drive efficient quantum complex networks design for quantum technologies \cite{Bianconi23}.  

In this work we study continuous variable quantum networks in the form of CV graph states with regular and complex topologies. CV quantum information describes quantum states living in infinite dimensional Hilbert spaces, protocols mainly rely on coherent (homodyne) detection which, differently from photon counting detectors, can be highly efficient at room temperature. Moreover, CV quantum networks can be generated deterministically with a large number of nodes \cite{chen2014experimental,cai2017multimode,yokoyama2013ultra,asavanant2019generation,larsen2019deterministic}, they can be easily reconfigured   \cite{nokkala2018reconfigurable, Renault23,kouadou2022spectrally,madsen2022quantum} and they have been also exploited in quantum advantage demonstrations \cite{madsen2022quantum}.

It is known that quantum feature of CV states can be lost because of losses and noise during transmission. Nevertheless substantial progress has been done in CV quantum states distribution  \cite{Suleiman2022} and CV quantum repeaters design \cite{Tillman22}.  Moreover CV quantum networks, that are easily reconfigurable and with a large number of components 
\cite{cai2017multimode, yokoyama2013ultra, larsen2019deterministic, madsen2022quantum},
can be easily exploited as local area quantum networks.

In this work we discuss  Gaussian graph states using mathematical tools from network science in order to estimate how the cost of their experimental implementation is affected by the topology and the size of the network.  In particular,  we derive  an equation providing the squeezing values required to experimentally build a graph state as a function of its graph spectrum. We then adopt a resource theory of squeezing to estimate the cost of expanding the network. 

Thereafter, we propose a CV architecture for the quantum internet based on the Gaussian network previously described. We simulate quantum communication protocols through the network by letting  the spatially separated agents present at each node perform a homodyne measurement on their optical mode and look for the optimal measurement strategy to maximize the logarithmic negativity --- an entanglement measure \cite{eisertphd,vidal2002computable,plenio2005logarithnmic} ---
of the entangled pair shared by the two users who want to communicate, Alice and Bob. We prove that when multiple entangled paths connect Alice to Bob the optimal measurement strategy allows to increase the logarithmic negativity in the final pair. This \textit{parallel enhancement of entanglement} can be used to increase the quality of quantum communications in some selected network topologies. 

Lastly,  we employ our previous findings to implement a heuristic routing protocol for distributing and boosting the entanglement between two arbitrary agents. The algorithm we provide, on the one hand, is much more efficient than directly checking all possible combinations of quadrature measurements and, on the other hand, it always provides higher logarithmic negativity than the classical scheme, which is directly employing the shortest path between Alice and Bob and neglect the parallel channels. 
\section{Results}
\subsection{Cost of quantum networks}\label{sec:cost}
Consider a  graph with $N$ vertices. 
It is fully defined by its adjacency matrix $A\in{\mathbb R}^{N\times N}$. 
A way to prepare the associated graph state is to prepare a mode in the vacuum state for each vertex $i$ and, 
whenever two vertices $i,j$ are connected by an edge --- when $A_{ij}\neq0$ ---, 
we apply an entangling get such as a CZ-gate of strength $A_{ij}$.
We then end up with a Gaussian graph state characterized the $2N\times2N$ covariance matrix $\sigma$:
\begin{equation}
    \sigma=\frac{1}{2}\begin{pmatrix}
\mathds{1} & A\\ A & \mathds{1}+A^2
    \end{pmatrix},
\end{equation}
where we have normalized the vacuum state variance to $1/2$.

These  Gaussian bosonic states  are of particular
significance in the theory of continuous variable quantum information. They are in fact resources for measurement based quantum computing  \cite{Menicucci06,gu2009quantum},  quantum simulations \cite{nokkala2018reconfigurable},   multi-party quantum communication \cite{cai2017multimode,Arzani19}, and  quantum metrology \cite{Pinel12,Gessner18}.
Furthermore, their graphical structure simplifies their study through \emph{graphical calculus},
a formalism introduced by Menicucci, Flammia and van Loock in Ref.\ \cite{menicucci2011graphical}.
Some elements of graphical calculus are summed up in Appendix \ref{app:GraphCalc}. 

The correlations between the quadrature measurements of Gaussian states are fully described by their covariance matrix $\sigma$.
Therefore, as usual in the literature, we will not further mention the first moments, which only describe a
deterministic shift of the measurements which can easily be compensated when known and are therefore irrelevant. 

Through the Bloch--Messiah decomposition (see Sec.\ \ref{sec:gqs}), one can see 
the eigenvalues $\lambda_i^\pm$ of the covariance matrix $\sigma$ 
represent the squeezed and antisqueezed variances of the uncoupled oscillators, e.g.\ the uncertainty of measuring the real and imaginary part of the electromagnetic field. 
Together, they form the \emph{squeezing spectrum}.   
The first result of this paper is the following analytical relation between the squeezing spectrum of the Gaussian state and the adjacency spectrum of the graph: 
\begin{align}\label{eq:spectrum}
    \lambda_i^{\pm}
    &=\frac{1}{2}\left(1+D_i^2/2 \pm  \sqrt{D_i^2+D_i^4/4} \right),
\end{align}
where $D_i$ are the eigenvalues of the adjacency matrix $A$.
Equation (\ref{eq:spectrum}) shows the interplay between the physical resources necessary to experimentally implement a CV graph state and the spectrum of the underlying graph. This implies that we can use spectral graph theory to characterize analytically the physical requirements of building Gaussian networks and thus predict which one will be easier to realize. 
A first crucial consequence is that different graph states whose underlying graphs are co-spectral, e.g.\ their adjacency matrices have the same eigenvalues, can be transformed into each other applying passive linear optics\footnote{In 
  general, any CV graph can be reshaped in any other graph via a symplectic transformation; in this case it is an orthogonal transformation, and its physical realization  involves only linear optics without any supplementary squeezing.}.  
The intrinsic connection between the squeezing of a Gaussian network and its topology was already put in evidence 
in the limit of large squeezing by Gu et al.\@ \cite{gu2009quantum}, who showed a relation between the squeezing required to produce a CV graph state and the singular value decomposition of the associated adjacency matrix. 
Our result is a generalization of theorems 2 and 3 of Ref.~\cite{gu2009quantum}, exact and valid for finite squeezing, 
i.e.\ in a regime accessible with current technology. 

Another  crucial consequence of equation (\ref{eq:spectrum}) is that for CV graph states  the number of independent squeezed modes in their Bloch--Messiah decomposition corresponds to the rank $\rk(A)$ of the associated adjacency matrix $A$.
This immediately translate ito the number of squeezers needed to contruct said state

Squeezing is the essential resource for building Gaussian entangled states.  A natural question is thus: what is the squeezing cost of producing a quantum state?
A general resource theory for Gaussian states is provided by Lami et al.\ in \cite{lami2018gaussian}, with the specific case of squeezing described in \cite{idel2016operational}, where Idel, Lercher and Wolf find an operational squeezing measure, \emph{the squeezing cost}. Its expression for any pure\footnote{The more generic expression for mixed states is 
  slightly more involved :
  $G(\sigma)=-\sum_{i=1}^N 10\log_{10}\left[\max\left(2\lambda_i^-(\sigma),1\right)\right]$. 
  For pure Gaussian states ---i.e.\ all the states considered in this work ---, 
  this expression is equivalent to Eq.\ (\ref{eq:SqCostSigma}) because 
  $\sqrt{\lambda^ +_i\lambda^-_i}=\frac12$.}
Gaussian state of covariance matrix $\sigma$ is
\begin{align}\label{eq:SqCostSigma}
    G&: \mathbbm{R}^{2N\times 2N} \rightarrow \mathbbm{R},&
    G(\sigma)&=\sum_{i=1}^N 10\log_{10}\left(2\lambda_i^+(\sigma)\right),
\end{align}
where we chose to express it in dB.

We employ this to classify networks topologies, depending on the scaling of their squeezing cost
with the size of the network. In the following, we will consider an initial set of $N$ modes vacuum states with no squeezing ($s=0$), as it can be easily proved that initial uniform squeezing only adds a constant factor to the final squeezing cost. We  also assume that the CZ-gate coupling strength $A_{ij}=g=1$ for every edge of the graph, keeping the effects of non-uniform correlations for future works. 

\begin{figure*}[htb]
\includegraphics[width=\linewidth]{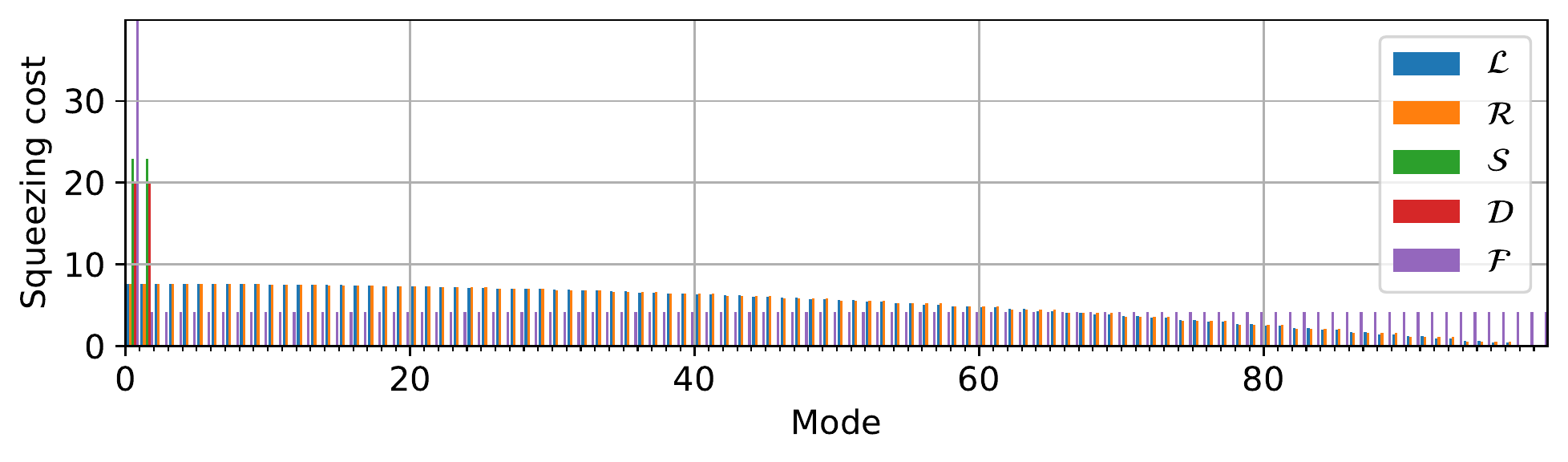}
\caption{\footnotesize \raggedright Squeezing cost distribution for 
the regular networks defined isn Sec.\ \ref{sec:RegNet}: linear $\mathcal{L}_N$, ring $\mathcal{R}_N$, star $\mathcal{S}_N$, diamond $\mathcal{D}_N$, fully connected $\mathcal{F}_N$ networks in the  $N=100$ supermodes,  $s=0,g=1$. All the networks present some squeezing in each mode except the $\mathcal{S}$ and $\mathcal{D}$ that have an equal amount of squeezing only in the first two modes. The $\mathcal{F}$ network has a large peak of squeezing in the first mode, while the remaining amount of squeezing is equally distributed in the other modes.}\label{fig:histoReg}
\end{figure*}{}

\subsubsection{Regular networks}\label{sec:costReg}

Let us first discuss some regular network structures by taking a look at the full squeezing spectrum of these topologies --- defined in Sec.\ \ref{sec:RegNet} --- for a fixed number of nodes $N=100$. This is shown in figure \ref{fig:histoReg}, where the values have been computed from the adjacency matrix spectrum of regular graphs \cite{cvetkovic1999spectra}.
We can see that, as expected, the linear and the ring graphs have a very similar spectrum, with small deviations induced by the periodicity of the latter that becomes negligible for large $N$. The star and the diamond networks 
only have two identically squeezed modes for all $N$. They are thus co-spectral up to a factor and can be transformed into each other with linear optics. 
Finally, the fully connected graph has one large eigenvalue that grows with $N$ 
and $N-1$ equal eigenvalues, independent of the size of the graph. 
These eigenvalues are computed analytically by diagonalizing the adjacency matrix $A$ and using Eq.\@ (\ref{eq:spectrum})
in Appendix \ref{app1}.

Let us now see how the total squeezing cost $G(\sigma)$ scales with the number of nodes $N$ for each of the network topologies presented above. This scaling is computed analytically using Eq.\ (\ref{eq:SqCostSigma}) in appendix \ref{app1}.
\begin{figure}[htb]
\centering
    \includegraphics[width=0.8\linewidth]{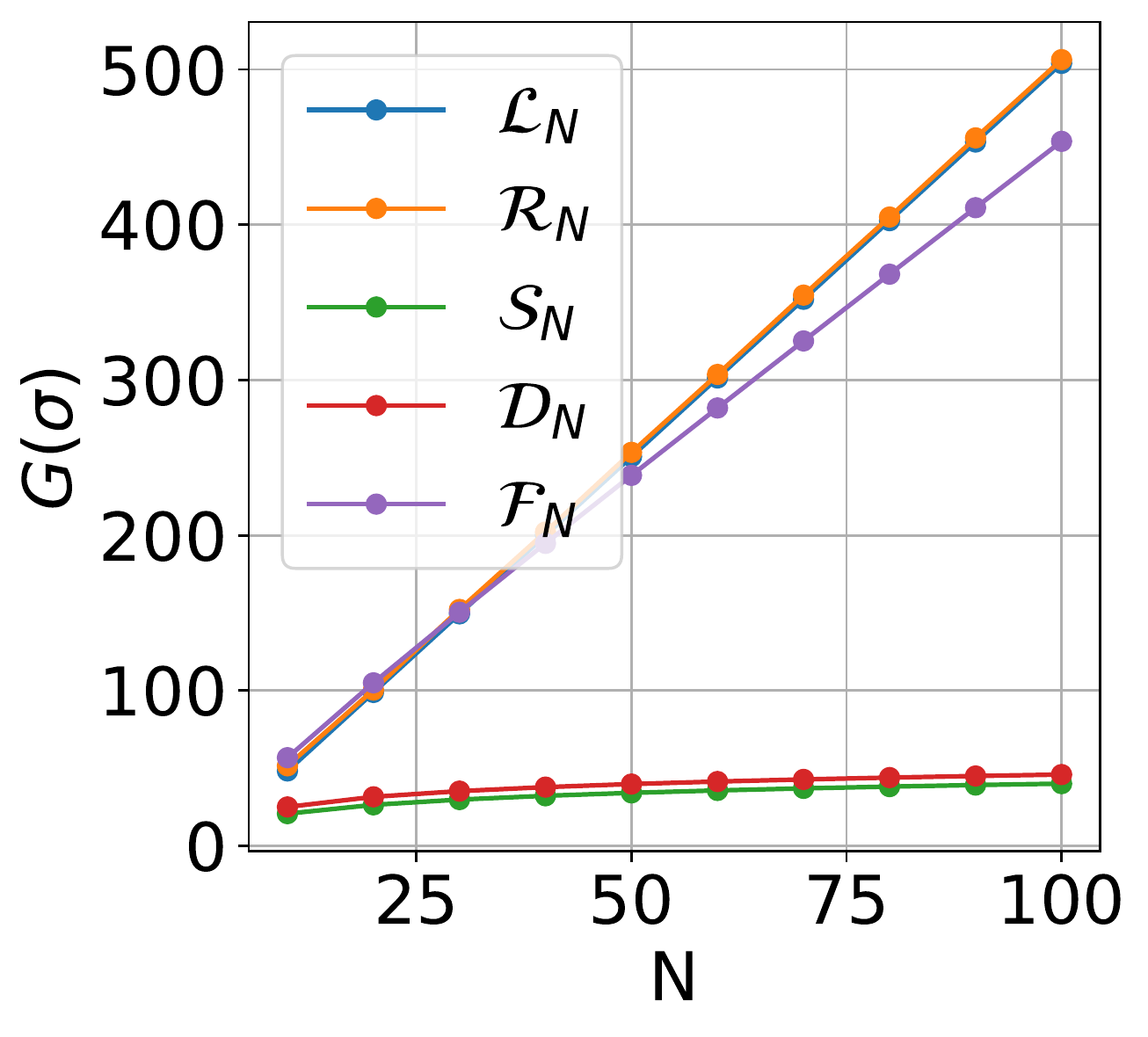} 
\caption{\footnotesize \raggedright Trend of the  squeezing cost  ${G}(\sigma)$ for the regular topologies: linear $\mathcal{L}_N$, ring $\mathcal{R}_N$, star $\mathcal{S}_N$, diamond $\mathcal{D}_N$, fully connected $\mathcal{F}_N$ networks for $N=100$ nodes. }\label{fig:GvNreg}
\end{figure}{}
In Fig.\ \ref{fig:GvNreg} we can see how the linear graph in blue and the ring graph in orange are superposed, sharing the same squeezing cost per node that, as shown in appendix \ref{app1}, is constant with $N$. The  cost of the star and diamond grows logarithmically with $N$ and in both cases has a simple expression
and present the lowest cost among the regular graphs we studied.
In all these cases, although the actual squeezing cost would be smaller for these networks, the amount of squeezing required in each mode is much larger so, depending on the experimental scenario, their implementation could be the easiest or the most challenging from an experimental point of view. In any case, these networks have interesting applications for quantum communications. In particular, the star graph can be used for secret sharing \cite{cai2017multimode}, while the diamond produce an effect that we called \textit{parallel enhancement of entanglement}, which will be explained in the next section. Finally, the fully connected graph,   has a  cost  that grows  linearly with $N$ and, despite having the largest number of edges and thus of squeezing increasing operations, for a large number of nodes is slightly cheaper than the linear graph.


A relevant application of these results would be to minimize the experimental difficulty
--- modelled by the squeezing cost and/or the number of squeezers ---
to prepare the Gaussian graph states used as resource for universal quantum computation.
To our knowledge, all such proposals rely on 2D-lattice structures
\cite{Menicucci06,gu2009quantum}, similar to the ones proved to be necessary in 
DV measurement-based quantum computation \cite{Van-den-Nest06}.
We show in appendix \ref{app1} the squeezing cost of a $D$ dimensional 
cubic lattice to scale linearly with the number of qumodes they contain. 
We conjecture the same proportionality holds for any $D$-dimensional regular lattice.
Furthermore, we conjecture this cost is proportional to the number of qubits used 
for the equivalent DV measurement-based quantum computation \cite{Van-den-Nest06}, 
which is itself 
proportional to the spacetime complexity of the corresponding quantum computation
in the circuit model.

\subsubsection{Complex networks}

\begin{figure*}[!htb]
\includegraphics[width=\textwidth]{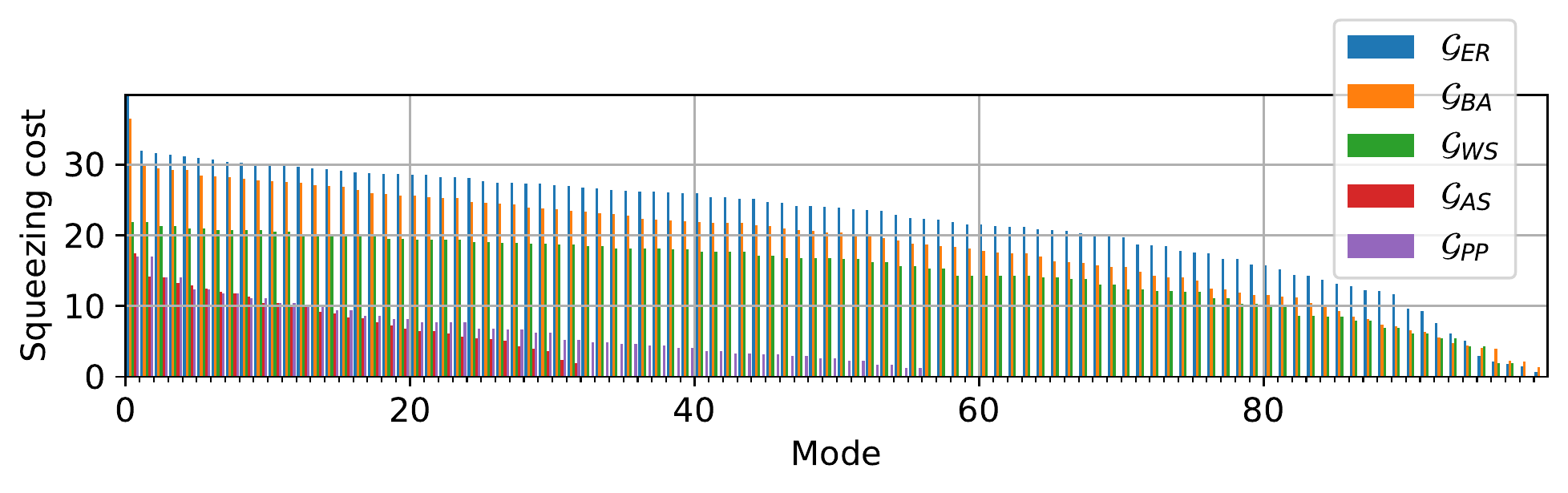}
\caption{\footnotesize \raggedright Squeezing cost distribution  for complex topologies: Erdős–Rényi  $\mathcal{G}_{ER}$,    Barabási--Albert $\mathcal{G}_{BA}$, Watts--Strogatz $\mathcal{G}_{WS}$, AS internet $\mathcal{G}_{AS}$ and Protein--Protein interaction $\mathcal{G}_{PP}$ networks in the $N=100$ supermodes. } \label{histoCX}
\end{figure*}

\begin{figure}[htb]
\centering
   \includegraphics[height=.75\linewidth]{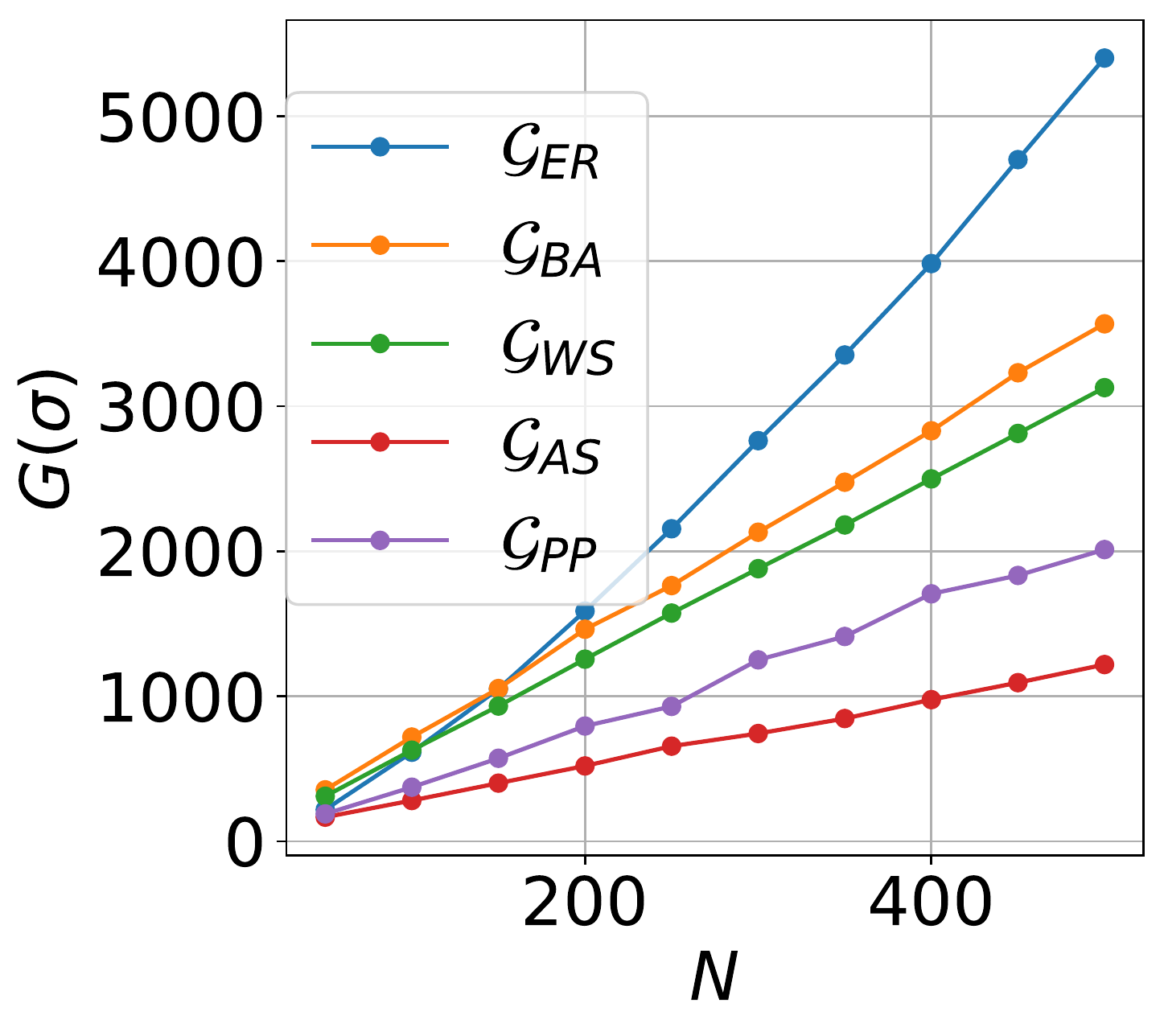}
\caption{\footnotesize \raggedright Trend of the  squeezing cost for complex topologies: Erdős--Rényi  $\mathcal{G}_{ER}$,    Barabási--Albert $\mathcal{G}_{BA}$, Watts--Strogatz $\mathcal{G}_{WS}$, AS internet $\mathcal{G}_{AS}$ and Protein--Protein interaction $\mathcal{G}_{PP}$ networks up to $N=500$ nodes. }\label{GvNcx}
\end{figure}

Complex networks are important in many natural and technological systems \cite{newman2003structure}; 
similarly
CV graph states with complex networks shapes are particularly relevant for simulating quantum complex networks environments \cite{nokkala2018reconfigurable,Mascherpa20} and to study future quantum information/communication networks mimicking the structure of the classical communication networks. It is then worth to study the scaling of the necessary squeezing resources for their implementation.  
In Fig.~\ref{histoCX} we show the   squeezing cost distribution for the various topologies of complex networks by showing the squeezing cost of  all the principal modes. Notice that, since complex networks are a subset of correlated random networks, 
the set of eigenvalues of the adjacency matrix is not deterministic. However, the eigenvalues follow a probability distribution
$f(x)$ --- well known in some selected cases \cite{farkas2001spectra} ---, 
which allows to derive the expectation value of the total squeezing cost as
\begin{equation}\label{eq:costCx}
    \langle G(\sigma)\rangle = 10 N \int f(x) \log_{10}[ \lambda^{(+)}(x)] \mathrm{d}x,
\end{equation}
where $\lambda^{(+)}(x)$ is the largest eigenvalue deduced from eq.\@~(\ref{eq:spectrum}),
where $x$ is the real random variable describing the eignevalues of the (random) adjacency matrix $A$. 
This gives us a straightforward recipe to compute $G$ from the
distribution $f$. Furthermore, it allows us do deduce the scaling of $G$ from the distribution of the eigenvalues and especially from the width of this distribution. In fact, if the distribution  does not change with $N$, then $G$ scales linearly with $N$, whereas any growth in the width of $f$ would imply a superlinear scaling of $G$.

In  Fig.~\ref{GvNcx} we report the trend of the total squeezing cost  as a function of the number of nodes for each of the above complex topologies. In this case, the networks were simulated and the values obtained were averaged over ten different samples.

From the plots we notice that the linear trend is the most common case. 
The spectral theory of real graphs is a much less established field and there is only a handful of results we can apply to actually make predictions. In particular, to the best of our knowledge there are only empirical results about the convergence of the  spectral density for the scale-free and the small-world models.  In Ref. \cite{farkas2001spectra}, however, Farkas et al.\@ show some crucial properties of the spectra of scale-free and small-world graphs, although further studies are required in order to have a deeper insight in the properties of complex quantum graph states. In particular, it is shown that, fixing the other network's parameters, the width of their distribution is constant with $N$, implying as we discussed a linear trend of $G$, consistent with the observed one. 

The protein--protein \cite{ispolatov2005duplication} and Autonomous System networks, 
which are scale free networks, have the additional property of having many nodes sharing the same neighborhood which implies a large kernel of their adjacency matrix and thus a slower growth of $G$ compared to the other complex graphs, which are essentially full rank. 
More specifically, the low rank of the protein--protein interaction graph of Ispolatov, Krapivsky an Yuryev \cite{ispolatov2005duplication}, is explained by the duplication process at the heart of its generation, which, by definition duplicates the neighborhood of vertices. For the Autonomous System graph --- a model of internet by Elmokahfi, Kvalbein and Dovrolis \cite{elmokashfi2008scalability} ---,  many of the nodes are leafs (clients) connected
 to a few client providers. This results in a very low --- but still linearly increasing with $N$ --- rank, explaining why its squeezing cost is the lowest of the line in Fig.~\ref{GvNcx}. 

The anomalous superlinear trend of the ER model is actually a direct consequence of the same theory.
As we show in appendix \ref{app12}, this $N\log N$ scaling is due to the widening with $N$ of the Wigner semi-circular law
\cite{wigner1958distribution} followed by the ER graph.  

Now that we have characterized the cost of implementing Gaussian quantum networks, we will describe how to use them as a substrate to perform quantum communications.

\subsection{Routing entanglement}

Quantum entanglement is a paramount resource for quantum information purposes. In particular, bipartite entanglement represents the fundamental requirement that a shared quantum channel should have in order to enable a truly quantum teleportation. In the framework of Quantum Communications, the networks previously described can be seen as distributed Gaussian \textit{quantum teleportation networks} \cite{van2000multipartite}, where each pair of nodes can employ the pre-established quantum correlations together with \textit{Local Operations} and \textit{Classical Communications} \textit{LOCC} to teleport a Gaussian quantum state from one node to the other.

In our framework, if two nodes A and B need to teleport a quantum state, they can be helped by the other nodes in the network who will perform either a  quadrature measurement in one of the two complementary canonical variables $\hat q$ or $\hat p$  in order to increase the strength of the entanglement in the final pair. As detailed in Sec. \ref{sec:teleportation}, these two measurements have the effect of either removing the target node and its connections from the network or to wire shorten all its neighbors, respectively.

\subsubsection{Regular networks}
\label{sec:RegNet}

\begin{figure}[tb]  \centering
 \includegraphics[width=\linewidth]{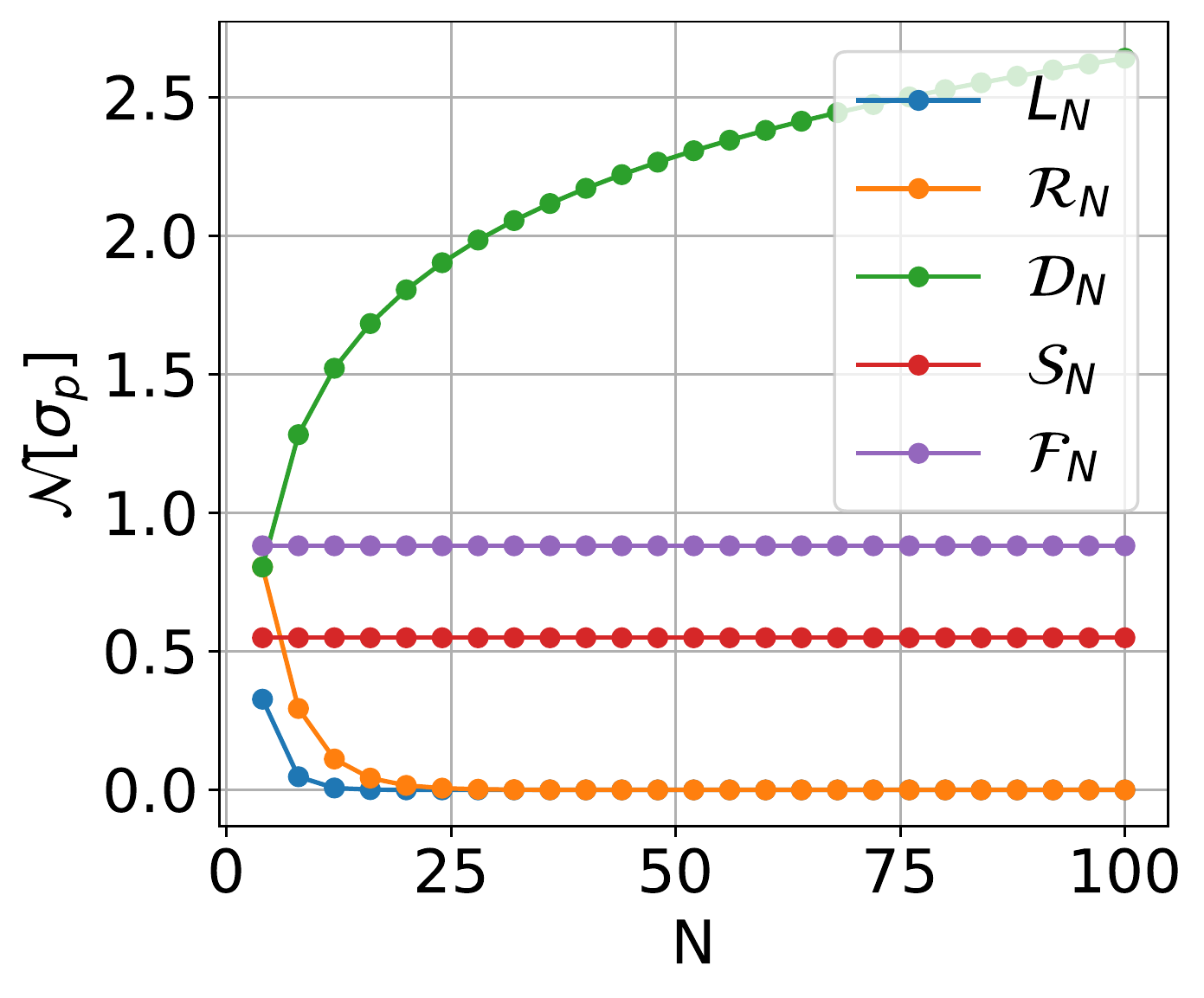}
 \caption{\footnotesize \raggedright Logarithmic negativity in the final two modes states after that all the other agents have locally measured their node for the regular topologies: linear $\mathcal{L}_N$, ring $\mathcal{R}_N$, star $\mathcal{S}_N$, diamond $\mathcal{D}_N$, fully connected $\mathcal{F}_N$ networks up to $N=100$ nodes.}\label{FvN}
\end{figure}

In  Fig.\ \ref{FvN} we compare the effect of different regular topologies of quantum networks with the purpose of distributing entanglement between two of the furthest nodes inside the network.

The decrease of entanglement with the size of the network seem to be typical in all configurations except the diamond graph, where all the central nodes are $\hat{p}$-measured. This behaviour  is quite counter-intuitive and might be expected to increase the fidelity of quantum communications. 
We show in Appendix \ref{app:ParallelEnhancement} the lowest symplectic eigenvalue of the partially transposed covariance matrix for this system goes like
\begin{equation}\label{eq:parallel}
    \left(\nu_-^{(\mathcal{D_N})}\right)^2=\frac{1}{1+2N Rg^2},
\end{equation}
Where $R=10^{s/10}$ is the inverse of the squeezing in $\hat p$, with squeezing factor $s$ in dB. 
Hence, the logarithmic negativity grows logarithmically with $NRg^2$ 
\begin{equation}
    \mathcal{N}^{(\mathcal{D_N})}=\log_2 (1 + 2NRg^2),
\end{equation}
and
the two modes become perfectly correlated in the limit of either infinite squeezing, infinite strength CZ-gate or infinitely many nodes in the network.

{}

\subsubsection{Complex networks}\label{sec:routingCX}

\begin{figure*}[tb]
\begin{center}
    \begin{tabular}{ccc}
         \includegraphics[width=0.3\linewidth]{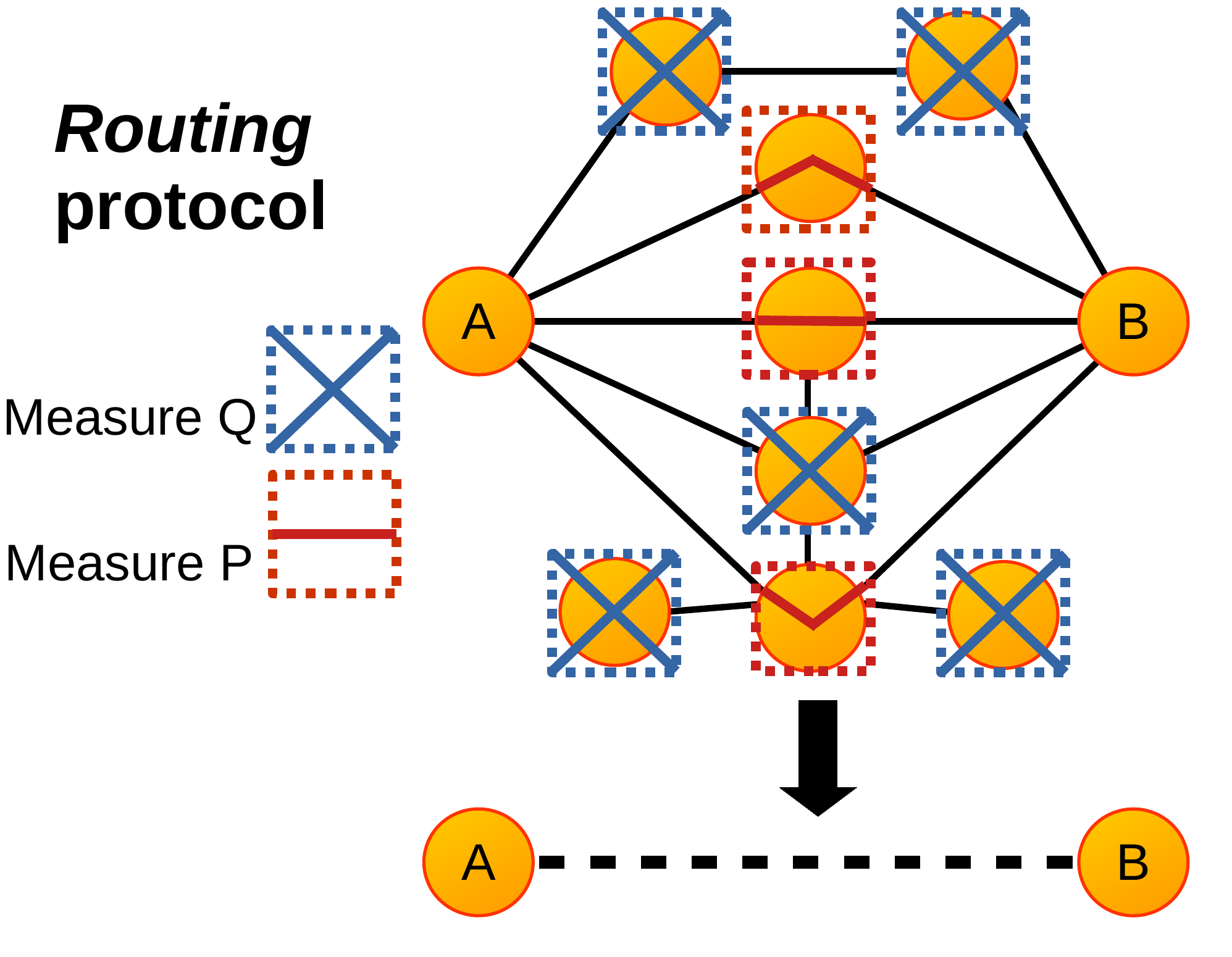}
 &           \includegraphics[width=0.3\linewidth]{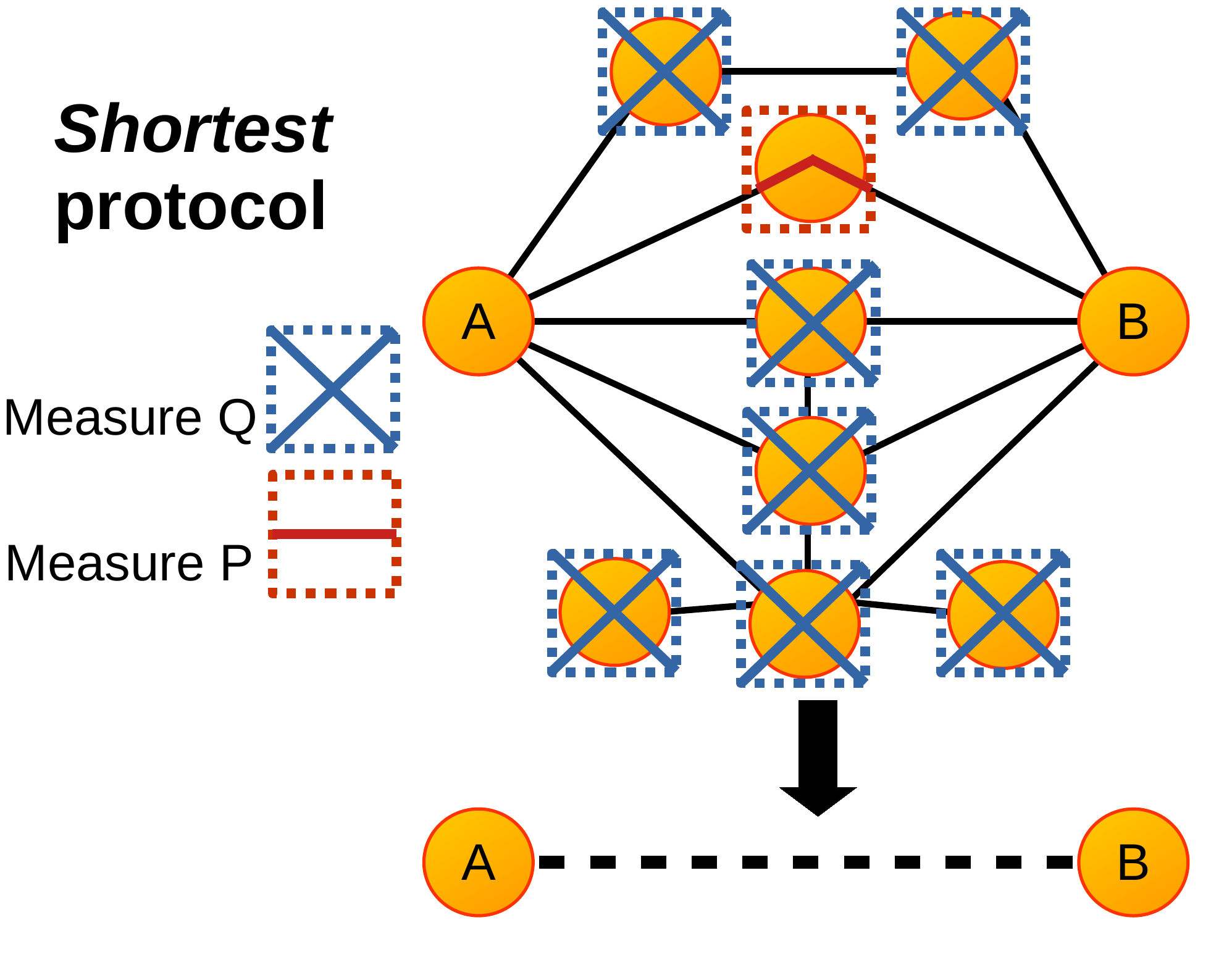}
 &         \includegraphics[width=0.3\linewidth]{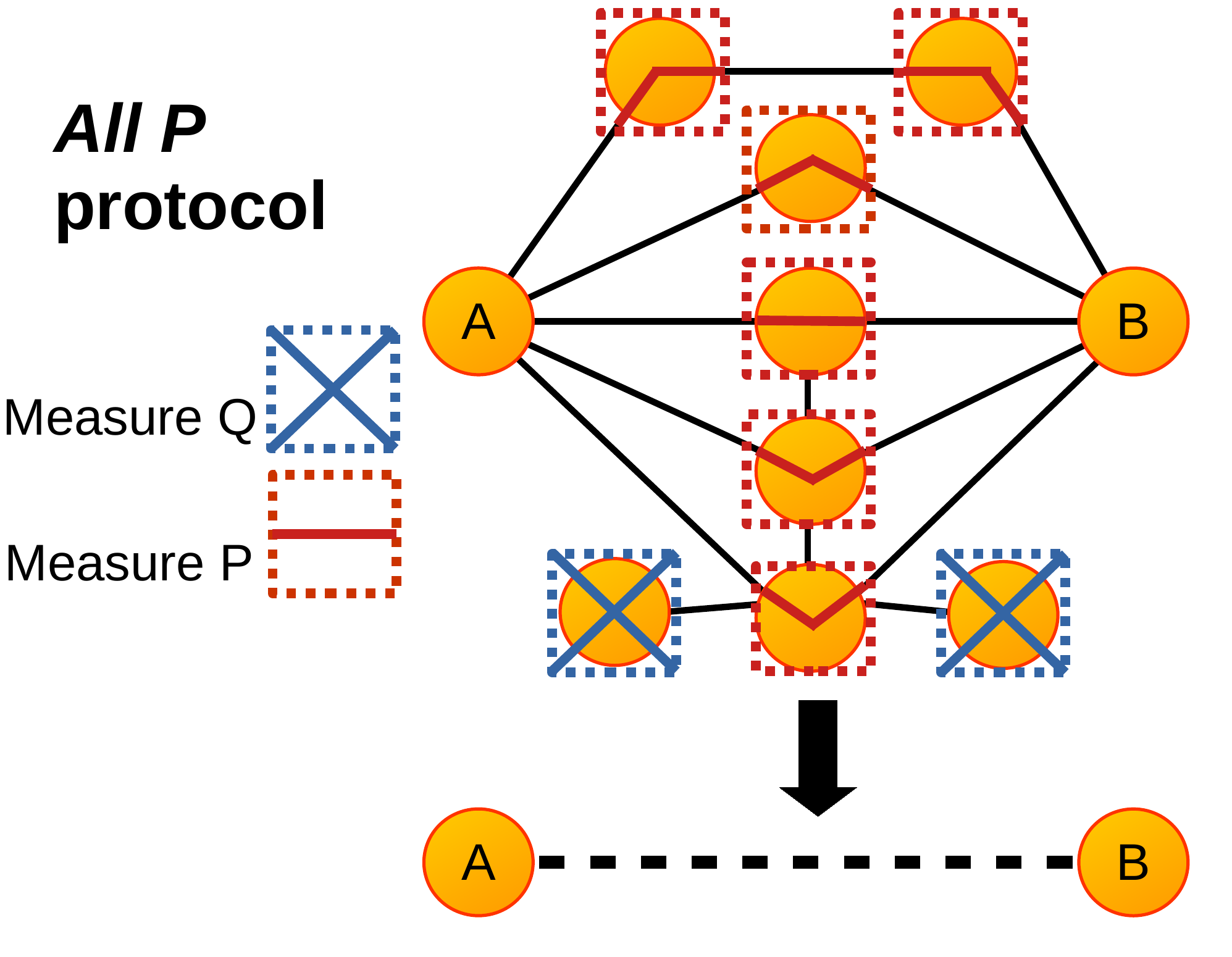}\\
 {\bf (a)} & {\bf (b)}  & {\bf (c)}

\end{tabular}
\end{center}
 \caption{\footnotesize \raggedright Scheme of the three protocols for the entanglement distribution: (a) the \textit{Routing} protocol takes a list of the shortest paths connecting A and B and measures in $\hat p$ those that increase the logarithmic negativity while the rest is measured in $\hat q$; (b) the \textit{Shortest} protocol only consider one of the shortest paths to be measured in $\hat p$ and the rest in measured in $\hat q$; (c) the \textit{All P} measures the nodes with only one connection in $\hat q$ and all the rest in $\hat p$. }\label{fig:protocols}
\end{figure*}

\begin{figure*}[tb]
     \includegraphics[width=\textwidth]{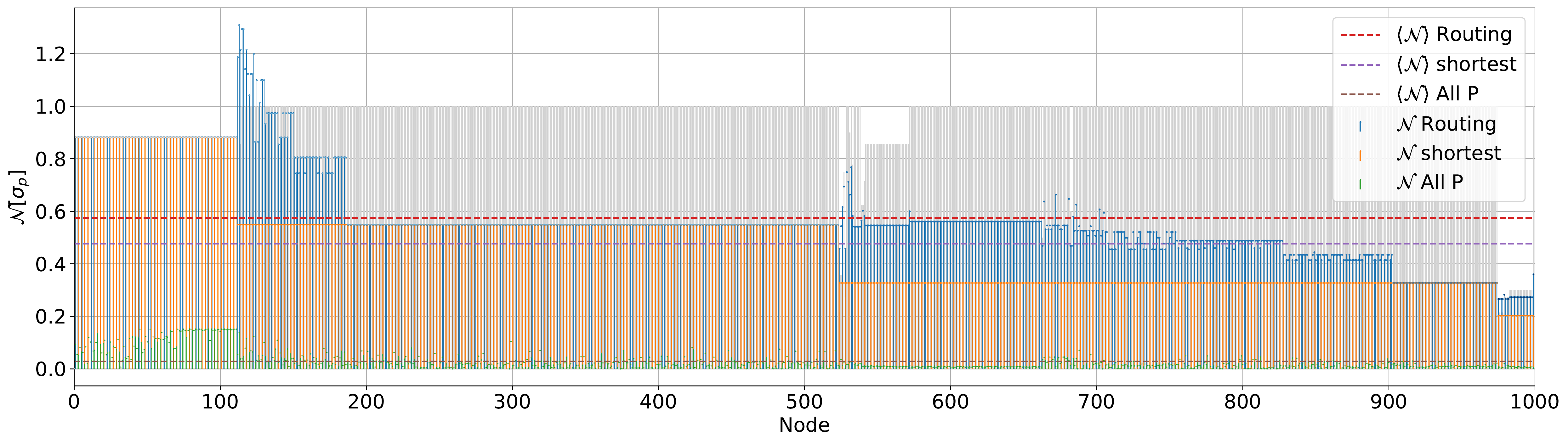}
 \caption{\footnotesize \raggedright Logarithmic negativity produced by the three different protocols applied to each node of the the Autonomous System $\mathcal{G}_{AS}(N=1000)$ network. The nodes are labeled in order of distance and of number of paths connecting to Alice. The blue, orange and green stems represent the logarithmic negativity of the final pair after the \textit{Routing}, \textit{Shortest} and \textit{All P} protocols respectively, while the dashed lines represent the mean value for all the nodes. The color of the marker indicates the distance of the node from A and the grey columns represent the ratio of paths that improved the entanglement in \textit{Routing}. }\label{fig:negRoute}
\end{figure*}

\begin{figure*}[tb]
     \includegraphics[width=0.6\linewidth]{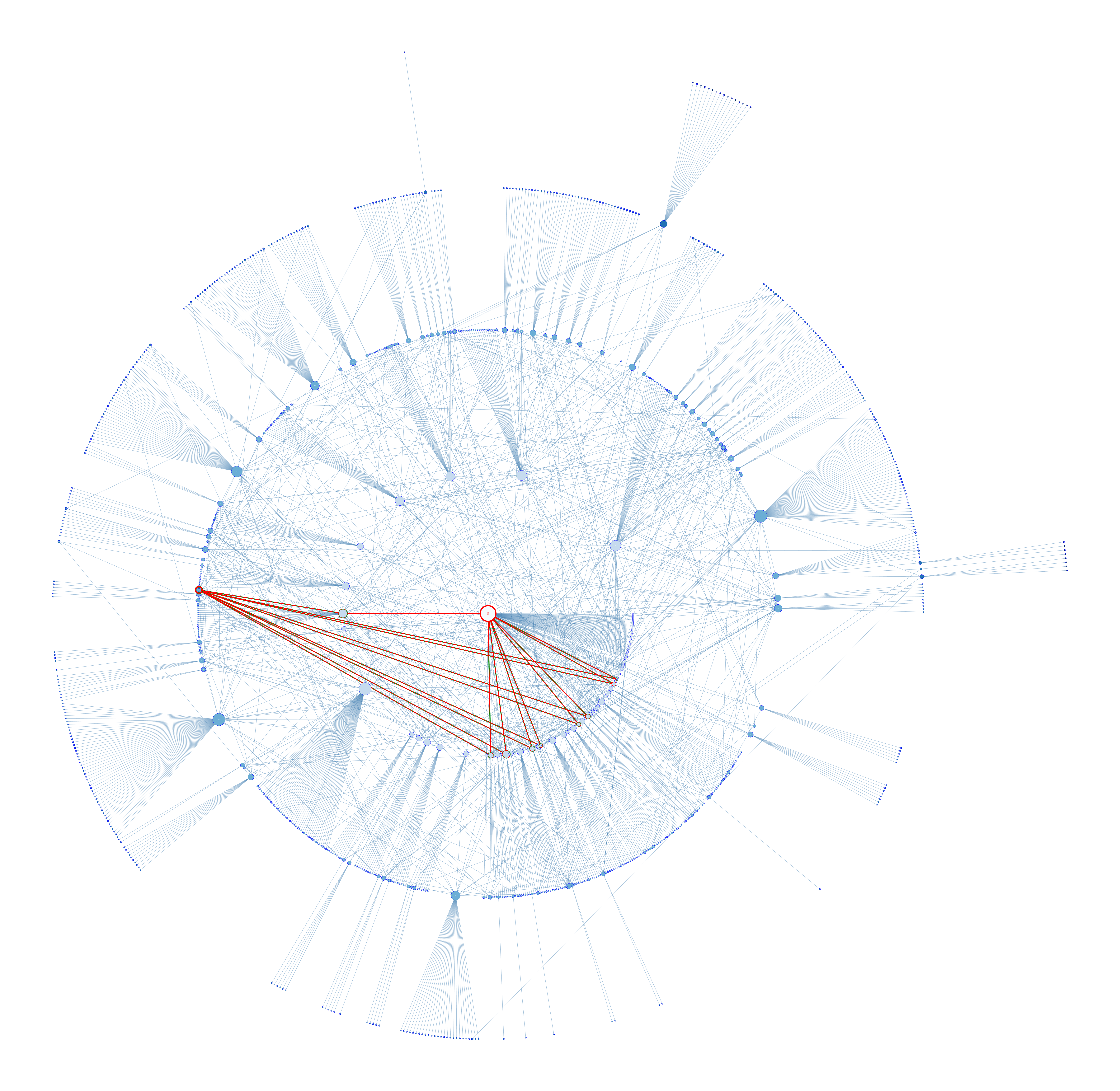}
          \includegraphics[width=0.39\linewidth]{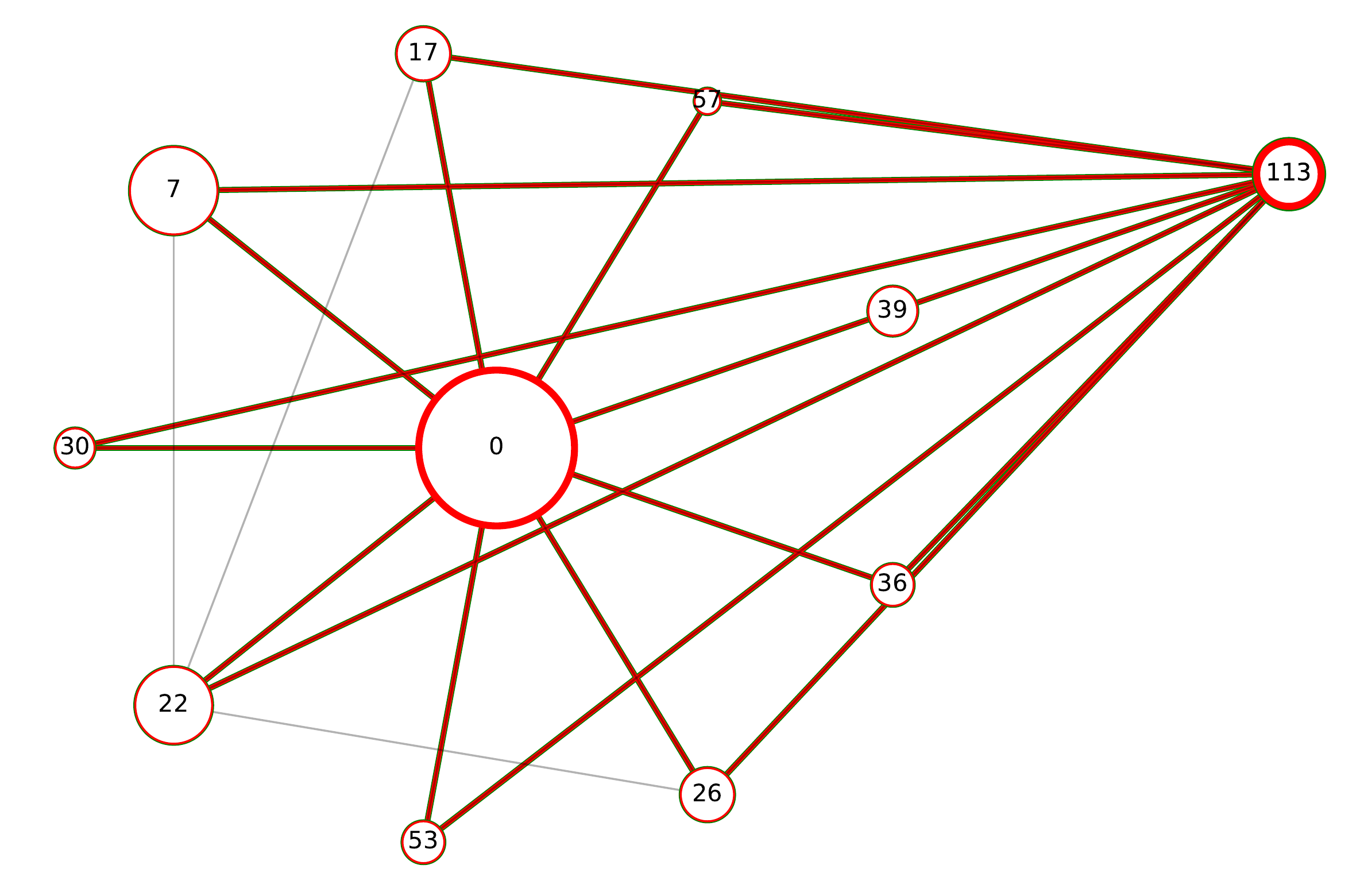}
 \caption{\footnotesize \raggedright Scheme of the $\mathcal{G}_{AS}(N=1000)$ network on which we performed the protocol and subgraph of the paths connecting to the node with highest logarithmic negativity. The nodes are set in circles according to their distance from Alice and their size is proportional to their degree.  }\label{fig:netRoute}
\end{figure*}

We present a naïve entanglement routing protocol that takes into account some of the properties studied in the previous section (notably, the parallel enhancement of entanglement) and we will apply it to complex topologies, to show that the enhancement of the entanglement with respect to the trivial protocol is, in principle, easily achievable. Imagine we have  a distributed network of entangled harmonic oscillators, where each node is honest and can perform classical communication and local homodyne measurement, and we want to establish an entangled pair between two nodes, Alice and Bob, that want to teleport a quantum state or perform QKD.  The trivial protocol --- called \textit{Shortest} in the following --- would be to find the shortest path between  them and measure in $\hat p$ all the qumodes along this path and in $\hat q$ all the others. A careful look at the inner structure of the network, however, might help us increase the strength of the correlation. For example if at any point, two nodes on the path are linked by multiple parallel routes, we can measure these in $\hat p$ to exploit the parallel enhancement of this Diamond-like subnetwork.

In order to show this in practice, we test the performances of three different routing protocols (shown in figure \ref{fig:protocols}) on various complex networks with the purpose of establishing a highly entangled pair \cite{Code}. We choose Alice to be one of the hubs of the graph and evaluate the efficiency of the protocol in delivering entanglement to all the other nodes.  The quantum protocol that we propose to exploit the parallel enhancement of entanglement will be simply called \textit{Routing}. 
\begin{itemize}
    \item \textit{Routing}: it takes as input the target node, Bob; it  lists all the shortest paths connecting it to Alice and measures all the nodes that are not in these paths in the $\hat q$ quadrature, so that they will not influence the protocol. Among the list of paths it checks one by one those to be measured in $\hat p$ in order to maximize the logarithmic negativity $\mathcal{N}$ of the final pair, while the rest will be measured in $\hat q$.
    \end{itemize}

In the \textit{Routing} protocol, in principle, we could have considered as well parallel paths of longer lengths  that might have contributed to improve the logarithmic negativity. However, in practice  the only observed effect was the slow down the performances while   the entanglement was not increasing for all the cases we considered.
The effect of the parallel paths can be appreciated when comparing the logarithmic negativity produced by \textit{Routing} with that produced by \textit{Shortest}.
\begin{itemize}
    \item \textit{Shortest}: the difference of the latter is that it only exploits one of the shortest parallel paths, directly measuring everything else in $\hat q$. 
\end{itemize}
In some cases the two protocols do not give a substantial difference, either because there are no parallel routes or because these do not help increasing the entanglement; however, in many instances the effects of parallel routing are significant. The last protocol we compare with is \textit{All P}.
\begin{itemize}
\item \textit{All P}: it measures  in $\hat q$ all the terminal nodes of degree 1 --- the leafs --- and the rest in $\hat p$.
\end{itemize}
 This protocol is  less effective than the first two but is always the quickest to simulate, whereas \textit{Routing} can be computationally very slow on regular networks, which are characterized by long distances and many parallel paths, but becomes very efficient on complex sparse networks. The following simulations can be reproduced using our python code, available at \cite{Code}.

One instance of this program is given in figure \ref{fig:negRoute} which shows the logarithmic negativity provided by the three different protocols for each node of an Autonomous System  network \cite{elmokashfi2008scalability} with 1000 nodes $\mathcal{G}_{AS}(N=1000)$.  At the beginning of the protocol, we pick Alice as the node, or one of the nodes, with the highest degree. The nodes are then sorted by their distance from Alice and, for the same distance, by the number of all the shortest paths connecting them to Alice.  Additionally, the grey column represents the fraction of parallel paths  useful to increase the entanglement. 
Notice that nodes at distance~1 cannot show any difference between the \textit{Routing} and the \textit{Shortest} protocols, however many nodes at distance~2 present a greater logarithmic negativity than those at shorter distance after the \textit{Routing}. This feature of quantum communication networks, e.g.\ that two nodes can benefit improved communication if they are at larger distance thanks to the parallel enhancement of entanglement, has no classical equivalent.

In figure  \ref{fig:netRoute} we show the graph of the network, where the nodes are again sorted by distance and number of parallel paths and the size of each node is proportional to its degree.  In this figure Alice is ‘0’ and has a thick red contour. The node with highest logarithmic negativity and all the paths that improved its entanglement are highlighted with red thick lines. 

Appendix \ref{App:RoutingComplex} presents the results of the same analysis 
we applied to other networks differing in size and topology, and for which we obtained different results. 

\subsection{Discussion}

 Our model aims at reproducing the existing photonic platforms 
 \cite{cai2017multimode, yokoyama2013ultra, larsen2019deterministic, madsen2022quantum}
 with realistic experimental constraints, such as limited amount of squeezing, but without taking into account propagation losses. This work is focused on the capabilities of pure CV quantum states to act as quantum networks and the resources needed for their generation. Generation losses can be very low, so that the hypothesis of pure states is a realistic one, whereas propagation losses can be mitigated by considering local (short distances) networks, and their effect on long distances will be included in future works. At the same time we probe  the capabilities of photonic platforms while the scaling of the network increases beyond the capacities allowed by the state-of-the-art technology. Since the main limitation to build these optical systems is finite squeezing,  we provide in section \ref{sec:cost} an analytical formula that allows to compute the amount of squeezing in each mode as a function of the spectrum of the underlying graph. We show the graph states' cost as a global measure of squeezing and number of squeezed modes that are necessary to build the network.  In particular we explore the cost of different networks topologies in term of number of needed squeezers at fixed number of nodes in the networks and of the global amount of squeezing.

 Although in entangled qubits networks the resource usage is always proportional to the number of links, we show that in CV Gaussian networks the trend of the squeezing cost  presents non-trivial scaling with the size of the network  and this is strictly dependent on its topology. We present as well a few instances of the full squeezing spectra --- i.e.\ the needed amount of squeezers with the required squeezing values --- of regular and complex networks, showing that some topologies are equivalent up to a linear optical transformation.

We then explore their potentialities to perform efficient quantum communications between two arbitrary nodes when assisted with a given class of Gaussian Local Operations and Classical Communication (GLOCC) by all the agents in the network.
A typical approach of quantum networking and routing consists in distributing  photonics states like single-photons, Bell pairs or Gaussian states and then use synchronous local operations that build the wanted entanglement structure between the agents
\cite{Leone21,Pant19,Meignant19,Meignant21,Zhang21,zhuang2020quantum,omar2019quantum,cuquet2011limited,harney2022analytical,solomons2022scalable}.  We remark that our type of communication quantum networks is inherently different from the typical qubits networks that are currently being deployed in different metropolitan areas \cite{joshi2020trusted}. In those cases, for example, each entanglement link is pairwise between two qubits and as a consequence each node of the network will have to receive, storage and measure as many quantum states as it has neighbors. Conversely, in a Gaussian quantum network the same qumode can be entangled with an arbitrary number of other nodes. Moreover, the production of such states, their manipulation to increase the entanglement among two nodes and their measurement to perform quantum teleportation can be achieved deterministically, unlike the discrete variables case. 
Nonetheless, qubits networks have been extensively studied over the last years, whereas Gaussian teleportation networks is a very recent emerging field. Our purpose is, thus, not to prove the superiority of the latter, but rather to explore its properties and the  differences from the DV schemes in order to get the best of both worlds.

We  consider the case where a preexisting CV multipartite entangled state is distributed among the players and then local operations reconfigure the entanglement connections, similarly to some protocol in the DV case \cite{Hahn19}. 
The choice is motivated by the fact that multimode entangled states can be directly generated via optical platform \cite{asavanant2019generation,larsen2019deterministic,roslund2014wavelength,chen2014experimental,yokoyama2013ultra} and their shape can be easily manipulated \cite{cai2017multimode,nokkala2018reconfigurable}. We  show, both analytically and numerically that, in  specific topologies, homodyne measurements along parallel paths allow to increase the entanglement between two distant nodes.

We discovered some outstanding properties of Gaussian networks. The most important is the parallel enhancement of entanglement in the diamond graph. If properly used, this feature can most certainly improve the routing of entanglement in regular and complex shaped networks.  On these grounds, the search for an optimal protocol that exploits all the qualities of these Gaussian networks is very desirable yet arduous, and will be subject of future investigations. 

Finally, we devise a routing protocol based on local quadrature measurements for reshaping  the network in order to perform a teleportation protocol between two arbitrary nodes of the networks. The \textit{Routing} protocol, which is based on wire-shortening over parallel paths among the nodes, improves the final entanglement between the two nodes in a considerable amount of cases, and it is particularly efficient in running-time for complex sparse networks.

 \section{Methods}

\subsection{Gaussian quantum states}\label{sec:gqs}

The generation of continuous variables multimode entangled states has been demonstrated in several optical setups. In such experiments we recover networks structures as naturally appearing entanglement correlations \cite{roslund2014wavelength}, reconfigurable Gaussian interactions  \cite{nokkala2018reconfigurable}, or imprinted cluster states \cite{yonezawa2004demonstration,yokoyama2013ultra, asavanant2019generation,larsen2019deterministic,cai2017multimode}.

These quantum states produced via parametric processes and linear optical transformations are characterized by Gaussian statistical distribution of the quadratures of the involved optical modes \cite{fabre2020modes}. The quadratures $\hat{q}_j$ and $\hat{p}_j$ of the $j^{\text{th}}$ mode are canonical conjugate variables, such that $[\hat{q}_j,\hat{p}_k]=i \delta_{j,k}$, associated to the quantum harmonic oscillator describing the light mode. In this work we adopt the following relation  with  creation and annihilation operators
$\mathbf{\hat{a}^{\dagger}}=(\mathbf{\hat{q}}-i\mathbf{\hat{p}})/\sqrt{2}$ and $\mathbf{\hat{a}}=(\mathbf{\hat{q}}+i\mathbf{\hat{p}})/\sqrt{2}$, such that the variance of the vacuum quadratures is normalized to $1/2$.



The produced states can then be completely characterized by the first two moments of the quadratures $\mathbf{\Bar{r}}=\Tr[\rho \mathbf{\hat{r}}]$ and $\mathbf{\sigma}=\Tr[\rho \{(\mathbf{\hat{r}}-\mathbf{\Bar{r}}),(\mathbf{\hat{r}}-\mathbf{\Bar{r}})^{T}\}]$, where $\rho$ is the density matrix of the Gaussian state and $\mathbf{\hat{r}}=( \hat{q}_1 ... \hat{q}_N, \hat{p}_1 ... \hat{p}_N )$ --- we follow here $qp$-ordering.

Parametric processes are described by quadratic Hamiltonians $\hat{\mathcal{H}}_I=\mathbf{\hat r} H \mathbf{\hat r}^T$, whose dynamics is implemented on the quadratures by $S_H=e^{\Omega H t}$, as
\begin{equation}\label{symp}
    \mathbf{\hat{r}}^{\prime}=S_H \mathbf{\hat{r}}_0
\end{equation}
where $\mathbf{\hat{r}}_0$ are quadratures of the initial state, $ \mathbf{\hat{r}}^{\prime}$ are the quadratures of the final state and $\Omega=\left(\begin{smallmatrix} 0 & \mathbbm{1}\\-\mathbbm{1}&0\end{smallmatrix}\right)$  is  a $2N\times 2N$ skew-symmetric matrix associated to the $N$ dimensional Hilbert space allowing to write the commutation relation of the canonical variables as \[
[\mathbf{\hat{r}},\mathbf{\hat{r}}^T]=i\Omega=i
\begin{pmatrix}
0 & \mathbbm{1}\\
-\mathbbm{1} & 0
\end{pmatrix}.
\] 

Since any pure Gaussian state can be obtained by the application of a unitary generated by a quadratic Hamiltonian $H$ to the vacuum, the most general pure Gaussian state covariance matrix is given by applying $S_H$ by congruence to the vacuum covariance matrix $\sigma_0=\mathbbm{1}/2$:
\begin{equation}
    \mathbf \sigma= S_H \mathbf \sigma_0 S_H^T= \frac{S_H S_H^T}{2}
\end{equation}

Singular value decomposition allows one to write the symplectic transformation in the so called Bloch-Messiah decomposition \cite{fabre2020modes} as a product of an orthogonal, a diagonal and an orthogonal matrices $S_H=O \Delta O^\prime$, which can be interpreted as a basis rotation, a squeezing in the diagonal basis and another rotation. The mode-basis in which the covariance matrix is diagonal and each component is independently squeezed is named the supermode basis. In \cite{cai2017multimode,roslund2014wavelength} where the pump and the phase matching function can be described by a Gaussian spectral profile, the supermode basis corresponds to  Hermite-Gauss spectral modes.
The squeezing values of $\Delta$ can be derived from the eigenvalues of the Hamiltonian $\hat{\mathcal{H}}_I$, while the orthogonal matrix $O$ can be interpreted as a measurement basis change or, equivalently, as a passive linear optical transformation. The other orthogonal matrix $O^\prime$ is simplified in the product $S_H S_H^T$ and can be disregarded:
\begin{equation}\label{covSym}
    \mathbf \sigma=  \frac{S_H S_H^T}{2}=\tfrac{1}{2}O\Delta^2O^T.
\end{equation}

The diagonal matrix $\Delta$ contains the information on the minimum number of squeezed modes in the system and their value of squeezing, which will later be used in the chosen resource theory. 
If we consider a single mode field, the squeezing operation is defined as a Gaussian transformation that reduces the variance of $\hat p$ by a factor $10^{-s/10}$, where $s$, measured in $dB$ throughout this article, is called \textit{squeezing factor}. Squeezing is represented by the local symplectic matrix
    \[
    S_{sq}(s)=\begin{pmatrix}
    10^{s/20} & 0\\ 0 & 10^{-s/20}
    \end{pmatrix}.
    \]

The multimode $\Delta$ matrix can then be written as  
\begin{multline}\label{eq:diagonalSq}
  \Delta= \diag\{10^{s_1/20},10^{s_2/20},...10^{s_N/20}, \\ 10^{-s_1/20},10^{-s_2/20},...10^{-s_N/20}  \} .
\end{multline}

This formalism can be used to visualize and manipulate Gaussian quantum states, that are readily available in most well-equipped photonics laboratories and, although the number of modes and their connections is still in large part limited, many efforts are employed to improve the capacities of these systems.

Targeted Gaussian quantum states, including the quantum networks of the next section, can be generated via the two following strategies: i) by tailoring Hamiltonians $\hat{\mathcal{H}}_I$ of multimode parametric processes  in order to get the decomposition of eq.\ (\ref{covSym}) corresponding to the desired covariance matrix \cite{cai2017multimode,chen2014experimental,roman2021continuous,christ2014theory,arzani2018versatile,gouzien2020morphing};  ii) by getting a number of single-mode squeezers equal to the the number of elements with $s_j \neq 0$ of $\Delta$ in eq.\ (\ref{covSym}) and producing the corresponding $s_j$ squeezed states, that are injected in a linear optic intereferometer corresponding to the orthogonal matrix $O$ in eq.\ (\ref{covSym}) \cite{yukawa2008experimental,barral2020versatile,arrazola2021quantum}.

\subsection{Graph states as quantum networks}\label{sec: Gra}

The above formalism can be employed to describe Gaussian graph states, that can be used as CV quantum networks. 
We at first recall that a network is mathematically described by a graph $\mathcal{G}(V,E)$, which is a set of vertices $V$ (or nodes) connected by a set of edges $E$. Labeling the nodes of the graph in some arbitrary order,
we can define a symmetric adjacency matrix $A = A^T $ whose
$(j,k)^{\text{th}}$ entry $A_{jk}$ is equal to the weight of the edge linking
node $j$ to node $k$ (with no edge corresponding to a weight of 0). 
Typically, the adjacency matrix is enough to completely characterize a graph, however we will see that in our case there are other degrees of freedom such as the squeezing of a node and its angle.

We can now describe the quantum networks we use in this work that are called graph- or cluster-states \footnote{While usually the name cluster is used when the graph shape allows for universal quantum computing, in this work we will use the terms cluster state and graph state as synonyms.} \cite{Menicucci06,menicucci2011graphical,gu2009quantum}.
Theoretically, they can be built by entangling a number of squeezed modes of light via  CZ-gates, which is a Gaussian operation implementing a correlation of strength $g$ between the $\hat q$ and the $\hat p$ of the two modes on which it acts. The corresponding symplectic matrix is
    \[
    S_{C_Z}(g)=\begin{pmatrix}
    1 & 0 & 0 & 0\\ 0 & 1 & 0 & 0\\ 0 & g & 1 & 0\\ g & 0 & 0 & 1
    \end{pmatrix}{}
    \]

The graph associated to the graph states identify edges as CZ-gates applied between nodes, that are the squeezed modes,  weighted with $g$.

In order to simplify the many degrees of freedom present in our networks, for the moment we shall assume that all the nodes will be squeezed in $\hat p$ by $s$ and all the edges have a correlation strength of $g$. 
If we apply a CZ-gate network with adjacency matrix $A$ to a multimode squeezed vacuum $\sigma_s$, with squeezing factor $s$ we obtain a Gaussian network with covariance matrix \cite{Wal21}
\begin{align}\label{eq:V}
    \sigma = \begin{pmatrix}\sigma_{qq} & \sigma_{qp} \\ \sigma_{pq} & \sigma_{pp}\end{pmatrix} &=  \begin{pmatrix}\mathds{1} & 0 \\  A & \mathds{1}\end{pmatrix} \sigma_s \begin{pmatrix}\mathds{1} &  A \\ 0 & \mathds{1}\end{pmatrix} \notag \\ 
    &=\begin{pmatrix}R\mathds{1} & R  A \\ R  A & R  A^2+\mathds{1}/R\end{pmatrix}\end{align}
Where $R=10^{s/10}$. 
The $2N \times 2N$ covariance matrix $\sigma$ is divided in four $N \times N$ blocks, where the blocks $\sigma_{qq}$ and $\sigma_{pp}$ represent the correlations among the different nodes' $q$- and $p$-quadratures, respectively, whereas the blocks $\sigma_{qp}$ and $\sigma_{pq}$ describes the correlations between $q$- and $p$-quadratures.

Bear in mind that the \textit{CZ-gate} operations that theoretically identify the edges of the networks 
are  seldom realized in any laboratory being very challenging to accomplish.  What is commonly done, as explained in the previous subsubsection, is the reduction of the covariance matrix of the graph state in (\ref{eq:V}) to the form of eq.\ (\ref{covSym}), that is also a recipe for bulding the graph states from a certain number of squeezed modes ($\Delta$) and linear optics transformations ($O$).
 
\subsection{Networks}
\subsubsection{Regular topologies}
Regular networks are generated through a deterministic mechanism.
We considered the following topologies:

\begin{itemize}

\item The \textit{linear graph} $\mathcal{L}_N$, with $N$ nodes and $N-1$ edges, is accomplished by connecting each node in series to the next.
\begin{center}
    \raisebox{-.5\height}{$\mathcal{L}_5$:} \raisebox{-.5\height}{\includegraphics[scale=0.2]{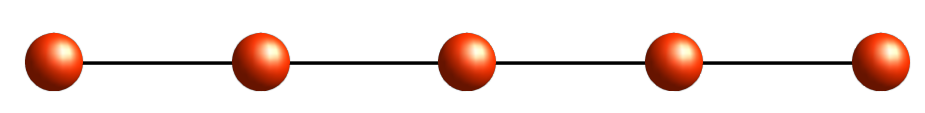}}
\end{center}

\item The \textit{ring graph} $\mathcal{R}_N$, with $N$ nodes and $N$ edges, is a linear graph with a closed loop.
\begin{center}
    \raisebox{-.5\height}{$\mathcal{R}_5$:} \raisebox{-.5\height}{\includegraphics[scale=0.2]{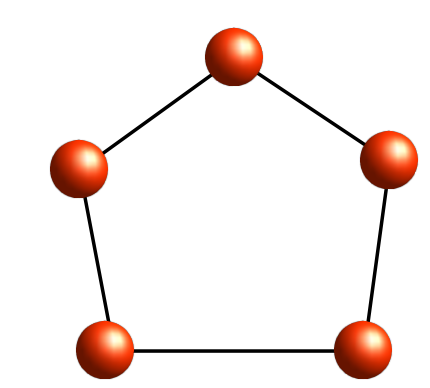}}
\end{center}
\item In the \textit{star graph} topology $\mathcal{S}_N$, with $N$ nodes and $N-1$ edges, every peripheral node is linked to a central node, called hub.
\begin{center}
    \raisebox{-.5\height}{$\mathcal{S}_5$:} \raisebox{-.5\height}{\includegraphics[scale=0.2]{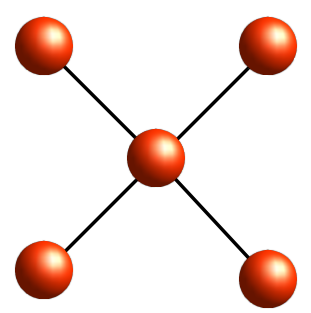}}
\end{center}
\item The \textit{diamond graph} $\mathcal{D}_N$, with $N$ nodes and $2(N-2)$ edges, has 2 hubs, each linked to all the $N-2$ central nodes of the network. It is isomorphic to the complete bipartite graph in which one of the subset has $2$ nodes and the other has $N-2$ nodes.
\begin{center}
    \raisebox{-.5\height}{$\mathcal{D}_5$:} \raisebox{-.5\height}{\includegraphics[scale=0.2]{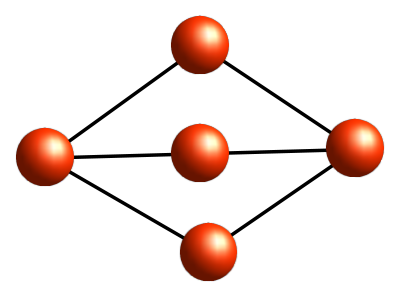}}
\end{center}
\item In the \textit{complete} (or \textit{fully connected}) \textit{graph} $\mathcal{F}_N$, with $N$ nodes and $\tfrac{N(N-1)}{2}$ edges, all nodes are interconnected.
\begin{center}
    \raisebox{-.5\height}{$\mathcal{F}_5$:} \raisebox{-.5\height}{\includegraphics[scale=0.2]{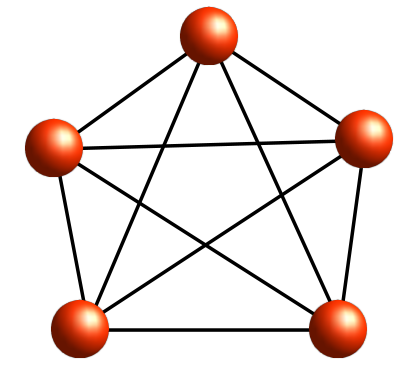}}
\end{center}

\end{itemize}{}

The experimental squeezing cost of these graphs was studied in \ref{sec:costReg}.

Notice that there is as well a simple relation between the adjacency matrix
$A$ of a graph state 
and the mean energy difference between the state and the vacuum. 
In fact, assuming that the energy $E$ of the state is the mean value of the harmonic oscillator Hamiltonian on the state
\begin{equation}
    E=\langle \hat{H}_{HO}\rangle= \frac{\hbar}{2}\sum_i  \omega_i(\langle \hat{x}_i^2 \rangle+\langle \hat{p}_i^2 \rangle)=\frac{\hbar\omega}{2} \text{Tr}(\sigma),
\end{equation}
where we also assumed that all the modes have the same  frequency  $\omega_i=\omega$. From this we can write
\begin{equation}
    \Delta E= E- E_0= \frac{\hbar\omega}{2}\text{Tr}(\sigma-\sigma_0)=\frac{\hbar\omega}{2}\text{Tr}(A^2).
\end{equation}

The quantity on the right hand-side is proportional to the  second moment of the eigenvalues distribution and it sets a fundamental lower bound on the energy necessary to implement such states. From a topological point of view, the trace of the $n$th power of the adjacency matrix equals the number of closed loops of length $n$ on the graph \cite{brouwer2011spectra}. Thus, for $n=2$, it corresponds to the number of edges in the network and implies that each independent application of the CZ-gate adds the same amount of energy. 
A resource scaling that is linear with the number of edges in the network is typical of DV networks (i.e.\ DV graph states), where each new edge requires a new Bell pair. The scaling of the squeezing in CV networks with the size of the network, as we have seen, depends instead  on the structure of the underlying graph and can be non-linear with the number of edges.
We want to stress here that squeezing --- and not energy --- is the technologically not-trivial enabling resource to be implemented for building the CV networks.  
If we look again at the example of the star and the diamond graph,  the number of edges in the two graphs are different, hence they would have different energies. As they have the same number of squeezed mode,  we can transform one into the other with passive optics, e.g.\ without spending energy. However, it has to be clarified  that the two main squeezed modes for the diamond have larger eigenvalues than the star network. Thus, if we transform a star graph into a diamond with linear optics, it would be equivalent to a graph state obtained by the application of CZ-gate with a weaker coupling $g$. But again this behaviour is totally accounted by the squeezing cost, which is the relevant quantity in designing CV networks. 

\subsubsection{Complex topologies}

\begin{figure*}[htb]
\begin{tabular}{ccccc}
    \includegraphics[width=0.18\linewidth]{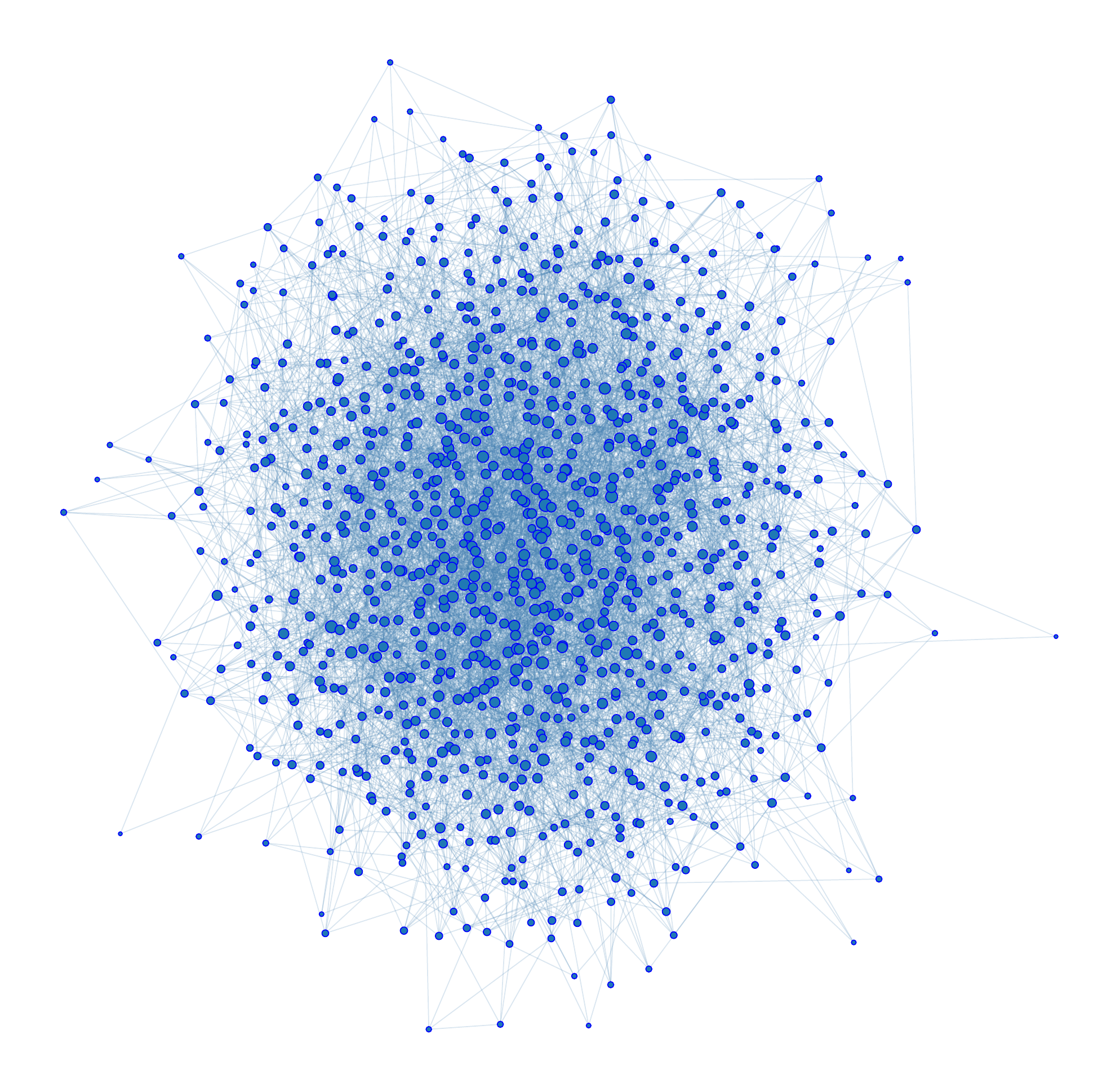}
     &     \includegraphics[width=0.18\linewidth]{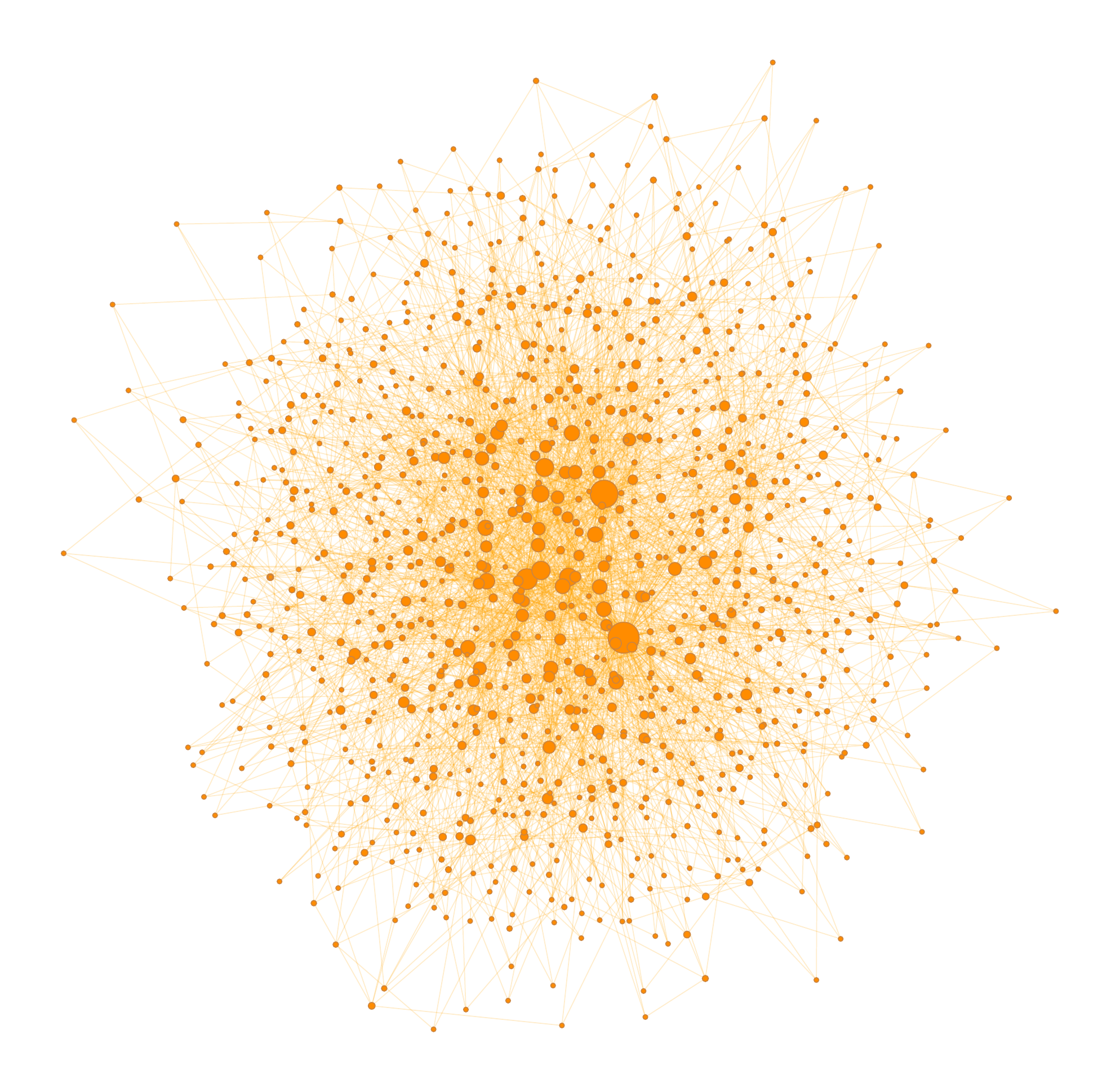}
 &       \includegraphics[width=0.18\linewidth]{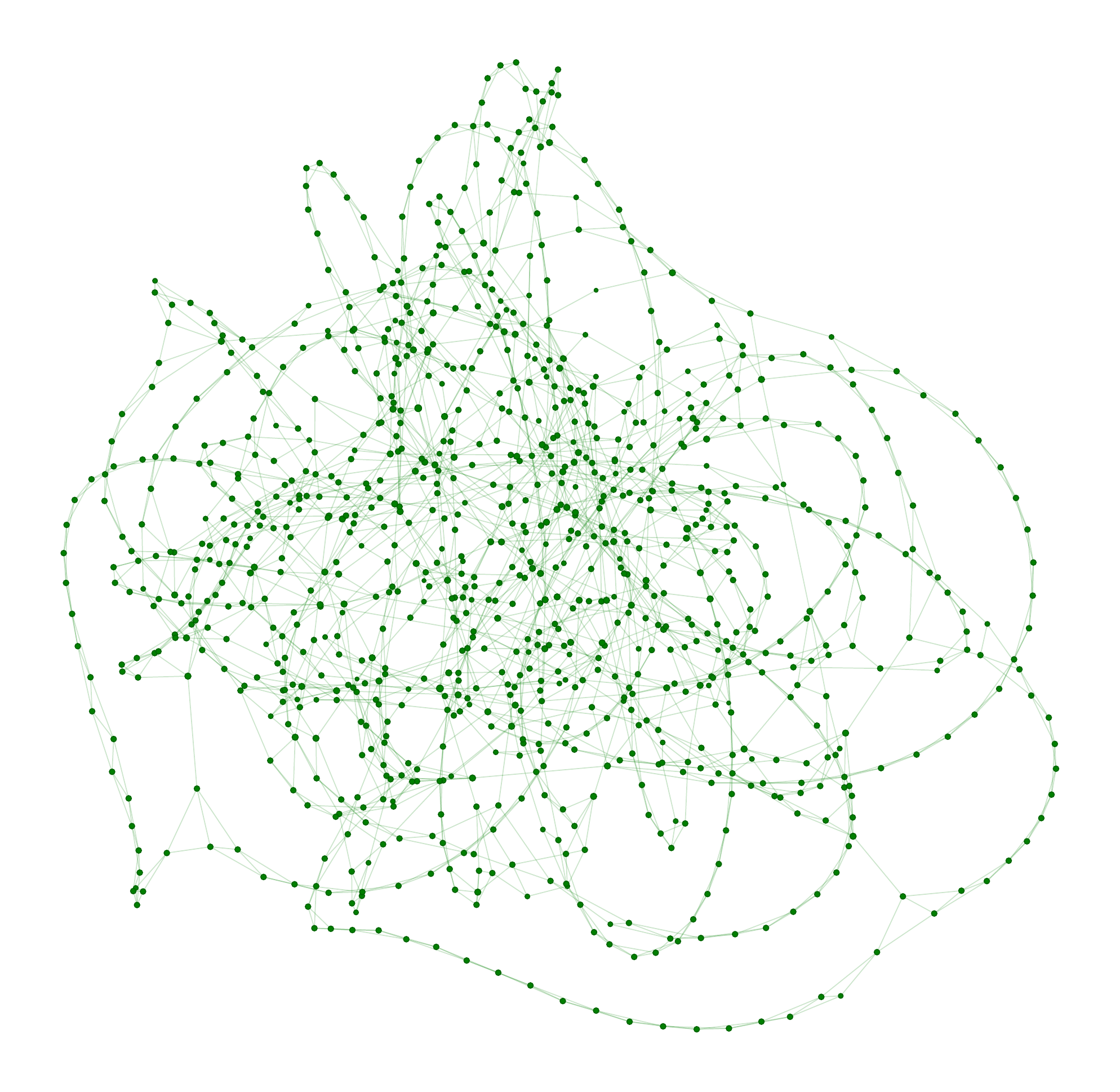}
&      \includegraphics[width=0.18\linewidth]{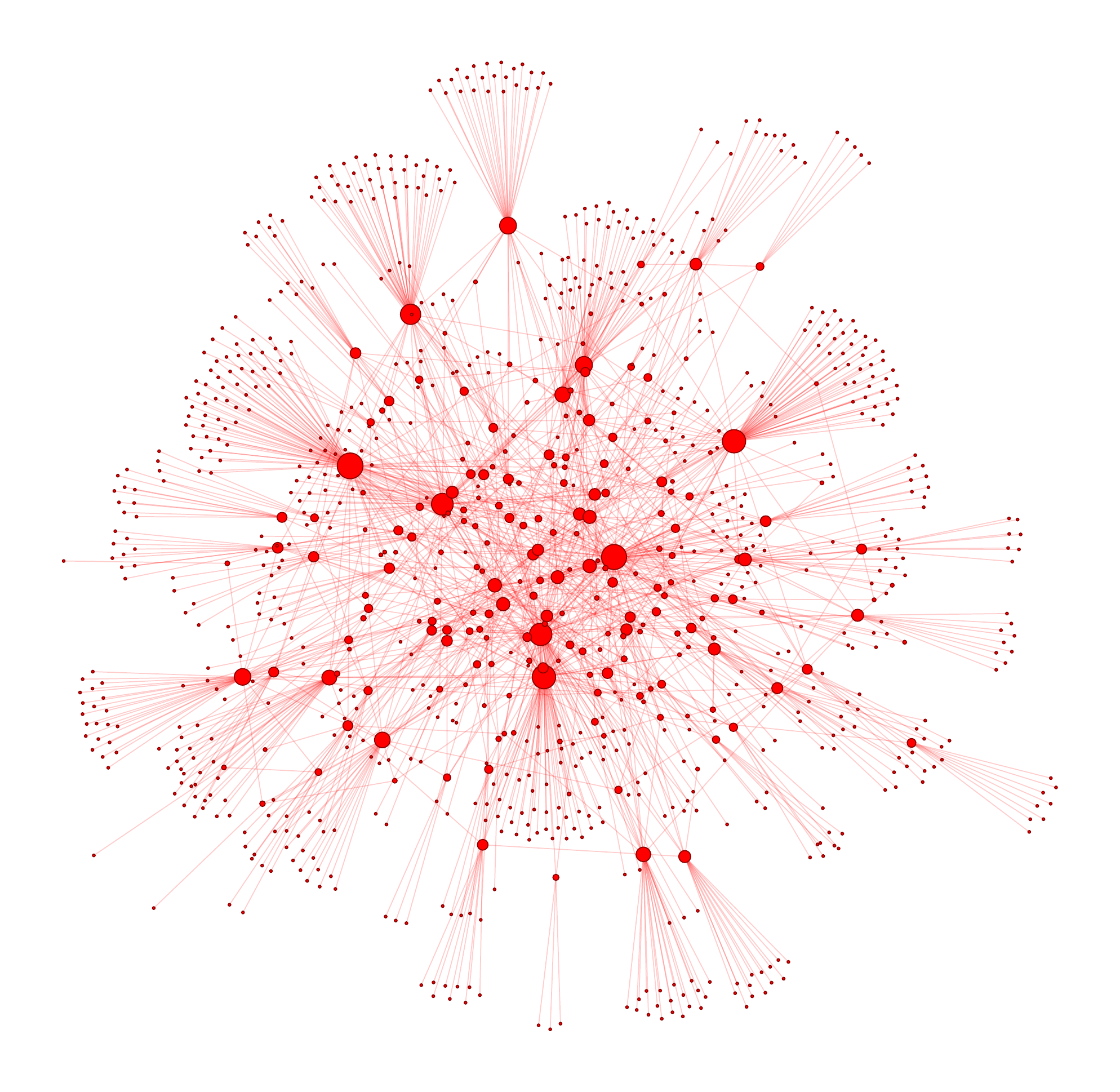}
&      \includegraphics[width=0.18\linewidth]{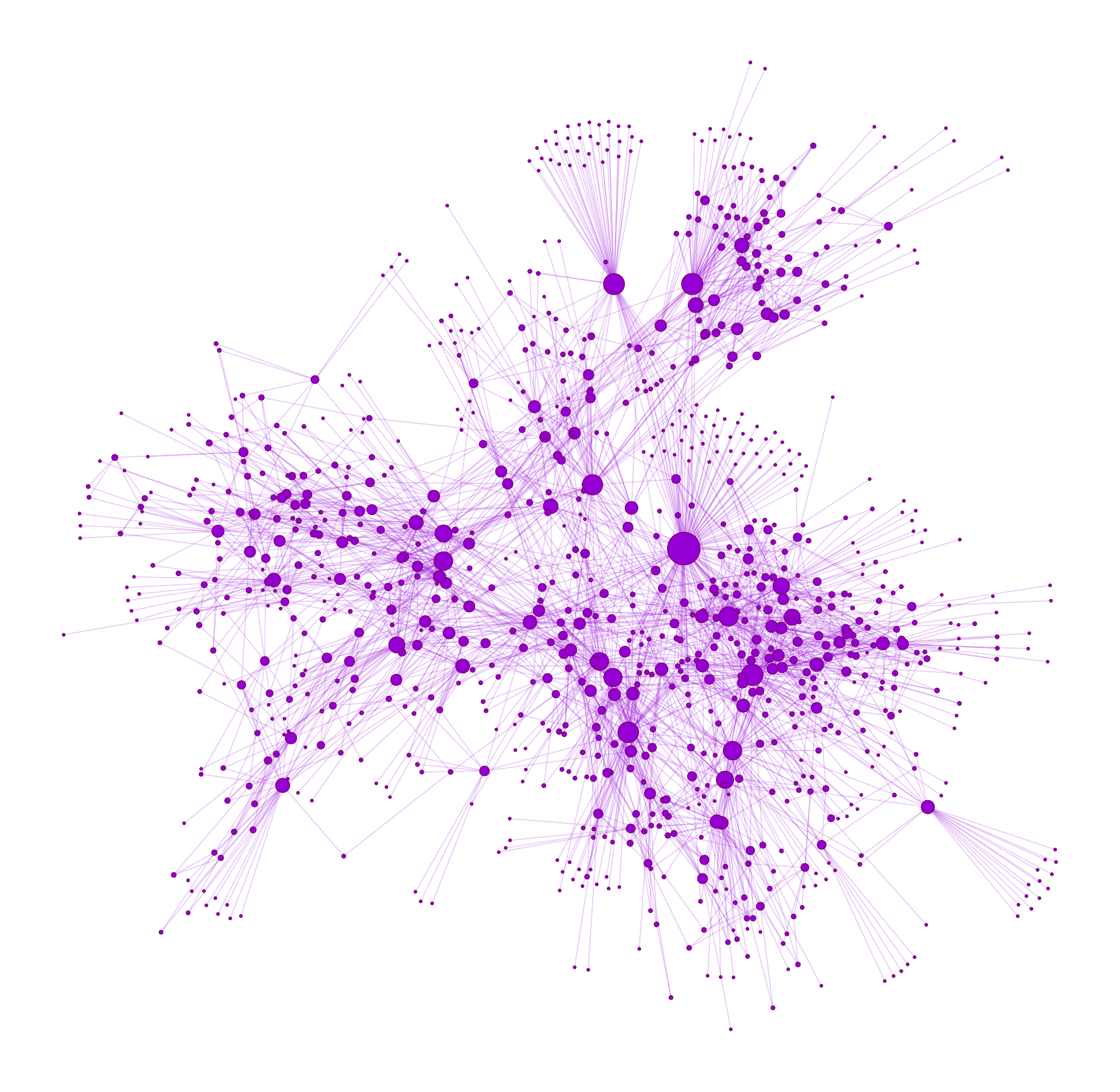}
\\[6pt]
    \includegraphics[width=0.18\linewidth]{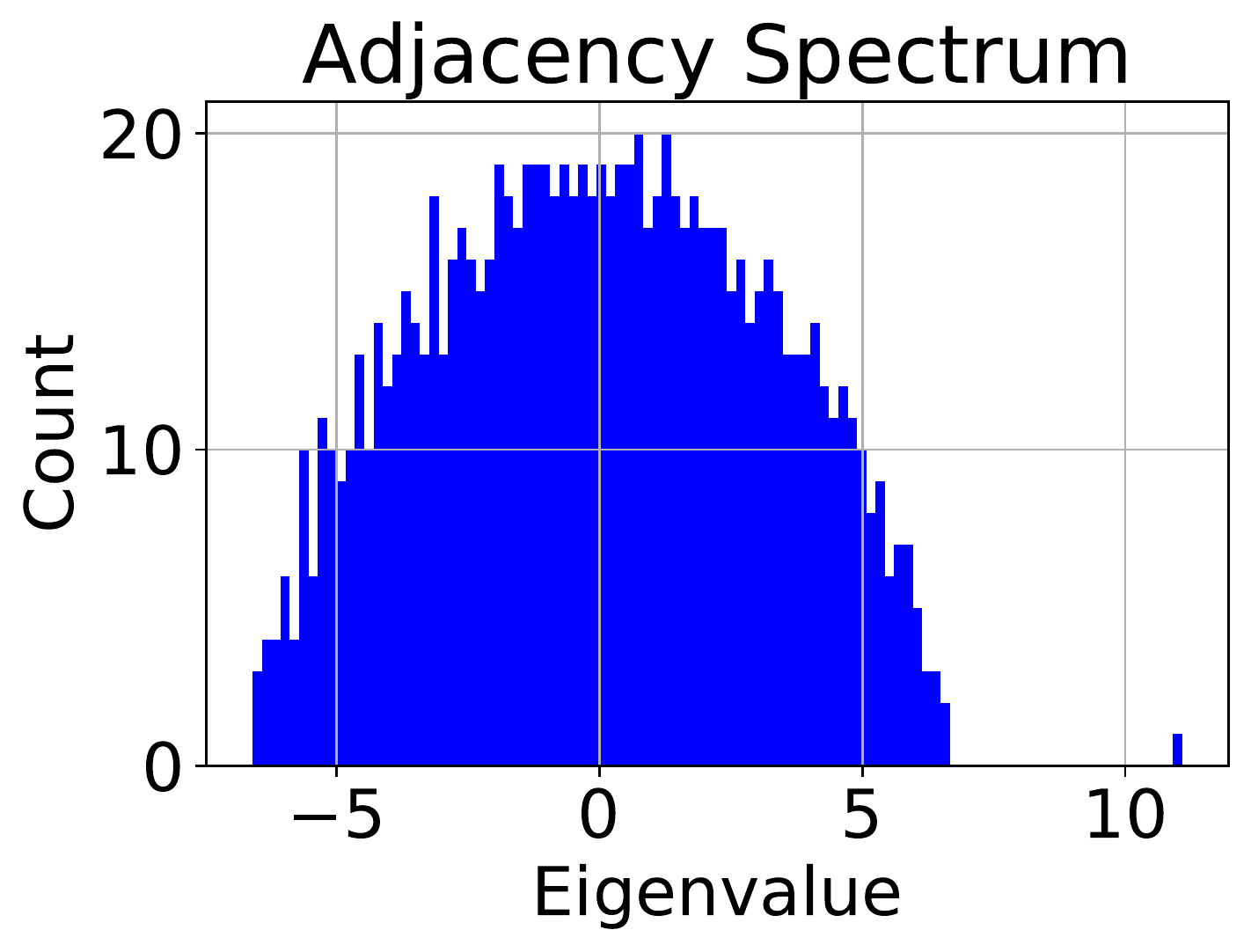}
     &     \includegraphics[width=0.18\linewidth]{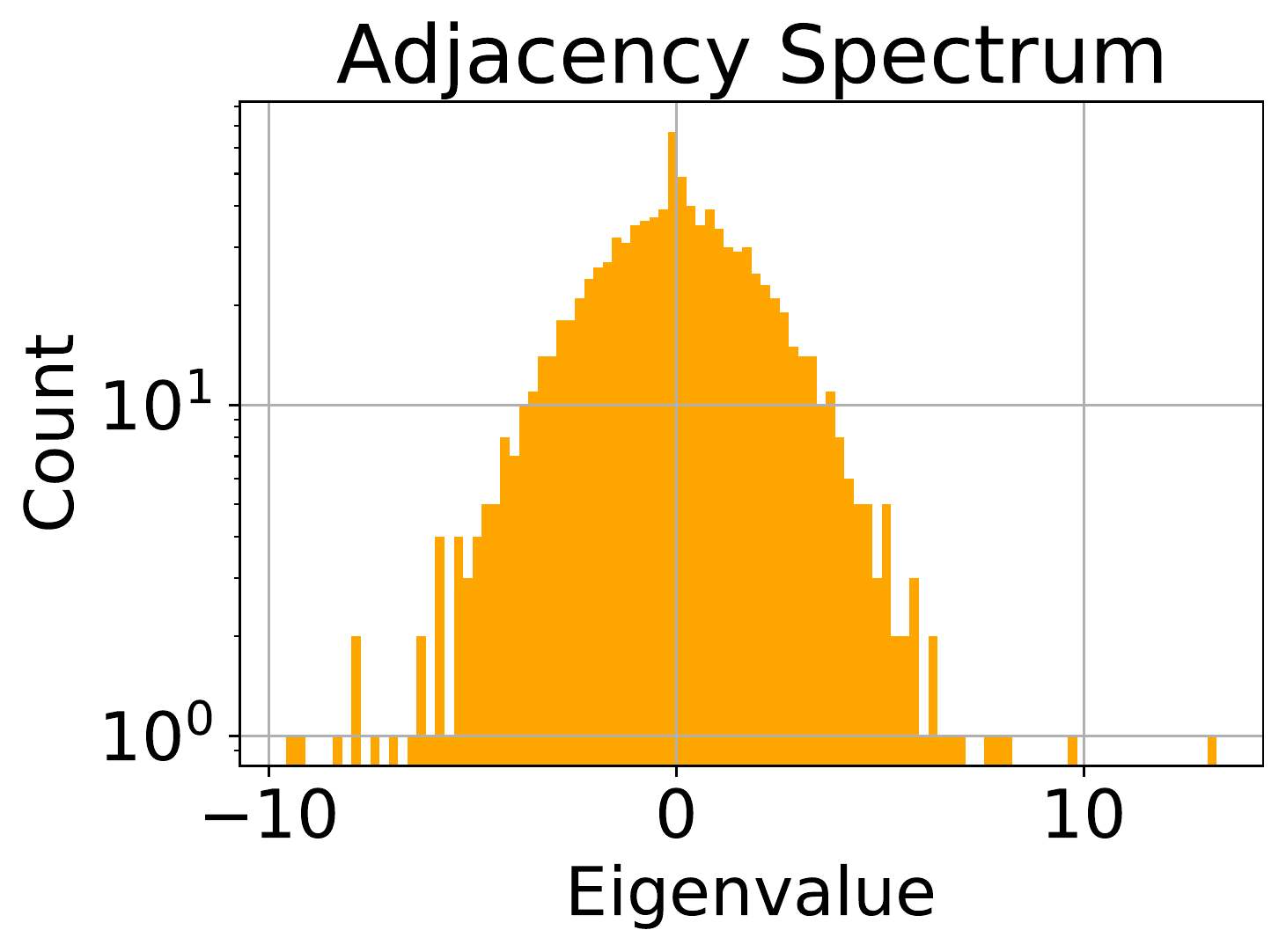}
 &       \includegraphics[width=0.18\linewidth]{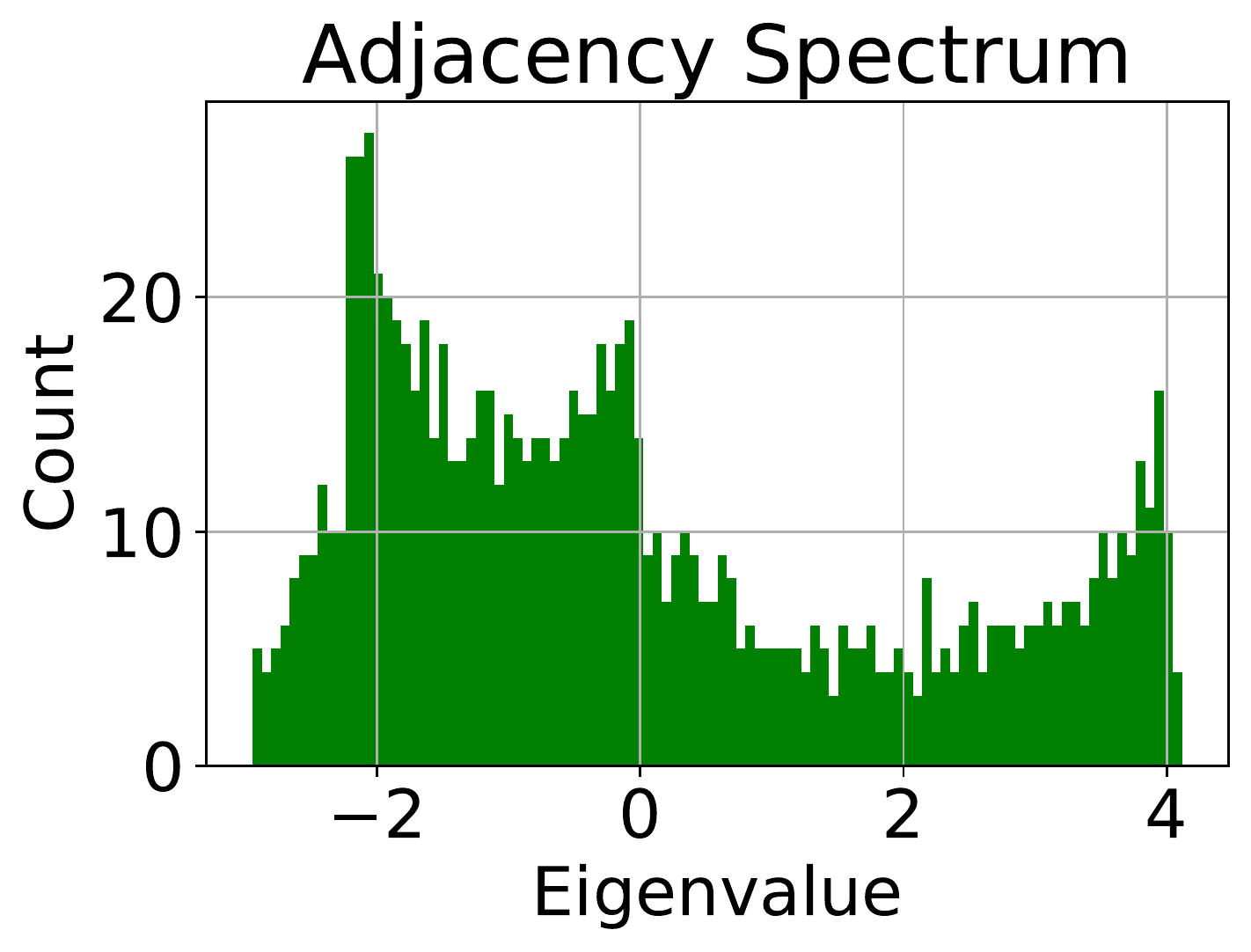}
&      \includegraphics[width=0.18\linewidth]{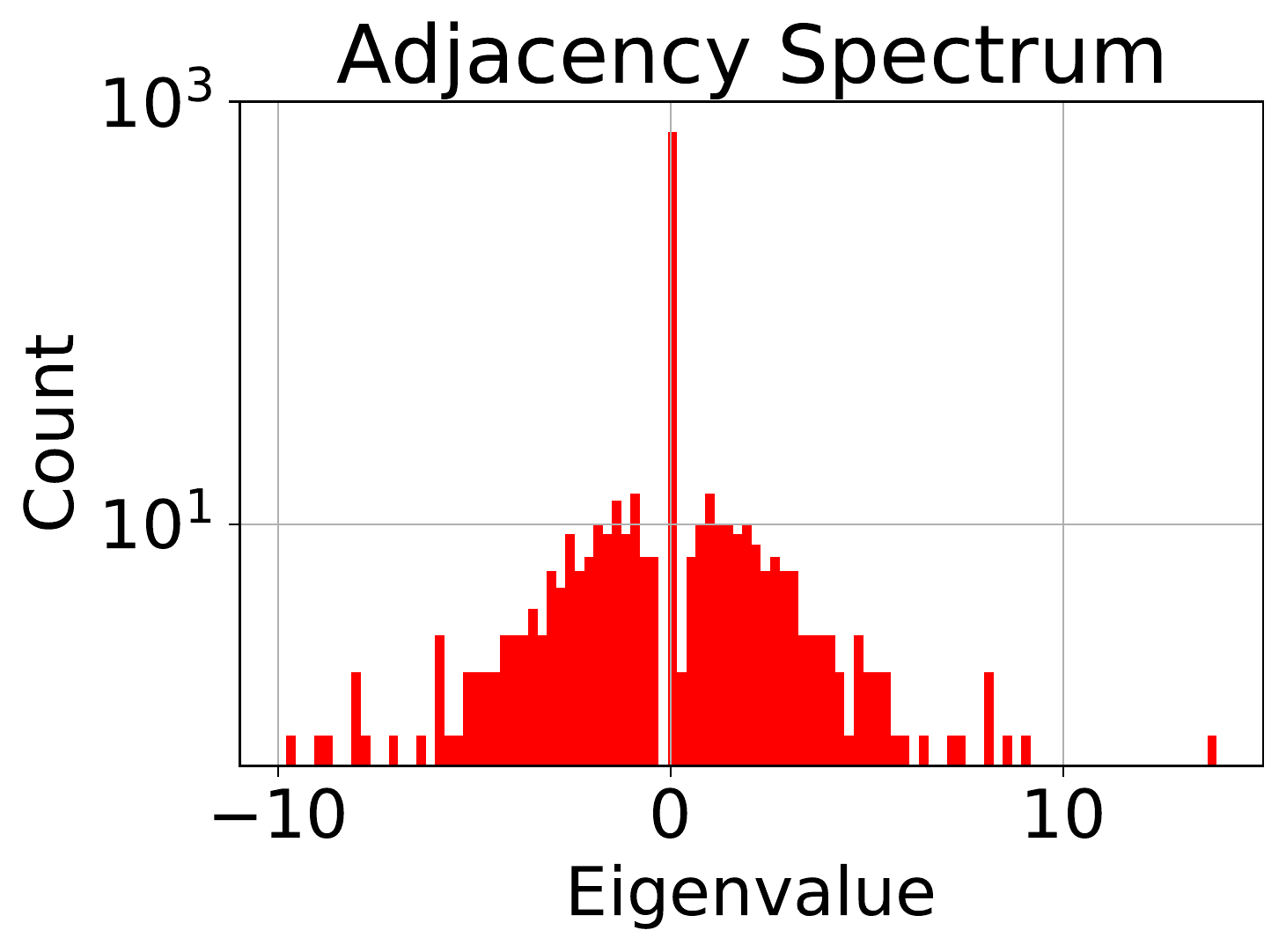}
&      \includegraphics[width=0.18\linewidth]{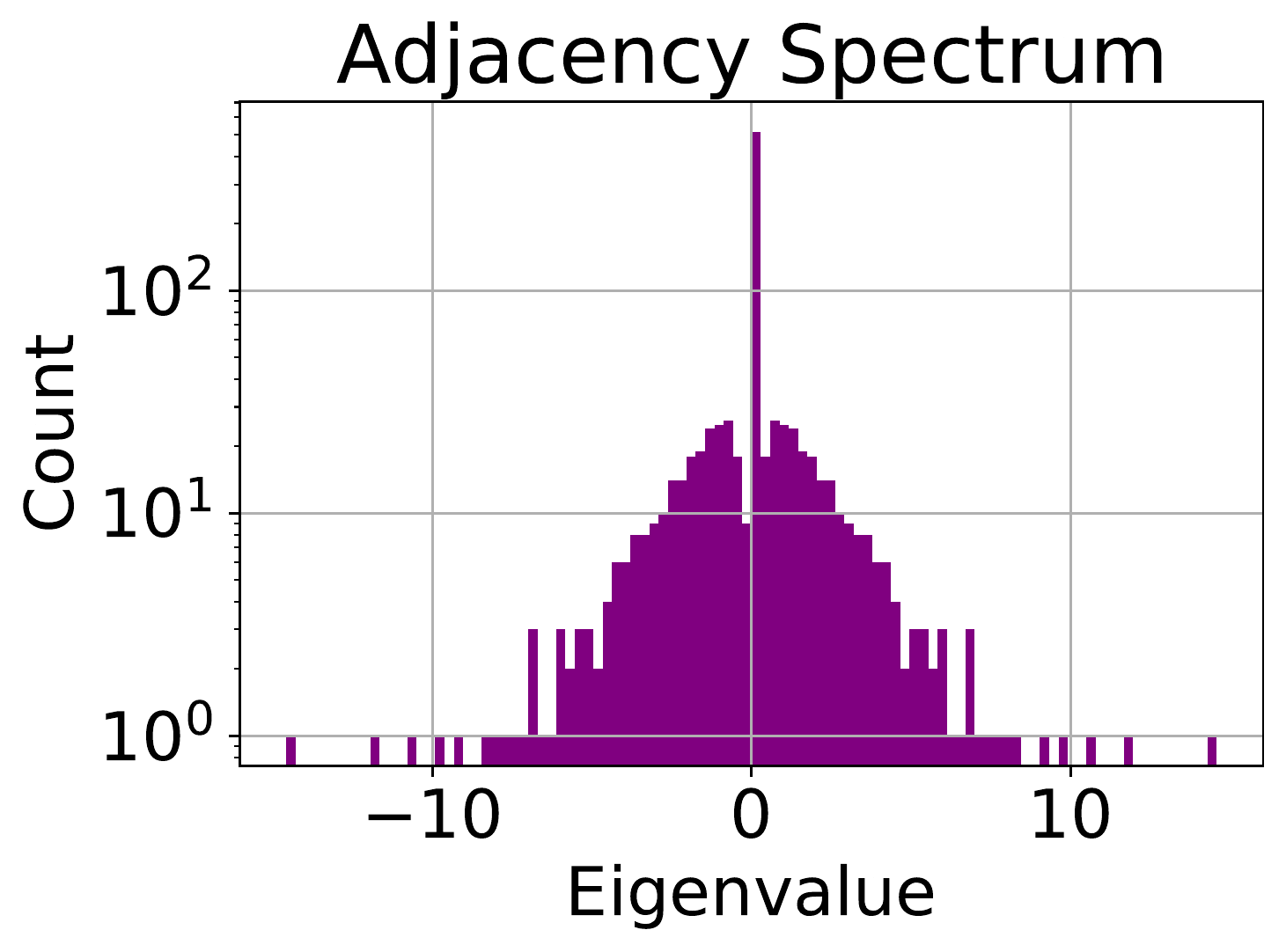}
\\[6pt]

{\footnotesize (a) Erdős–Rényi  $\mathcal{G}_{ER}$ } & (b) {\footnotesize Barabási--Albert $\mathcal{G}_{BA}$} & {\footnotesize (c)  Watts--Strogatz $\mathcal{G}_{WS}$} & {\footnotesize (d) AS internet $\mathcal{G}_{AS}$} & {\footnotesize(e) Proteins interaction $\mathcal{G}_{PP}$ } \\
{\footnotesize
$\rk(A_{\mathcal{G}_{ER}})=1000$} & {\footnotesize $\rk(A_{\mathcal{G}_{BA}})=997$} & {\footnotesize $\rk(A_{\mathcal{G}_{WS}})=1000$} &{\footnotesize $\rk(A_{\mathcal{G}_{AS}})=278$} &{\footnotesize $\rk(A_{\mathcal{G}_{PP}})=492$
}
\end{tabular}
\caption{ \footnotesize \raggedright Some complex networks, their adjacency matrix eigenvalues distribution and the rank of their adjacency matrix for $N=1000$ nodes. In the distributions of the BA, AS and PP the y-axis is in log-scale. }\label{fig:cxgraphs}
\end{figure*}

In the previous section we described graphs that are built through a deterministic algorithm, though we can also construct a graph based on  statistical models \cite{Newman18,Barabasi16}. This is the difference between regular and random networks. An exemplary standard for random networks is the Erdős--Rényi model $\mathcal{G}_{ER}(N,p)$, in which each pair of the $N$ nodes have a probability $p$ to be linked; the network thus has $\binom{N}{ 2} p$ edges on average \cite{erdos1960evolution}.

Most of the network properties observed in nature, however, simply cannot be described by regular or random graphs. For this reason, a youthful branch of scientific research is committed to the study of \textit{complex networks}. In the field of network theory, complex networks are a type of graphs with non trivial topological features, that are  shared by neither regular nor completely random graphs, but are rather akin to networks modeling real systems \cite{newman2003structure}.

An important class of complex networks is characterized by the \textit{small world} property. These networks exhibit the peculiarity of having a low average path length, which is the mean distance between two arbitrary nodes, and a high clustering, which is a measure of the degree to which nodes in a graph tend to cluster together. The emblematic network presenting these features is the the Watts--Strogatz model $\mathcal{ G}_{WS}(N,Q,\beta)$ \cite{watts1998collective}. In this model, we first construct a regular periodic graph with $N$ nodes and $\tfrac{NQ}{2}$ edges where each node has exactly $Q$ neighbors, then with probability $\beta$ we rewire each edge with another node chosen uniformly at random while avoiding self loops and link duplications. 

The second relevant class of complex networks present the typical aspect of being \textit{scale-free} and having \textit{long-tailed} structures. Scale-free networks show a power law in the degree distribution $P(k)\propto k^{-\gamma}$ for some $\gamma>0$, which is self-similar at all values of $k$ in the \textit{tail} of the distribution, unlike the ER and WS models that go to zero very quickly and have no tails. This fractal like attribute is well reproduced by the Barabási--Albert model $\mathcal{G}_{BA}(N,K)$, which can also emulate growth and preferential attachment in networks \cite{albert2002statistical}. The graph is built sequentially by adding one node at a time and wiring it to $K$ other nodes with a probability that is proportional to the number of links that the target node already has. While adding one node at a time, the first $K$ nodes initially will not be linked to anything, hence this graph will result having $(N-K)K$ edges, mostly connected to a great hub. This type of graph is the canonical example to reproduce some properties of the \textit{World Wide Web}. A particular case of the Barabási--Albert network is the \textit{Tree-graph} $\mathcal{G}_{T}(N)=\mathcal{G}_{BA}(N,1)$.

Another notable class of complex networks is constituted by \textit{technological networks}, artificial networks designed typically for distribution
of some merchandise or resource, such as electricity or 
information. The most famous example in this category is the Internet Autonomous System (AS) \cite{elmokashfi2008scalability} $\mathcal{G}_{AS}(N)$, the physical global computer data network. In order to study this topology we will base on the work put forward in ref.~\cite{elmokashfi2008scalability}.

All the complex networks mentioned so far are meant to emulate man-made structures, however complex topologies appear in nature in the most surprising ways. There is a large variety of \textit{biological networks}, among these we will consider specifically the protein--protein interaction network model $\mathcal{G}_{PP}(N,\sigma)$ 
developed in Ref.~\cite{ispolatov2005duplication}.

\section{Quantum teleportation Gaussian Networks}\label{sec:teleportation}

In a naïve strategy, the teleportation between two arbitrary nodes can be implemented simply by ignoring all the other nodes and exploiting the residual bipartite entanglement together with classical communications. This strategy is a
direct extension of the standard teleportation protocol from two to more stations and is called \textit{non-assisted} protocol \cite{pirandola2006quantum}.

Another set of strategies is based upon a cooperative behavior, where all the other nodes assist the teleportation between the chosen pair (Alice and Bob) by means of LOCC. In fact, if the external nodes perform suitable local measurements and then classically communicate their outcomes to Bob, the latter can use this additional classical information to improve the process via modified conditional displacements. These strategies are called \textit{assisted} protocols and are the ones that determine what we call routing protocol in this Article.

According to Gu et al.\@ \cite{gu2009quantum} quadrature measurement on a mode of a Gaussian network like the ones we considered so far can be described by two simple rules:
\begin{itemize}
    \item \textit{Vertex Removal:} a $\hat{q}$-measurement on a qumode removes it from the network, along with all the edges that connect it.
    \item \textit{Wire Shortening:} a $\hat{p}$-measurement on a qumode is just a $\hat{q}$-measurement after a Fourier Transform, which corresponds to a phase rotation of $\pi/2$:  $S_F=S_R(\theta=\frac{\pi}{2})$. The node will thus be removed but the phase shift will induce correlations between the neighboring edges.  Thus, measurements in the momentum basis allow us to effectively “shorten” linear graph states.
    \end{itemize}
These rules are exposed more formally in Appendix \ref{app:GraphCalc}, using the graphical calculus
introduced by Menicucci, Flammia and van Loock \cite{menicucci2011graphical}.
    
If two nodes A and B need to teleport a quantum state, they can be helped by the other nodes in the network who will perform these operations in order to increase the strength of the entanglement in the final pair. A typical measure of entanglement is the \textit{logarithmic negativity} \cite{eisertphd,vidal2002computable,plenio2005logarithnmic, WeedbrookRMP}
\begin{equation}
    \mathcal{N}=-2\log_2\nu_-
\end{equation}
Where $\nu_-$ is the smallest symplectic eigenvalue of the partially transposed covariance matrix of the pair. Partial transposition is a necessary operation for the PPT criterion \cite{simon2000peres} and is easily implemented in Gaussian states by changing the sign of the momentum of one of the two subsystems.

The symplectic eigenvalues $\nu_\pm$ of a two-mode system can be computed through the invariants of the covariance matrix \cite{WeedbrookRMP}. More specifically, we can define the \textit{seralian} 
\begin{equation} \label{eq:seralian}
    \Delta=\det \sigma_A+\det \sigma_B+2 \det \sigma_{AB},
\end{equation} where  $\sigma_A$ and $\sigma_B$ are the local covariance matrices of the single-mode sub-systems A and B, and $\sigma_{AB}$ represents their correlations. From this we can compute the symplectic eigenvalues as:
\begin{equation}\label{eq:sympEig}
  \nu^2_{\pm}=\frac{\Delta\pm\sqrt{\Delta^2-4 \det \sigma}}{2}
\end{equation}

\section{Conclusion and Outlook}
In this work we have investigated Gaussian multimode graph states with regular and complex topologies and studied their potential application for quantum communication protocols.

  Firstly, we have shown analytically and numerically that the cost of the networks is  in general nonlinear with the number of edges and nodes and there are particular (regular and complex) graph shapes that optimize the cost and the number of squeezers over number of nodes and edges in the networks. Among regular networks the diamond and the star graph need only two squeezed nodes to be built, independently from their number of nodes. Among the complex networks shapes, the Internet Autonomous System model is the most convenient in number of needed squeezed states \cite{elmokashfi2008scalability}. 
  
    Then, we have studied the assisted teleportation protocol in Gaussian entangled networks, where couple of nodes are assisted in the teleportation by local measurement in all the other nodes. This naturally define a routing protocol in Gaussian networks. In particular we have considered $\hat q$ and $\hat p$ homodyne quadrature measurement that allow respectively for vertex-removal and wire-shortening.
  We showed analytically the effect of parallel enhancement of entanglement to improve the quality of quantum communication in the diamond network.
     
   Finally, inspired by this quantum effect,  we have devised a routing protocol that exploits wire shortening in parallels paths and we have applied it to complex networks graphs. The protocol named \textit{Routing} is compared with \textit{Shortest}, where wire shortening is done only in the shortest path, and \textit{All P}, which removes all the terminal nodes while it wire-shorten all the others. In most cases, the \textit{Routing} improves the entanglement compared with \textit{Shortest}. Also, in terms of computational complexity, the \textit{Routing} is much slower than \textit{All P} in regular networks, where there are long distances between nodes and several parallel paths, but it is very efficient in complex sparse networks.

The devised \textit{Routing} protocol is very general so that it can be applied to arbitrary networks, and it is particularly efficient for  sparse not regular networks. Our simple graph exploration approach would be improved in computational efficiency  by real graph-based algorithms, especially if we allow for  approximate solutions \cite{acosta2012frequent}.  
Also it would  be interesting to allow for non uniform distributions of squeezing $s$ and CZ-gate strength $g$ or more general homodyne measurements, i.e.\ going beyond the two $\hat p$ and $\hat q$ cases and considering measurements along $\hat q_{\theta}=\cos(\theta)\hat q+\sin(\theta)\hat p$. In addition, it could be interesting to examine a scenario in which the intermediate nodes are dishonest and do not cooperate to perform the routing, exploiting, for example, non-Gaussian operations. 
Finally the routing protocol has been implemented to  solve the particular task of creating a perfect EPR pair between two nodes, future protocols will consider general reshaping in arbitrary multiparty states.

\section{Acknowledgements}
We thank Dimitrios Tsintsilidas for helpful discussions.
This   work   is   supported   by   the   European   Research Council under the Consolidator Grant COQCOoN (Grant No.\ 820079),
by the European Union under the project 101080174 CLUSTEC,
by the French state from
the Plan France 2030 managed by the Agence Nationale de la Recherche through the ANR-22-PETQ-0006 NISQ2LSQ and the 
OQuLus projects, by the ERC AdG CERQUTE and by the Government of Spain (Severo Ochoa CEX2019-000910-S and TRANQI), Fundació Cellex, Fundació Mir-Puig, Generalitat de Catalunya (CERCA program). 

\bibliographystyle{ieeetr}

\bibliography{biblio}

\section{Competing Interests}
The Authors declare no Competing Financial or Non-Financial Interests.

\section{ DATA AVAILABILITY}
The numerical data  that support the finding are available from the authors upon request and can be reproduced using the repository \cite{Code}.

\section{Author contributions}
F.C. operated the numerical simulations and analyzed the data. All authors contributed to the theoretical analysis and to writing the manuscript. V.P and F.C. supervised the work. V.P conceived the project.

%
%
\appendix

\section{Squeezing spectra of regular graphs}\label{app1}

Given any graph's spectrum we can employ equation (\ref{eq:spectrum}) to compute the amount of squeezing in each mode required to build the Gaussian network. In the following calculations we used the spectra of regular graphs known in literature. A detailed reference on the spectra of regular graphs and how to obtain them can be found here \cite{jones2013spectra,brouwer2011spectra}.

\begin{figure}[tb]
\centering
    \includegraphics[width=1\linewidth]{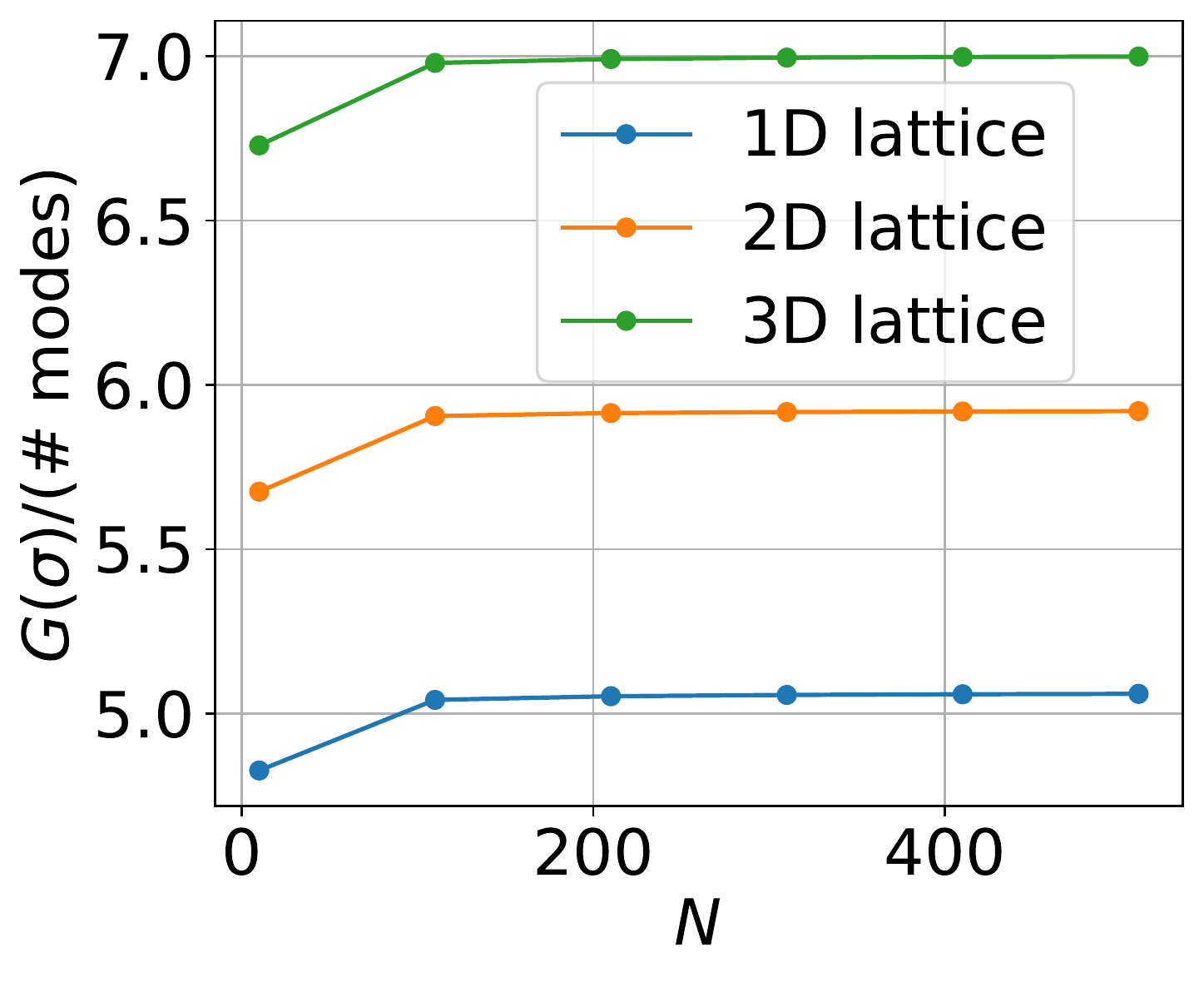} 
\caption{\footnotesize \raggedright  Trend of the average squeezing cost per mode $\bar{G}(\sigma)$ for 1D, 2D and 3D lattices, with $N$, $N^2$ and $N^3$ nodes respectively. Notice that the cost becomes asymptotically constant, as predicted by the theory. }\label{fig:GvNlat}
\end{figure}

We use the linear graph as a benchmark to see how the squeezing cost scales with the number of nodes and links. In fact,  single mode squeezing  and the CZ-gate  both require a fixed amount of squeezing to be implemented, so we would expect $G(\sigma)$ to scale linearly with the number of links and nodes.  This is a direct consequence of the spectral distribution of the linear graph, which is
\begin{equation}\label{eq:linearSpectrum}
    D_k(\mathcal{L}_N)=2g\cos{\frac{\pi k}{N+1}}, \; \{ k=1,...,N\}.
\end{equation}
We can use equations (\ref{eq:spectrum}) and (\ref{eq:SqCostSigma}) to compute the squeezing cost exactly for any given $N$. For $N\gg 1$, the sum in Eq.\ (\ref{eq:SqCostSigma}) allowing to compute
the average squeezing cost per mode $\bar{G}=G/(\#\text{squeezed modes})=G/\rk(A)$ can be seen as a Riemann integral, which converges to
\begin{multline}
    \bar{G}\simeq 10\int_{\mathrlap0}^{\mathrlap{1}} 
      \log_{10}\left[1+\frac{g^2\cos^2{\pi y}^2}{2} +\vphantom{\sqrt{\frac{g^4}{4}}}
      \right.\\\left.
         \sqrt{g^2\cos^2{\pi y}+\frac{g^4\cos^4{\pi y}}{4}}\right] \mathrm{d}y
\end{multline}
This can be easily generalized to the case of a  $D$-dimensional cubic lattice $\mathcal{L}^{(D)}_N$, considering that adding a new dimension would just add a new set of eigenvalues of the form (\ref{eq:linearSpectrum}), as shown in section 2.6 of \cite{cvetkovic1999spectra}. As a consequence, the  squeezing cost  of the $\mathcal{L}^{(D)}_N$ is $G=O(N^{D})$, whereas the average cost per mode would be again constant with the number of nodes $N$, as shown in Fig.~\ref{fig:GvNlat}. 

Similarly, we can use the eigenvalues expression of the circular graph (or its generalization with $Q>1$ nearest neighbors) 

\begin{equation}
D_k(\mathcal{G}_{C_{(Q)}})=g\frac{\sin[(Q+1) k\pi/N ]}{\sin[k\pi/N ]}-g,\; k=\{0,....N-1\}.
\end{equation}
From this we can see why the linear and circular graph  have the same scaling. In fact for $Q=1$, we have  $D_k(\mathcal{G}_{C_{(Q1}})=2g\cos(k\pi/N)-g$.

The spectra of the star and diamond graph have only two non-null eigenvalues \cite{jones2013spectra}
\begin{align}
    \{D_k(\mathcal{S}_N)\}&=\left\{ g\sqrt{N-1}, 0^{\otimes(N-2)},-g\sqrt{N-1} \right\},\\
    \{D_k(\mathcal{D}_N)\}&=\left\{ g\sqrt{2N}, 0^{\otimes(N-2)},-g\sqrt{2N} \right\}.
\end{align}

As a consequence, the  cost of the star and diamond Gaussian networks grows logarithmically with $N$ and have the following expressions.
\begin{widetext}
    \begin{align}
    {G}(\sigma_{\mathcal{S}_N})&=20 \log_{10} 
       \frac{2-g^2+g^2N+\sqrt{g^4N^2-2(2g^4-g^2)N+g^4-4g^2}}2
       \label{eq:GsigmaNg}\\
    &=20 \log_{10} N + 20\log g^2 + O\left(\tfrac1N\right);
\end{align}
\end{widetext}

\begin{align}
    {G}(\sigma_{\mathcal{D}_N})&=20 \log_{10} \left(1+g^2N+g\sqrt{N(g^2N+4)} \right) \\
        &=20 \log_{10} N +20 \log_{10} 2g^2+ O\left(\tfrac1N\right).
\end{align}

Finally, the spectrum of the fully connected graph is
\begin{equation}
    \{D_k(\mathcal{S}_N)\}=\{g(N-1),-g^{\otimes(N-1)} \},
\end{equation}
yielding a squeezing cost that can be expressed as the sum of two contributions
\begin{align}
       G(\sigma_{\mathcal{F}_N})=&10(N-1)\log_{10}\tfrac{2+g^2+g\sqrt{g^2+4}}{2}\notag\\
       +&10\log_{10}\left[1+\tfrac{1+\sqrt{1
                            +\tfrac4{g^2(N-1)^{\mathrlap2}}}}2g^2(N-1)^2\right]\\\allowbreak
       =&10(N-1)\log_{10}\tfrac{2+g^2+g\sqrt{g^2+4}}{2} 
       \notag\\&
       + 20 \log_{10} N +20 \log_{10} g+ O\left(\tfrac1N\right),
\end{align}
which sums up to a cost growing essentially linearly in $N$.

{}

\section{Squeezing spectrum of Erdős--Rényi graphs}\label{app12}

\begin{figure}[tb]
\includegraphics[height=.75\linewidth]{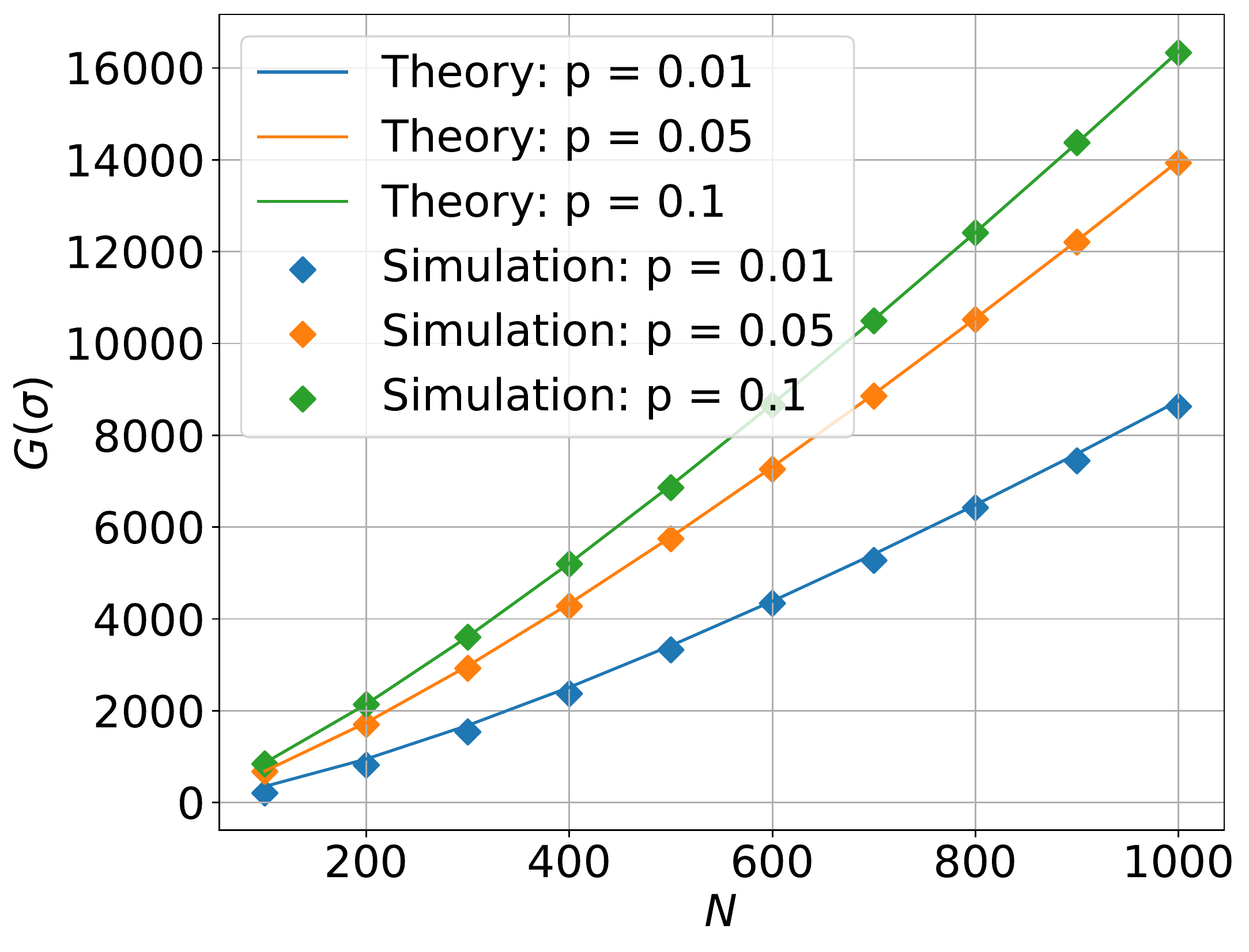}
\caption{\footnotesize \raggedright Comparison between the theoretical prediction of equation (\ref{eq:costCx}) and the experimental numerical simulation of the squeezing cost of the Erdős--Rényi graph state, for different values of $p$. } \label{fig:ThVsimER}
\end{figure}

From Fig.~\ref{GvNcx}   we  notice that the most expensive growth belongs to the ER topology, which is the only one among the topologies we studied whose trend is superlinear. This behaviour is actually the easiest to predict from random matrix theory: the Wigner semi-circular law for the distribution of the eigenvalues of a random graph
\cite{wigner1958distribution} gives their probability distribution in the form $f(x)=f_{ER}(x)=\frac{2}{\pi R^2}\sqrt{R^2-x^2}$ (see Fig.~\ref{fig:cxgraphs}a), where $R=2g\sqrt{Np(1-p)}$, 
with a supplementary large eigenvalue $D_1$ such that $\lim_{N\rightarrow\infty} D_1/gN=p$, with probability 1. Casting it in equation (\ref{eq:costCx}) gives
\begin{align}
    \label{eq:GsigmaER}
    \langle  G(\sigma_{ER})\rangle=&10 \log_{10}[2\lambda^{+}(gpN)]\\
    &+ \tfrac{40 N}{\pi R^2} \!\int_{\mathrlap{0}}^{\mathrlap{R}} \sqrt{R^2-x^2} \log_{10}\left(2\lambda^{+}(x)\right)\mathrm{d}x,\notag
\end{align}
where we used the parity of $f_{ER}(x)$, and the fact that the support is in $[0,R]$.   Figure \ref{fig:ThVsimER} shows the comparison between the theoretical behaviour of the squeezing cost of the ER graph and the numerical experiments from the simulations. 
It shows a superlinear $\propto N\log N$ increase of the squeezing cost, which is explained by the widening of the support of $f$ due to the increasing values of $R$.
More formally, making the variable change $x=Ry$, we have 
\begin{align}
    \log_{10}\left(2\lambda^{+}(Ry)\right)
      &=2\log_{10} R + \log_{10} \tfrac{y^{\mathrlap{2}}}{2} +  O\left(\tfrac1{R^2y^2}\right).
\end{align}
The second term in eq.~(\ref{eq:GsigmaER}) then becomes
\begin{align}
    \tfrac{40N}{\pi}&\int_{\mathrlap{0}}^{\mathrlap{1}} \sqrt{1-y^2} \log_{10}\left(\lambda^{+}(Ry)\right)\mathrm{d}y\notag\\
    &\phantom{\int}= \tfrac{80N \log_{10}(R)}{\pi}\int_{\mathrlap{0}}^{\mathrlap{1}} \sqrt{1-y^2} \mathrm{d}y + O(N) \notag\\
    &\phantom{\int}= 10 N \log_{10}(g^2Np(1-p))+ O(N),
\end{align}
and is the dominant term in the squeezing cost.

\section{Graphical Calculus}
\label{app:GraphCalc}

In Ref.\ \cite{menicucci2011graphical} Menicucci, Flammia and van Loock is provide a unified graphical calculus for all Gaussian pure states which is particularly suited for describing highly multimode Gaussian networks.

In this framework, a $N$ mode Gaussian state is completely described, up to displacements, by a $N\times N$ complex valued adjacency matrix:
\begin{equation}\label{eq:Z}
    Z=V+iU
\end{equation}

Where the real and imaginary part of $Z$, $V$ and $U$ respectively, are related to the covariance matrix through the following unique decomposition
\begin{equation}\label{eq:Zsigma}
    \sigma=\frac{1}{2}\begin{pmatrix}
    U^{-1} & U^{-1}V\\
    VU^{-1} & U+VU^{-1}V
    \end{pmatrix}
\end{equation}

Gaussian graph states have a particular simple graphical representation, being
\begin{equation}
    Z=A+iD
\end{equation}
Where $A$ is the weighted adjacency matrix of the graph and $D$ is a diagonal matrix that represents momentum squeezing, i.e.\ for $D=10^{-2s/10}\mathbb{1}$ the momentum variance of all modes is reduced by $2s$ decibels.

All symplectic operations can be reproduced in this language, however, since we already know how to represents the resource graph states, we only need to implement the quadrature measurements in $\hat{x}$ and $\hat{p}$. We can express the state as 
\begin{equation}
    Z=\begin{pmatrix}
    t & R^T\\
    R & W
    \end{pmatrix}
\end{equation}
Where the scalar $t$ is the target mode we want to measure, 
$W\in{\mathbb R}^{(N-1)\times(N-1)}$ is the  subgraph of the untouched modes and $R\in{\mathbb R}^{(N-1)\times1}$ their correlations with the target mode.
We have the following two rules:
\begin{itemize}
    \item $Z\longrightarrow Z_q=W$ after a $\hat{q}$ measurement.
    \item $Z\longrightarrow Z_p=W-\frac{RR^T}{t}$ after a $\hat{p}$ measurement.
\end{itemize}

Thus, for a measurement in $\hat{q}$ we simply remove the node and its links from the graph, whereas 
for a measurement in $\hat{p}$ we also change the correlation between its neighbors, since 
$(RR^T)_{ij}\neq 0$ iff both $i$ and $j$ are in the node's neighborhood.

\section{Parallel enhancement of entanglement}\label{app:ParallelEnhancement}
\begin{figure}[tb]
 \includegraphics[scale=0.3]{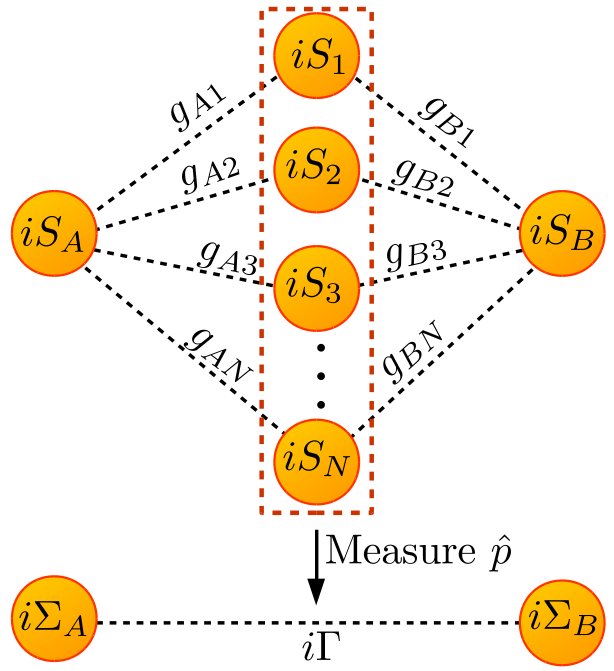}
    \caption{\footnotesize \raggedright Graphical representation of the diamond graph and its parallel enhancement of entanglement.}
    \label{fig:graphCalc}
\end{figure}

We can use the rules described in Appendix \ref{app:GraphCalc}\ to prove eq.\ (\ref{eq:parallel}),
which expresses analytically the power of parallel enhancement of entanglement in the diamond network when measuring the central nodes in $\hat{p}$. 
Let us assume that the nodes A and B are squeezed by a factor $S_A$ and $S_B$ respectively, there are $N$ central nodes and the $k$th mode has squeezing $S_k$ and is correlated with A and B through a CZ-gate with strength $g_{Ak}$ and $g_{Bk}$. It can then be easily showed that the final pair will have a purely imaginary adjacency matrix of the form
\begin{equation}
    Z_{AB}=i\begin{pmatrix}
    \Sigma_A && \Gamma \\
    \Gamma && \Sigma_B
    \end{pmatrix}
\end{equation}
Where $\Sigma_A=S_A+\sum_k  \frac{g_{Ak}^2 }{S_k} $, $\Sigma_B=S_B+\sum_k  \frac{g_{Bk}^2 }{S_k} $ and $\Gamma=\sum_k \frac{g_{Ak}g_{Bk} }{S_k}  $. These result can be derived by direct application of the rule for measuring $\hat{p}$ in the graphical calculus formalism, schematized in figure \ref{fig:graphCalc}. 

Employing eqs.\ (\ref{eq:Z}) and (\ref{eq:Zsigma}) and noticing that $V=0$, we can reconstruct the covariance matrix of the final pair: 
\begin{equation}
    \sigma_f=\begin{pmatrix}
    \frac{\Sigma_B}{\Sigma_A\Sigma_B -\Gamma^2}&& -\frac{\Gamma}{\Sigma_A\Sigma_B -\Gamma^2 }&& 0 && 0 \\
    -\frac{\Gamma}{\Sigma_A\Sigma_B -\Gamma^2} && \frac{\Sigma_A}{\Sigma_A\Sigma_B -\Gamma^2}&& 0 && 0 \\
     0 && 0 && \Sigma_A && \Gamma\\
      0 && 0 && \Gamma && \Sigma_B
    \end{pmatrix}
\end{equation}
Notice that this state differs from a graph state by a local phase.

By computing the seralian --- defined in Eq.\ (\ref{eq:seralian}) --- of the partially transpose covariance matrix of the pair $\Tilde{\sigma}_f$ and applying formula (\ref{eq:sympEig}), 
we obtain the general lowest symplectic eigenvalue of the partial transpose of the state
\begin{equation}
    \nu_-^2=\frac{(\sqrt{\Sigma_A\Sigma_B}-\Gamma)^2}{\Sigma_A\Sigma_B-\Gamma^2}
\end{equation}
Finally, if we assume that all the modes are equally squeezed in $\hat{p}$ of a factor $R^{-1}=10^{2s/10}$ and all the CZ-gate correlations have a strength $g$, we arrive to Eq.\ (\ref{eq:parallel}).

\begin{figure}[tb]
    \centering
        \begin{tabular}{cc}
 \includegraphics[width=0.59\linewidth]{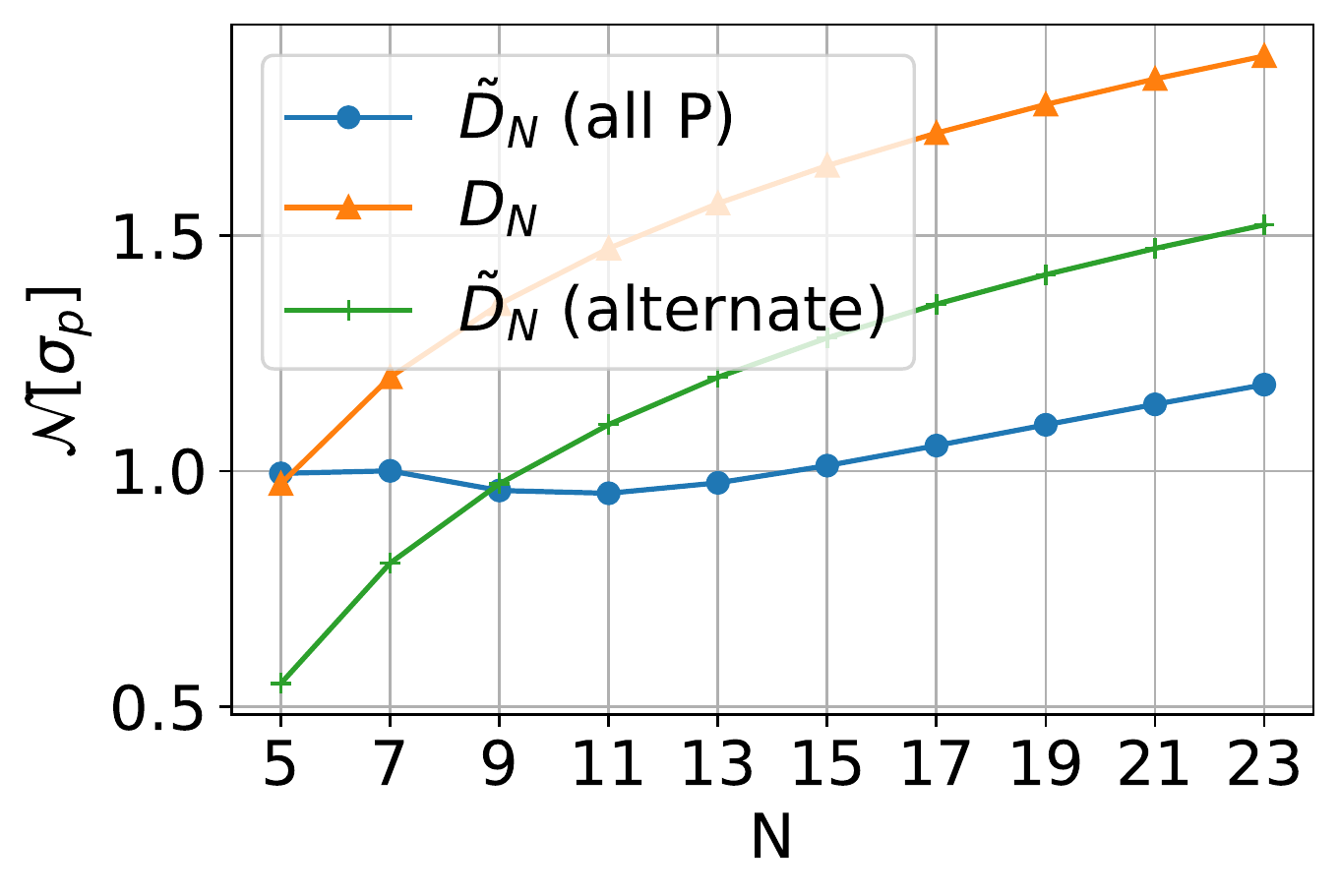} &
  \includegraphics[width=0.39\linewidth]{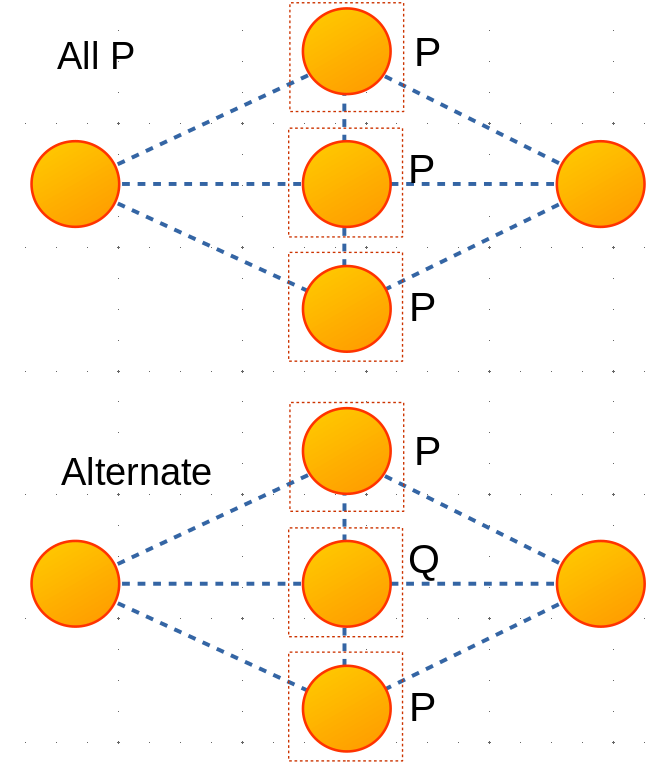}
      \end{tabular}
    \caption{\footnotesize \raggedright Different measurement strategies for two types of diamond network: the standard $\mathcal{D}_N$ we have seen so far and the $\Tilde{\mathcal{D}}_N$, in which the central nodes are connected to their neighbors. We apply two different strategies to $\Tilde{\mathcal{D}}_N$: one is to measure all the central nodes in $P$ and the other is to alternate a $p$ and a $q$ measurement. We can see that measuring always in $p$ is not necessarily the optimal strategy. On the right side you can see a scheme of the $\Tilde{\mathcal{D}}$ network and the two different measurement strategies.}
    \label{fig:LSD}
\end{figure}
\begin{figure}[tb]
        \centering
    \begin{tabular}{cc}
\includegraphics[width=0.48\linewidth]{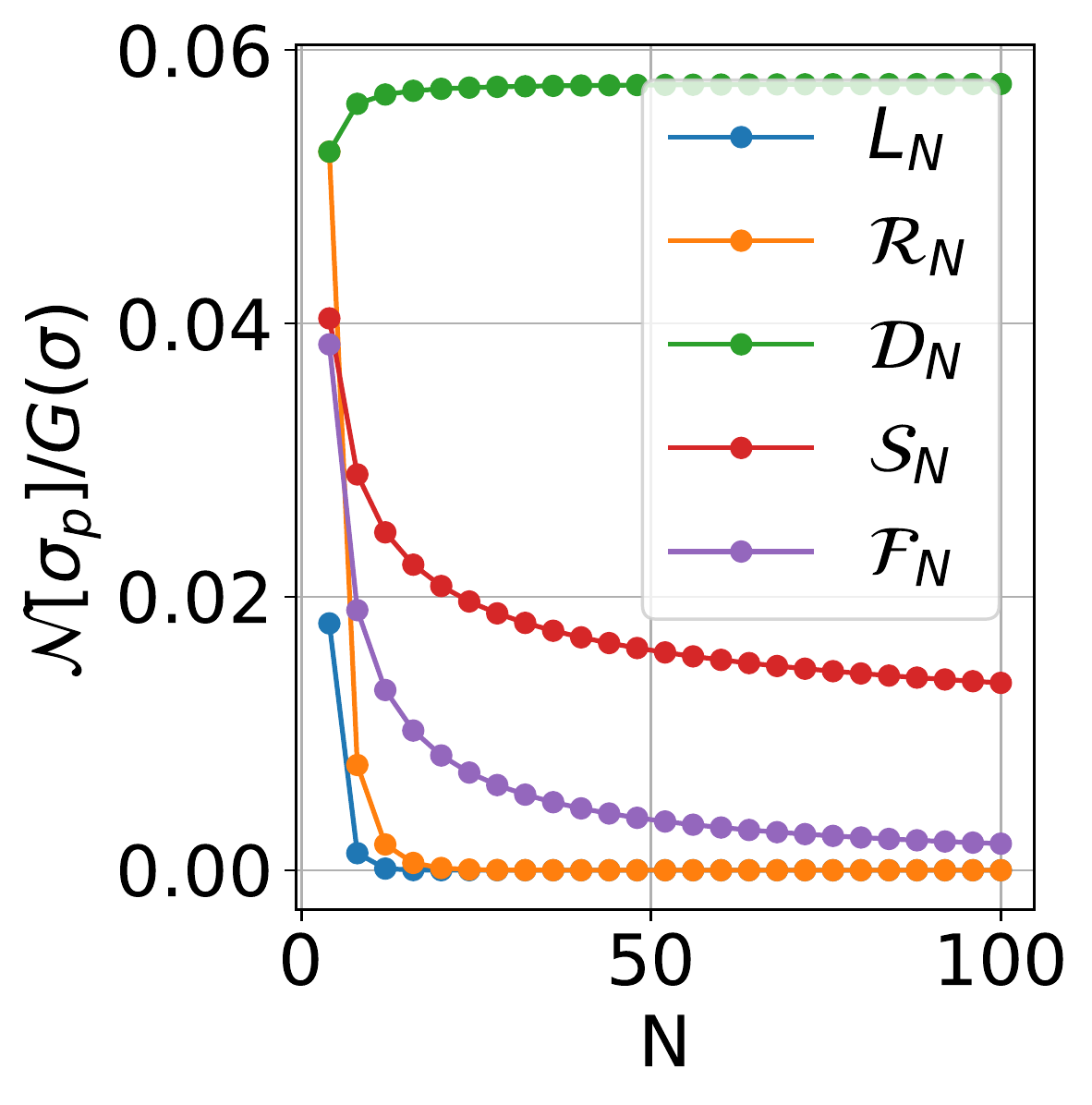}
&  \includegraphics[width=0.465\linewidth]{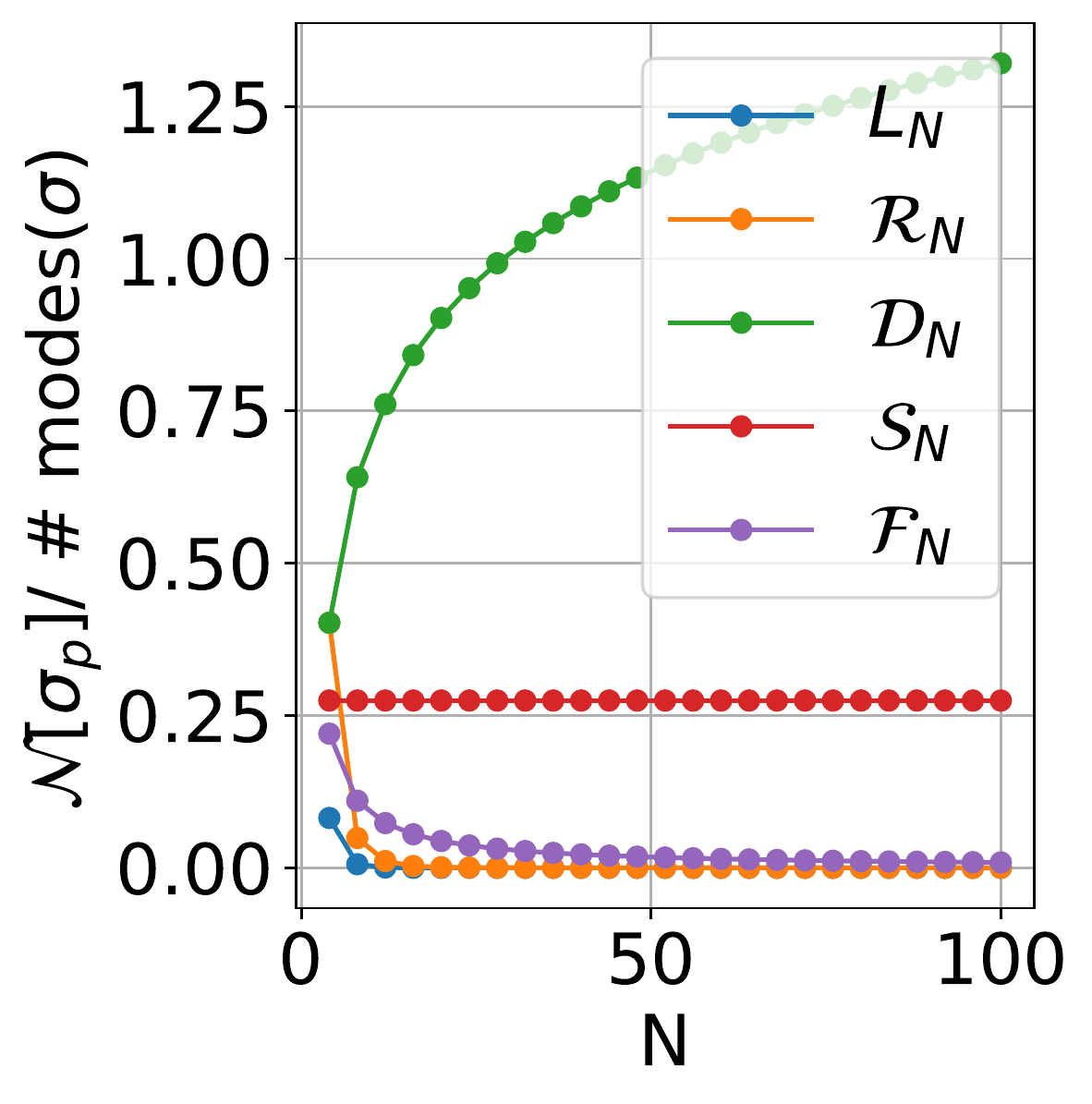}
 \\
     {\bf (a)}    & {\bf (b)}
    \end{tabular}
    \caption{\footnotesize \raggedright Trend of the ratio between the logarithmic negativity of the final state and (a) the squeezing cost of the initial state or (b) the total numer of modes in the initial state for regular topologies:  linear $\mathcal{L}_N$, ring $\mathcal{R}_N$, star $\mathcal{S}_N$, diamond $\mathcal{D}_N$, and fully connected $\mathcal{F}_N$ networks up to $N=100$ nodes. }
    \label{fig:FsuGvNreg}
\end{figure}  
This property of the Diamond network, however, is not easily generalized to all graphs that present parallel connections and the quest for the optimal measurement strategy in order to improve the final entanglement is by no means trivial.
 This is the case, for example, of the $\Tilde{\mathcal{D}}$ graph shown in Fig.\ \ref{fig:LSD}, generated by taking the diamond network and add a CZ-gate link between adjacent central nodes. We can see that for $N>9$ always measuring $\hat{p}$ in this network is not the optimal strategy, whereas a better strategy is to alternate a $\hat{p}$ and $\hat{q}$ measurement in order to restore a (smaller) diamond network.

Another important figure of merit is the entanglement per squeezing cost, shown in  Fig \ref{fig:FsuGvNreg} (a).

We see that the diamond is the only one that gives the a ratio of entanglement per cost of the network that becomes constant for large $N$. However, the linear graph is the one that links two nodes that are the furthest away from each other. Conversely, figure \ref{fig:FsuGvNreg} (b) shows the logarithmic negativity in the final pair divided by the number of modes in the initial state. Once again, the diamond structure is particularly convenient, yielding the highest logarithmic negativity while keeping a constant number of independent squeezers.

\begin{figure}[h]
    \centering
 \includegraphics[width=.5\linewidth]{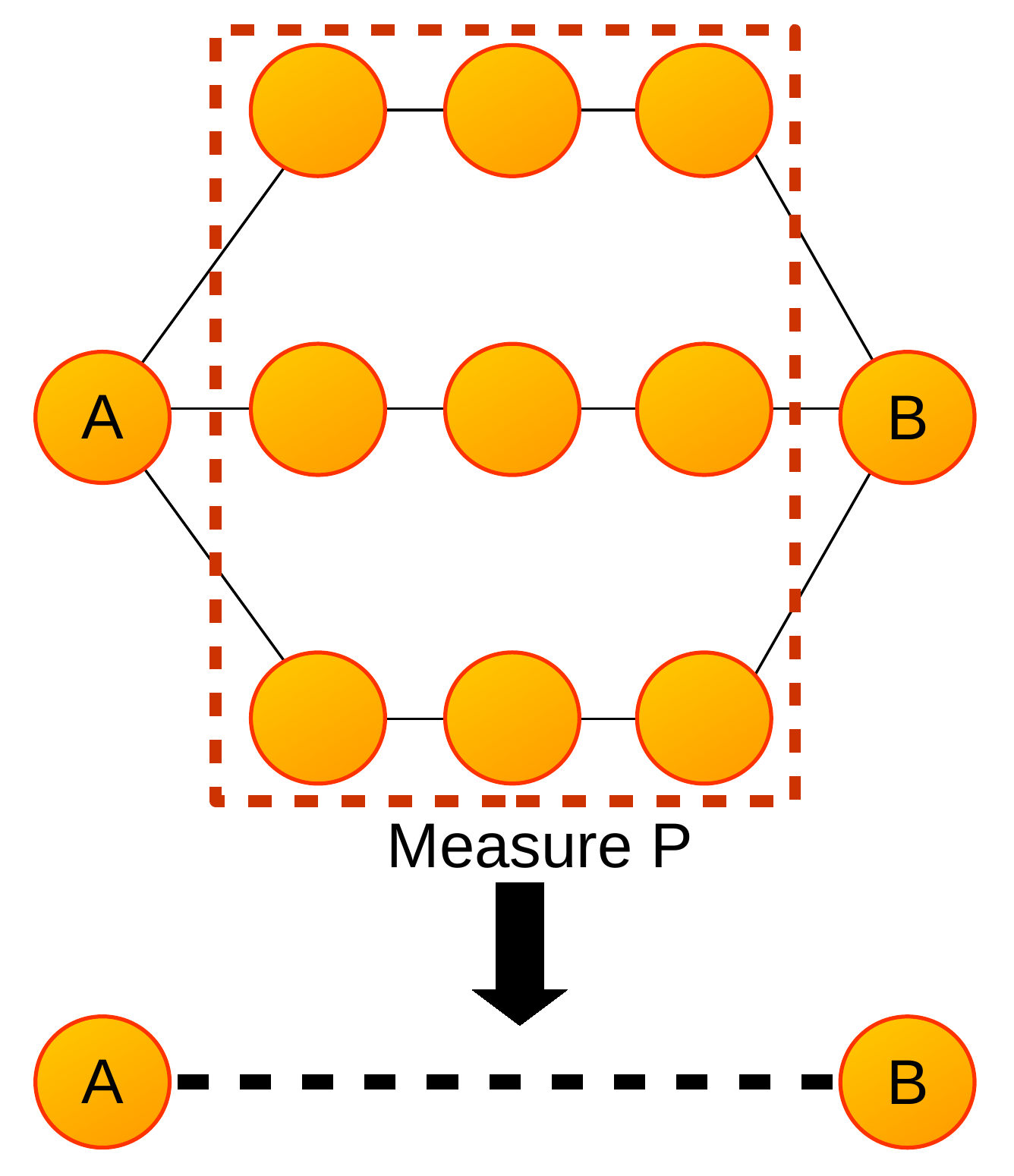}
 \caption{\footnotesize \raggedright Scheme of an entanglement routing protocol in a diamond chain with K=3. All the central nodes are measured in P in order to concentrate entanglement between Alice and Bob.}\label{fig:DC}
\end{figure}

In order to give a fair comparison between the capacity of the linear network to bridge distant nodes and that of the diamond to increase the final entanglement we need to generalize the diamond graph to a diamond chain graph, $\mathcal{DC}_{K,N}$, where $K$ is the number of parallel branches linking the two hubs that want to perform quantum communications as in figure \ref{fig:DC}.

\begin{figure}[t]
    \centering
    \begin{tabular}{cc}
             \includegraphics[width=0.473\linewidth]{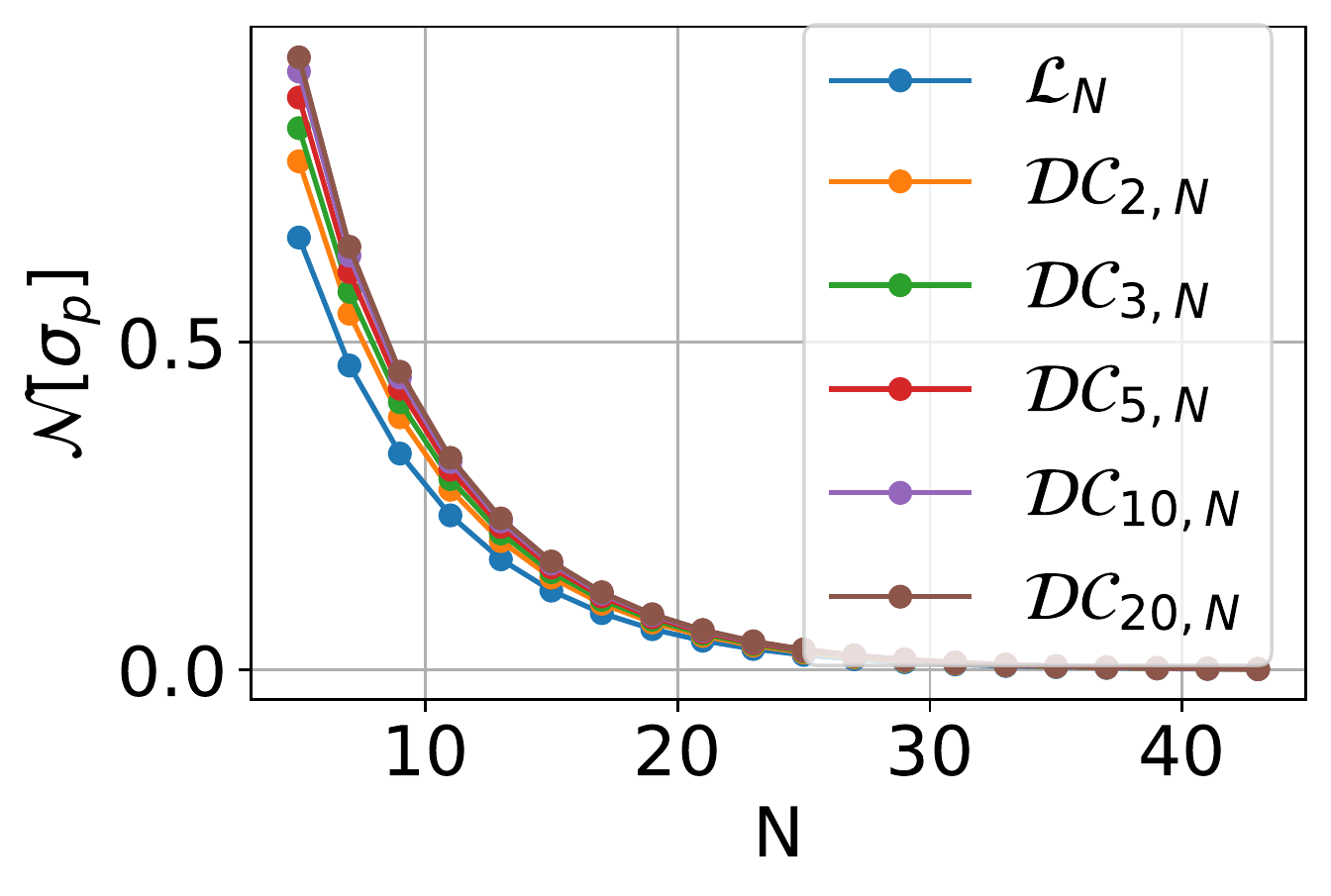}
 &       \includegraphics[width=0.49\linewidth]{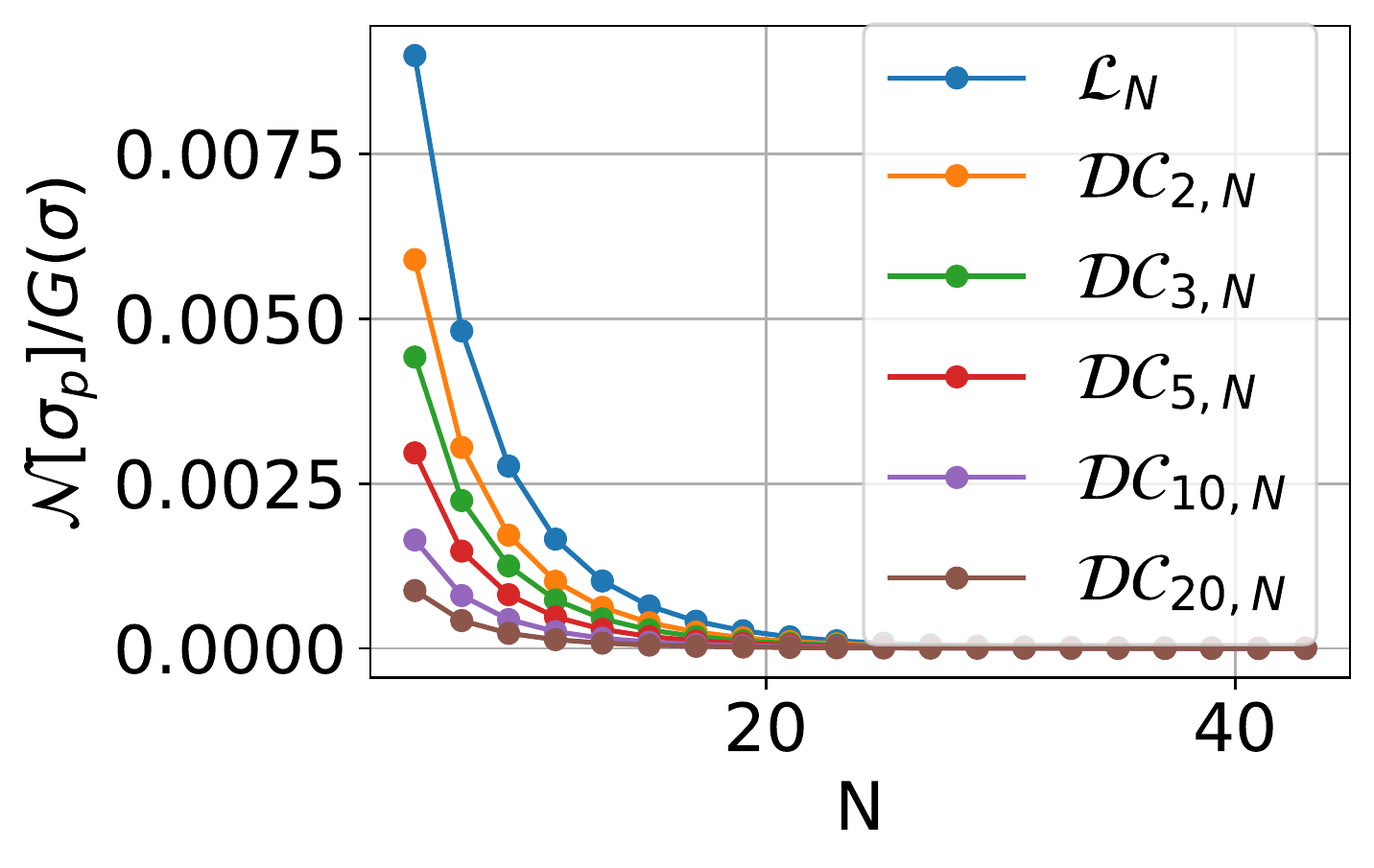}
 \\
     {\bf (a)}    & {\bf (b)}
    \end{tabular}
 \caption{\footnotesize \raggedright (a) Trend of the logarithmic negativity of the output state for the diamond chain network, for various values of the number of branches K (K=1 is the linear network). (b) Trend of the ratio between the logarithmic negativity of the final state and the squeezing cost of the initial state for the diamond chains. }\label{FvNdiam}
\end{figure}

We can then compare the entanglement concentrated using multiple path strategies to link two nodes far away from each other. We can see in figure \ref{FvNdiam} that the presence of parallel links has indeed the desired effect, despite the quality of the final pair, which still decreases exponentially with the distance.  On the other hand, notice that the parallel links can help concentrating more entanglement until the system reaches a plateau and even the additional channels will not allow to increase the logarithmic negativity. Moreover, the value for effort of this networks, specifically the ratio between the entanglement of the pair after the protocol and the squeezing cost before the protocol, is maximized by the linear graph.

\begin{figure}[tb]
    \centering
     \includegraphics[width=0.66\linewidth]{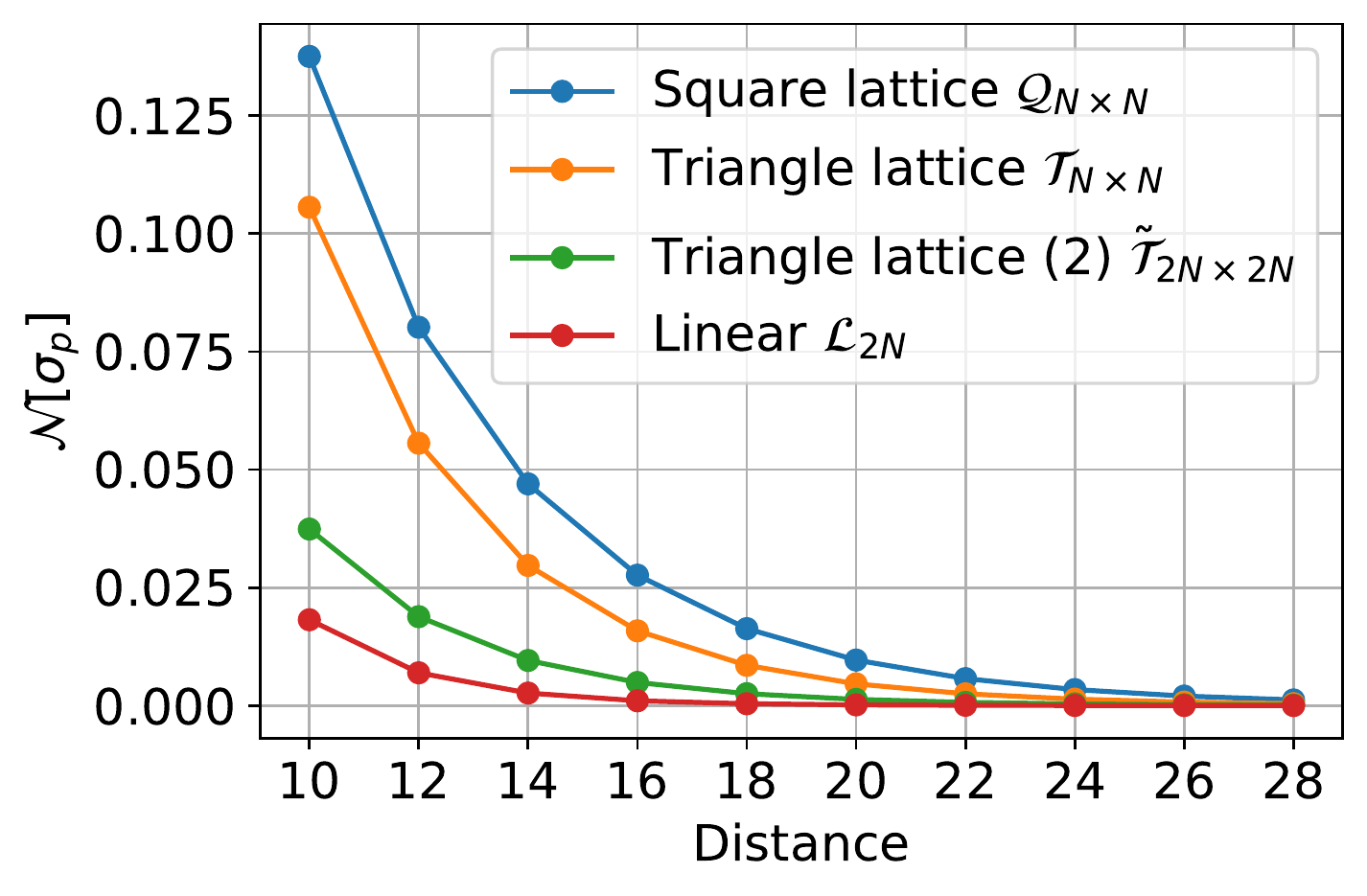}
     \includegraphics[width=0.32\linewidth]{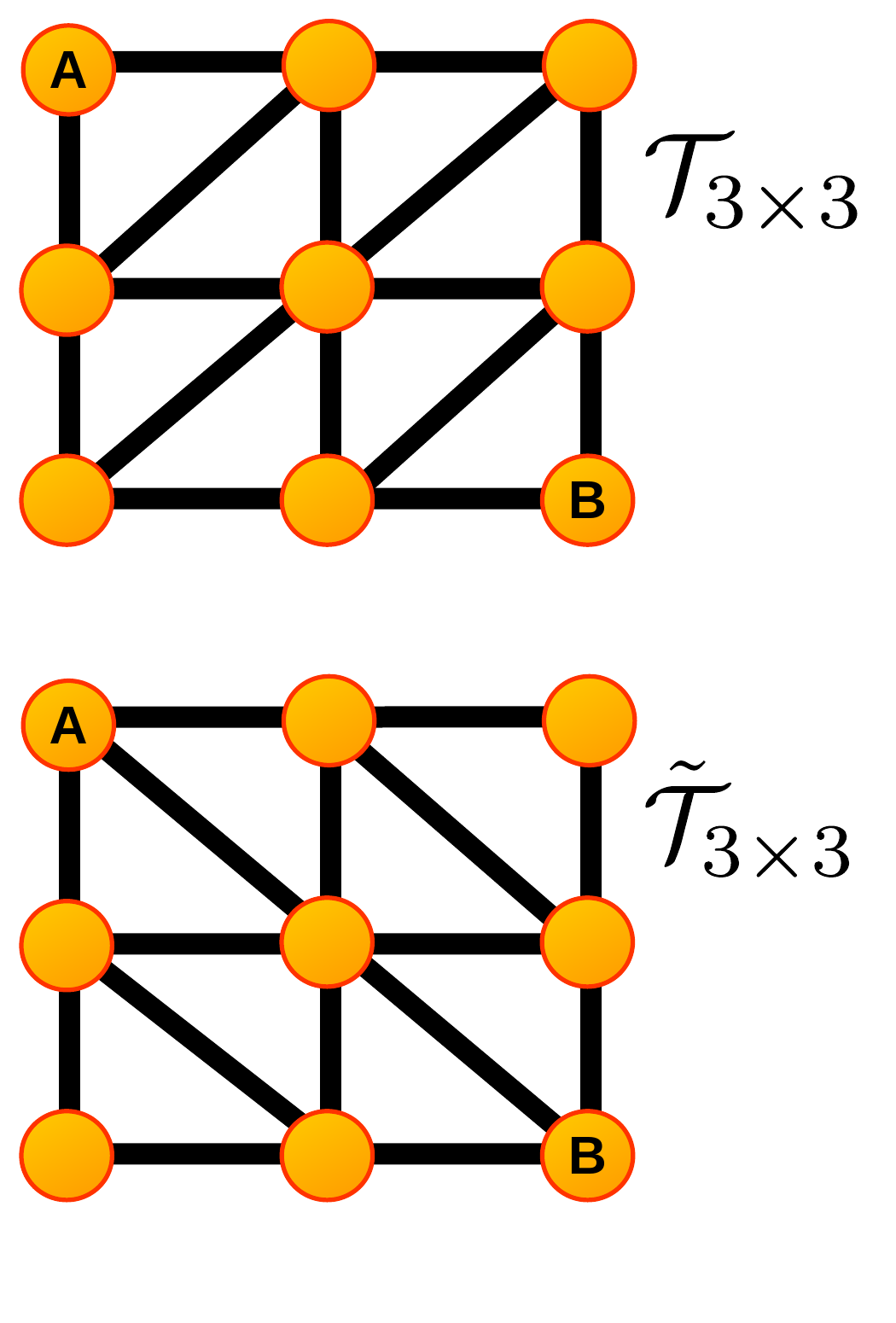}
 \caption{\footnotesize \raggedright Comparison between the entanglement capacity between two nodes at the same distance of three lattices graphs, the square lattice $\mathcal{Q}_{N\times N}$ and the two triangles $\mathcal{T}_{N\times N}$, formed from the square by adding edges on the diagonals in such a way that the distance between A and B is the same, $\Tilde{\mathcal{T}}_{N\times N}$ formed by adding edges to the diagonals so that the distance is the same as the linear graph, and the linear graph $\mathcal{L}_N$.  In order to compare the networks with the same distance we doubled the size of the $\Tilde{\mathcal{T}}$ and the $\mathcal{L}$ graphs. }\label{lattice}
\end{figure}

Another important class of networks, notably for measurement based quantum computation, is constituted by grid cluster states that belong to graph shapes that allow for  universal quantum computation \cite{Van-den-Nest06}. Similarly to the diamond network, the presence of ancillary nodes between the emitter and the receiver can improve the quality of the quantum link with respect to the linear network. This, however, is not a general rule and sometimes the presence of additional links can be detrimental. This is the case of the triangular lattice, generated from the square lattice by adding a  link between the nodes in the  diagonal. There are two ways of generating the triangular and only one of the two, $\Tilde{\mathcal{T}}$, decreases effectively the distance between Alice and Bob. In both cases the result is detrimental, however $\mathcal{T}$ is slightly better than $\Tilde{\mathcal{T}}$, while the square lattice $\mathcal{Q}$ seems to be the most effective. This result is shown in figure \ref{lattice}.

\section{Routing in complex networks}
\label{App:RoutingComplex}

\begin{figure*}[htb]
\begin{tabular}{cc}
   \begin{tabular}{c}
\includegraphics[width=0.7\linewidth]{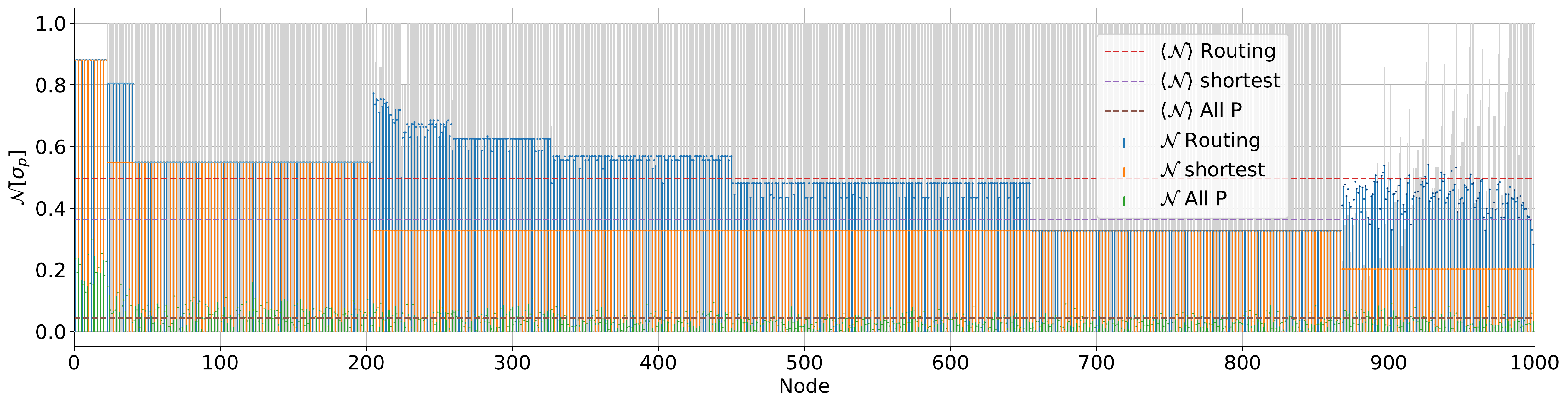}
 \\
\includegraphics[width=0.14\linewidth]{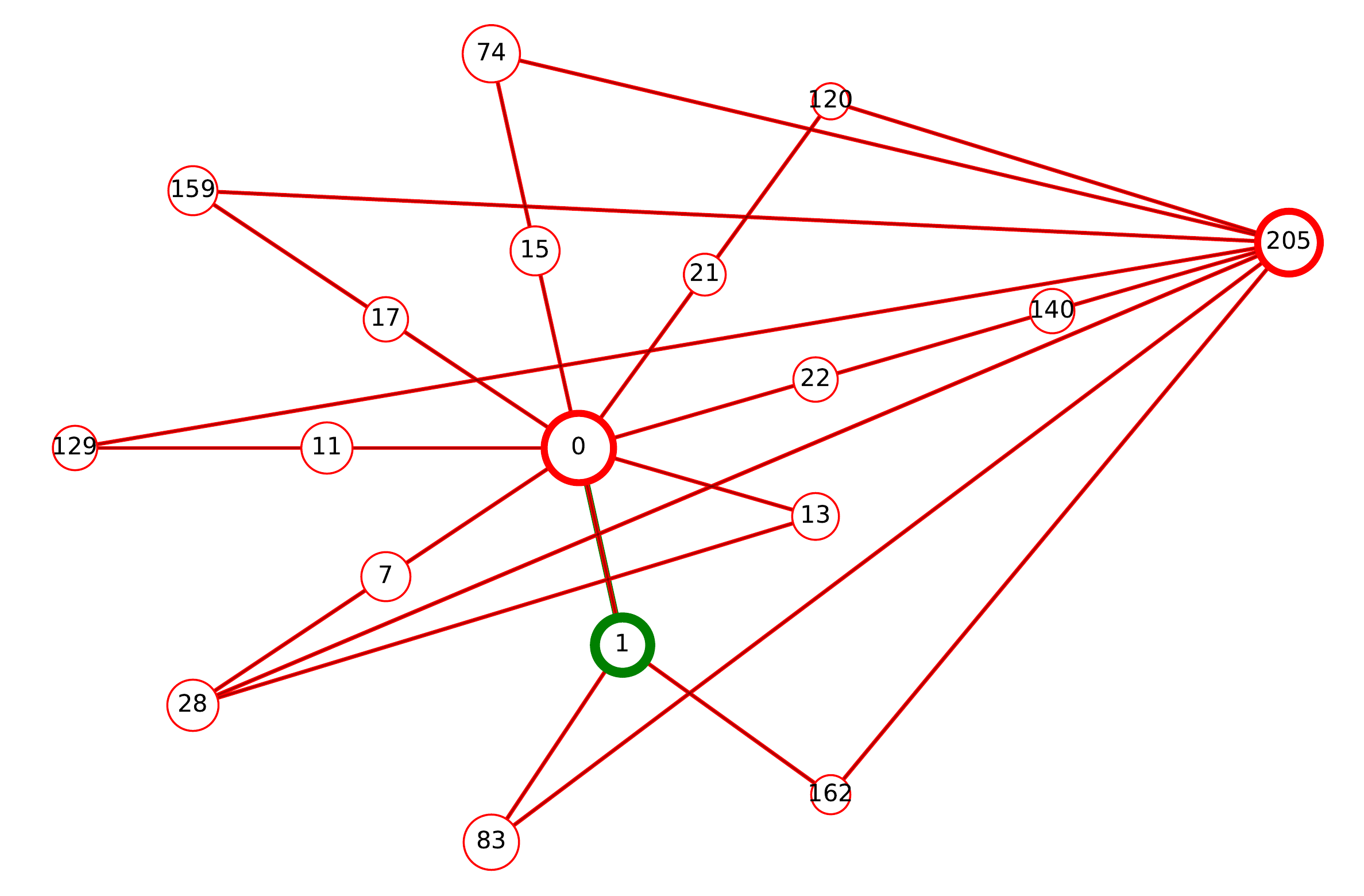}
   \end{tabular}
   &
\raisebox{-0.4\totalheight}{\includegraphics[width=0.275\linewidth]{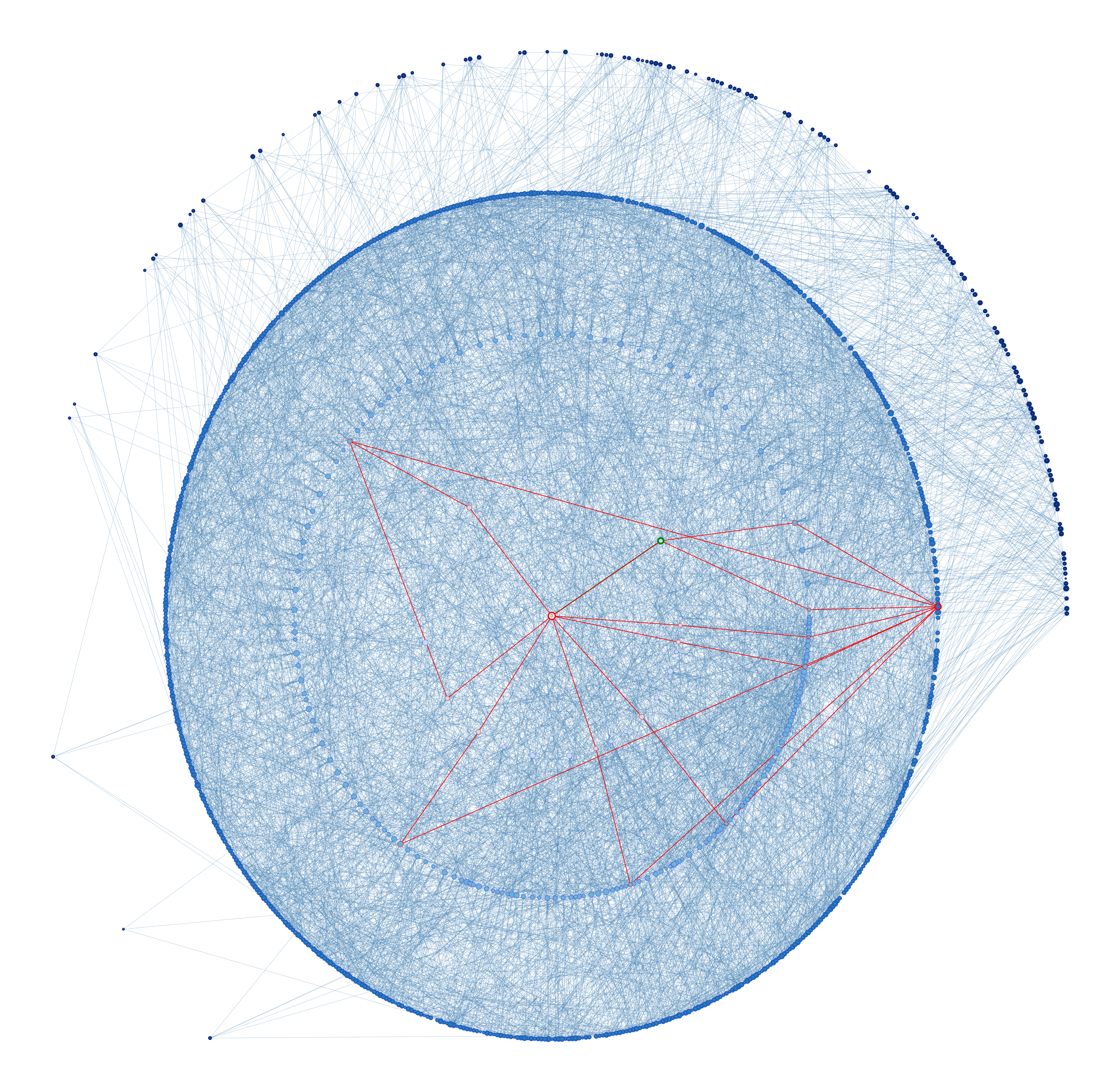}}
\end{tabular}
 \caption{\footnotesize \raggedright Logarithmic negativity produced by the three different protocols applied to each node of the the $\mathcal{G}_{ER}(N=1000, p=0.4)$ network. }\label{fig:netRouteER}
\end{figure*}

\begin{figure*}[htb]
\begin{tabular}{cc}
   \begin{tabular}{c}
     \includegraphics[width=0.7\linewidth]{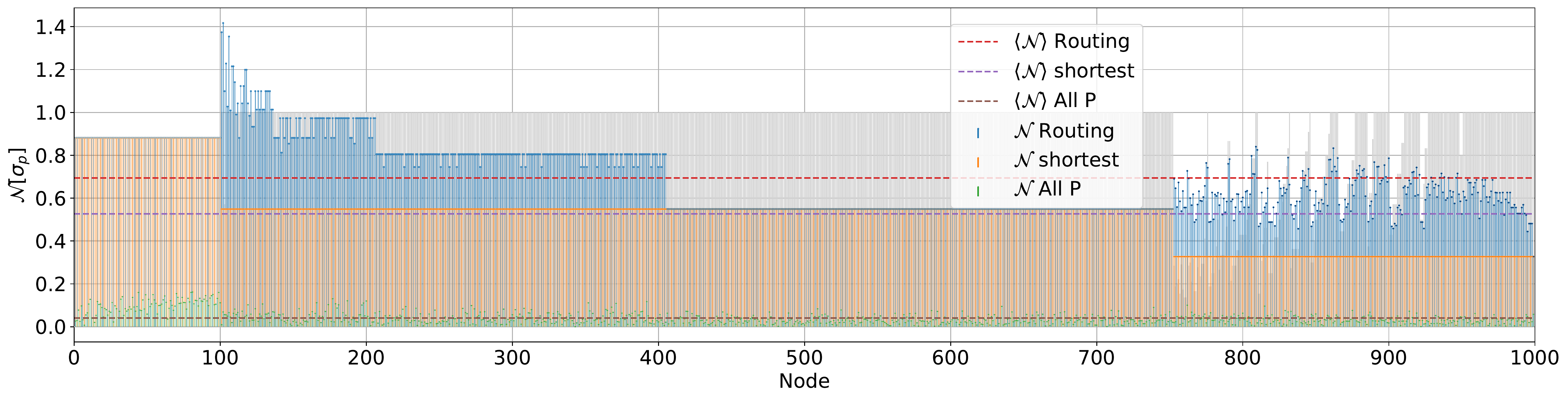}
 \\
\includegraphics[width=0.14\linewidth]{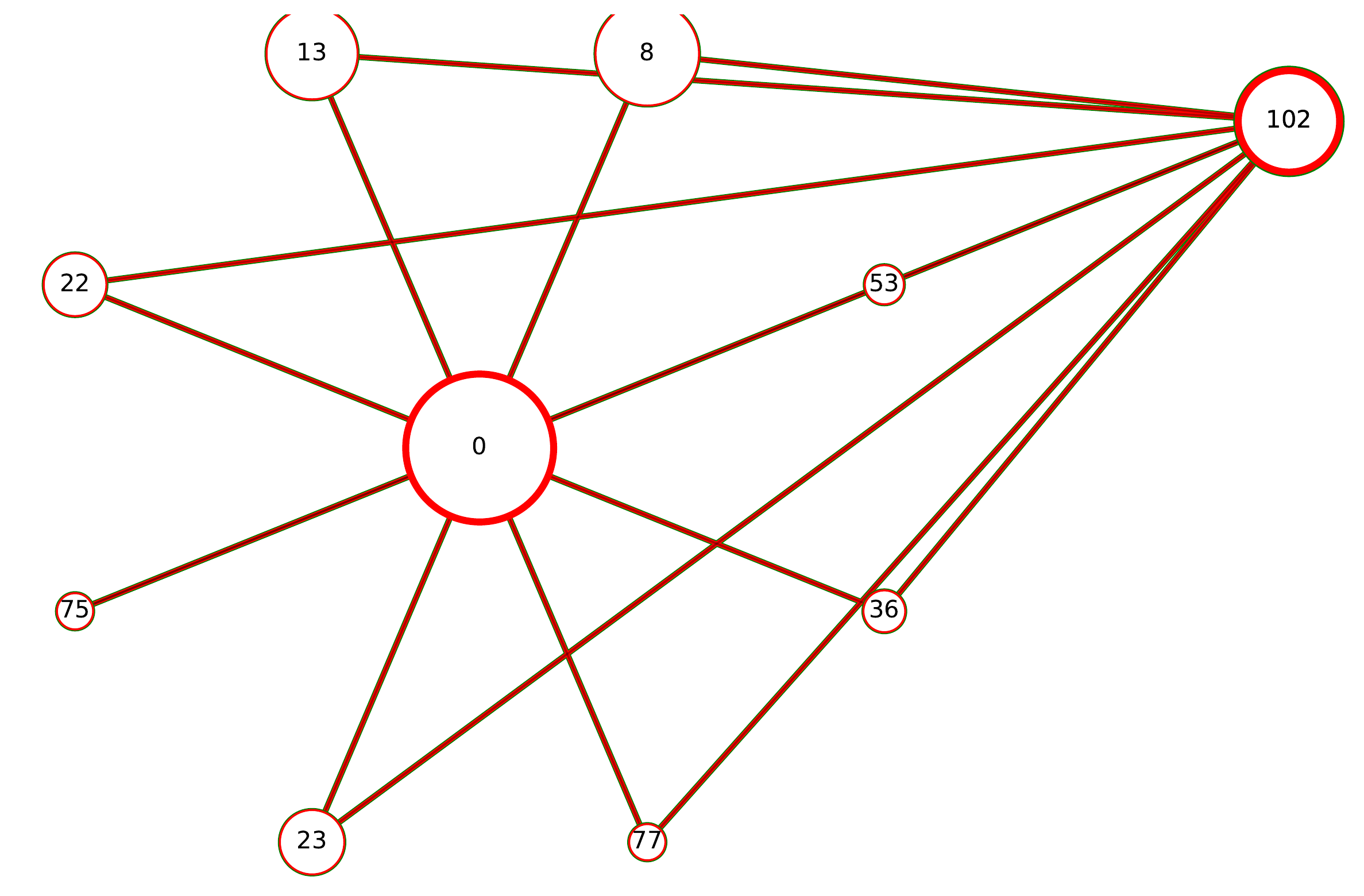}
   \end{tabular}
   &
\raisebox{-0.4\totalheight}{\includegraphics[width=0.275\linewidth]{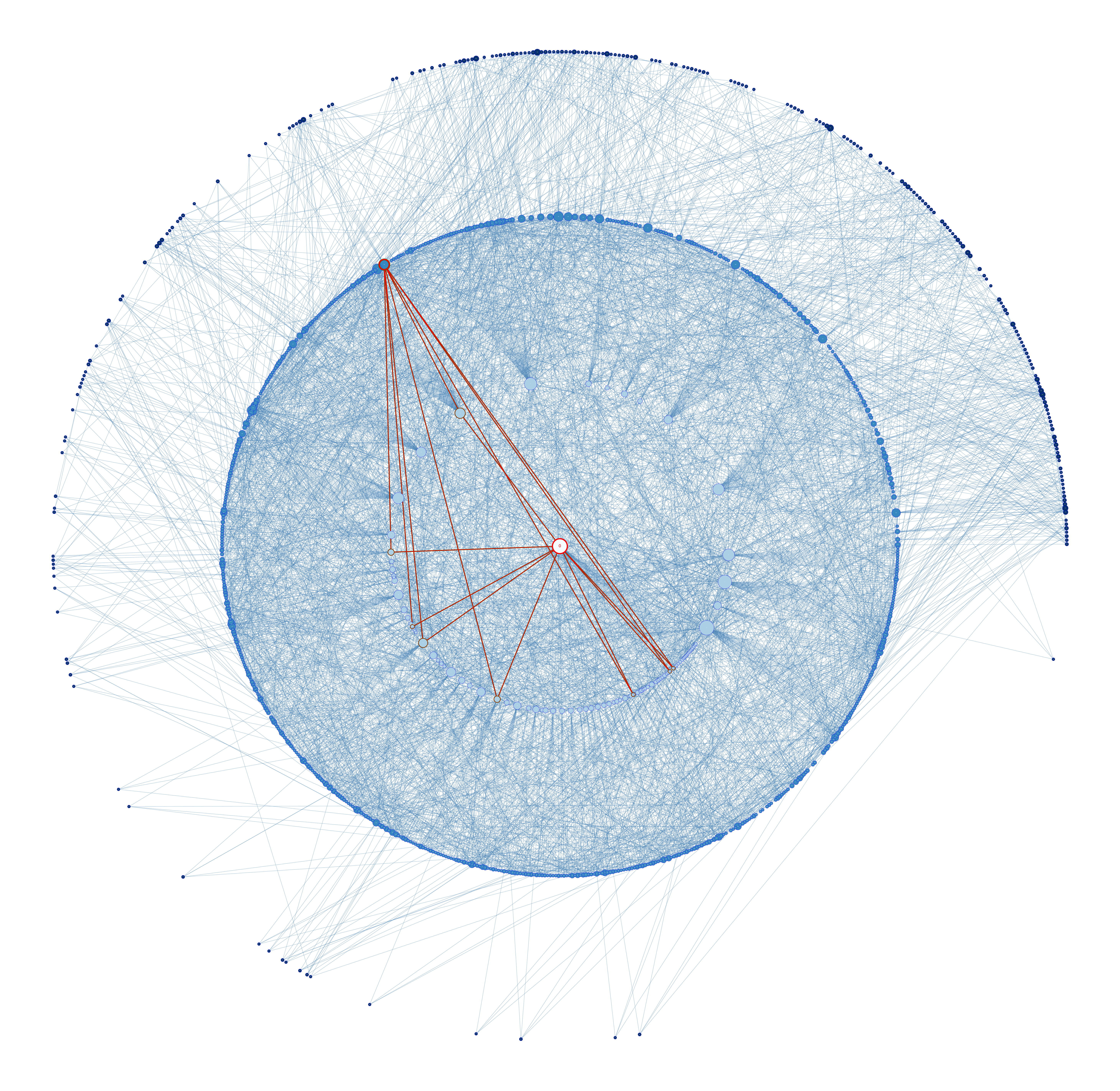}}
\end{tabular}
 \caption{\footnotesize \raggedright Logarithmic negativity produced by the three different protocols applied to each node of the  $\mathcal{G}_{BA}(N=1000, K=4)$ network. }\label{fig:netRouteBA}
\end{figure*}

\begin{figure*}[htb]
\begin{tabular}{cc}
   \begin{tabular}{c}
     \includegraphics[width=0.7\linewidth]{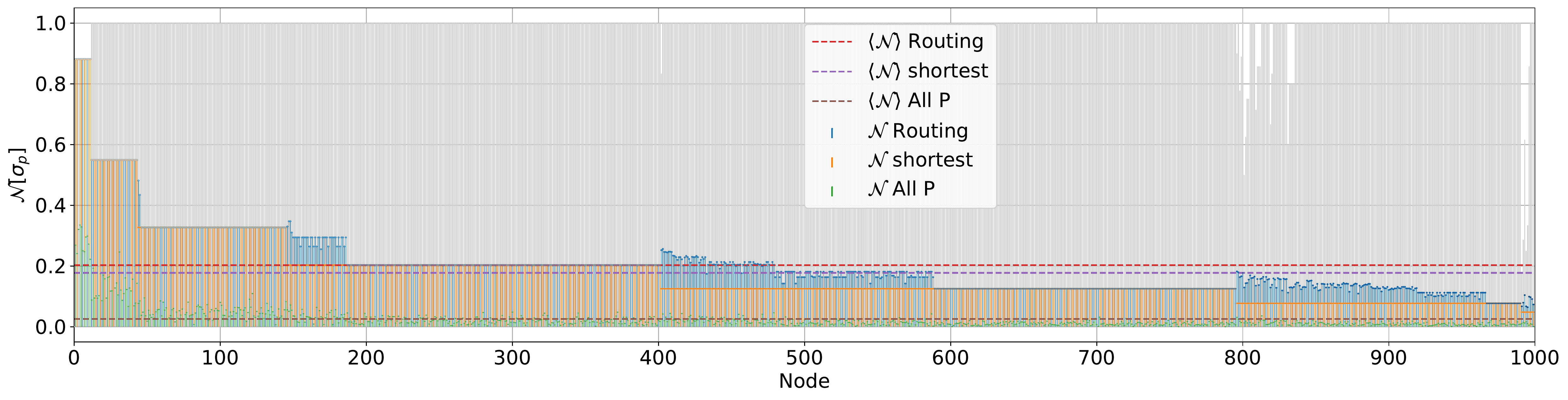}
     \\
\includegraphics[width=0.14\linewidth]{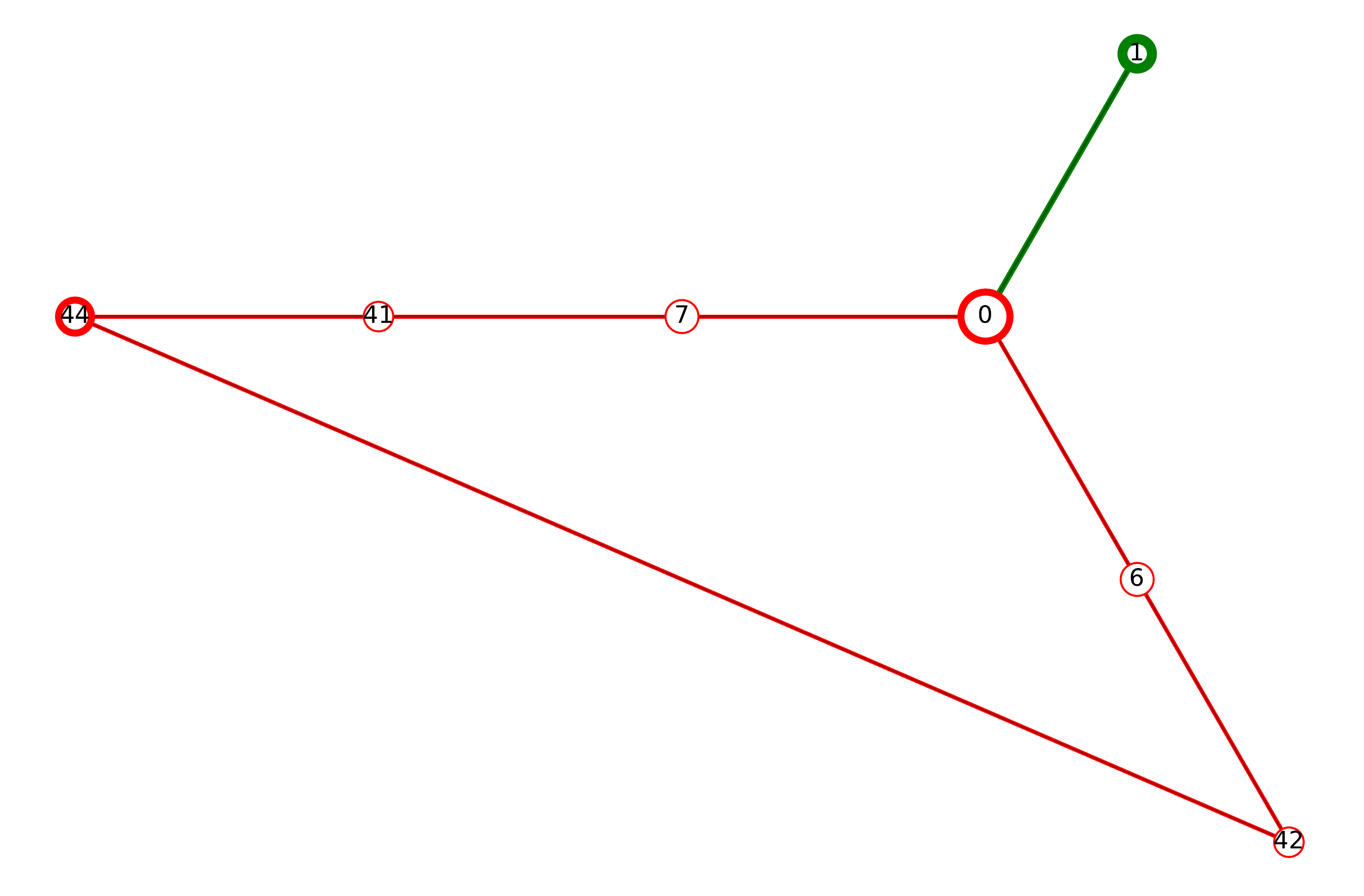}
   \end{tabular}
   &
\raisebox{-0.4\totalheight}{\includegraphics[width=0.275\linewidth]{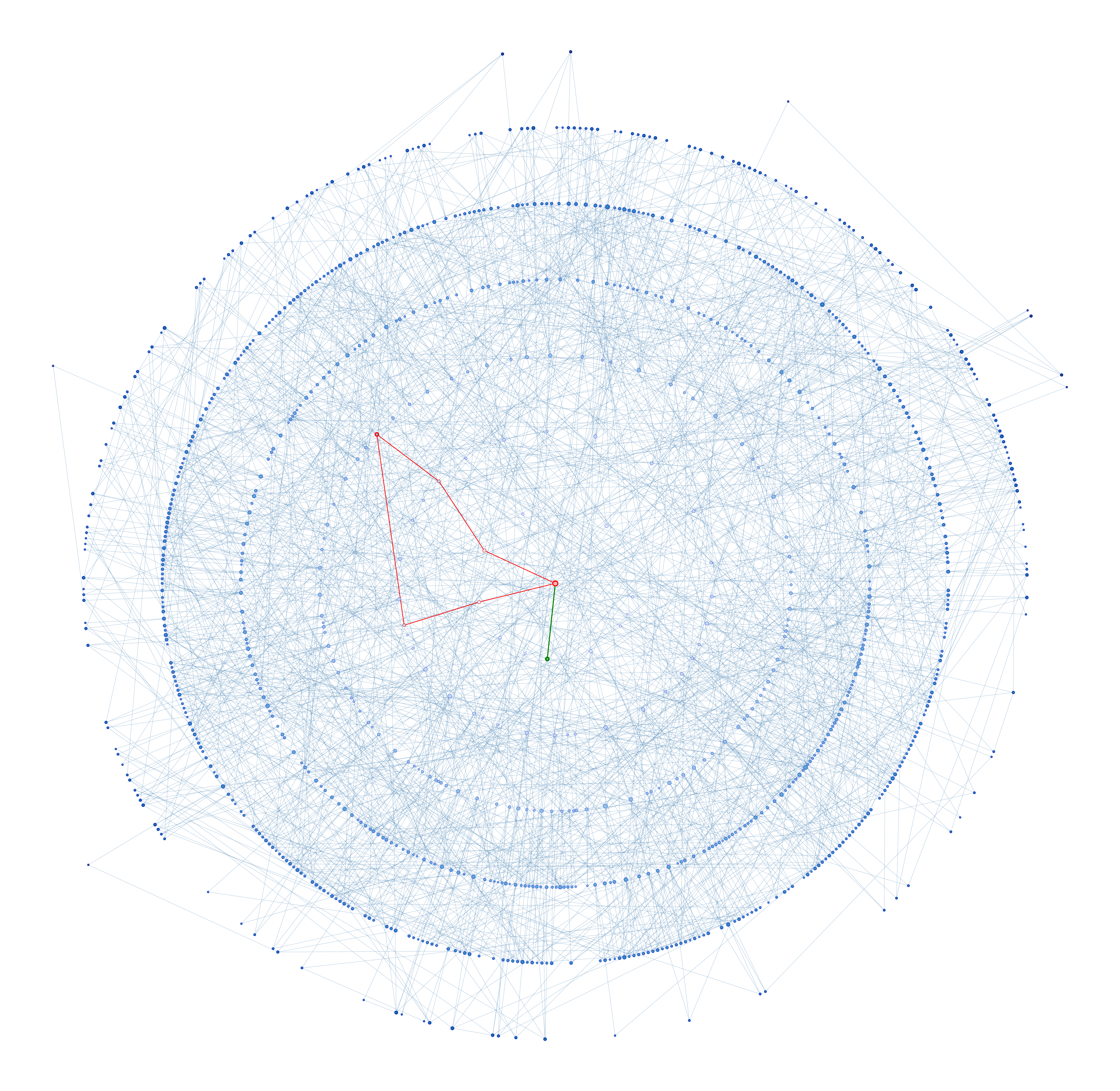}}
\end{tabular}
 \caption{\footnotesize \raggedright Logarithmic negativity produced by the three different protocols applied to each node of the $\mathcal{G}_{WS}(N=1000, Q=4, \beta=0.9)$ network. }\label{fig:netRouteWS}
\end{figure*}

\begin{figure*}[htb]
\begin{tabular}{cc}
   \begin{tabular}{c}
     \includegraphics[width=0.7\linewidth]{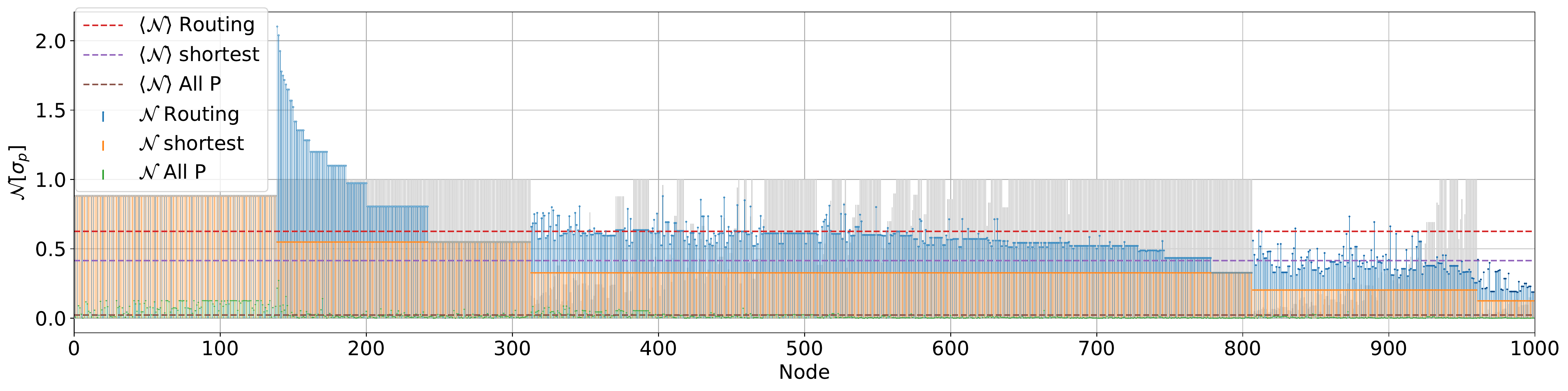}
 \\
\includegraphics[width=0.14\linewidth]{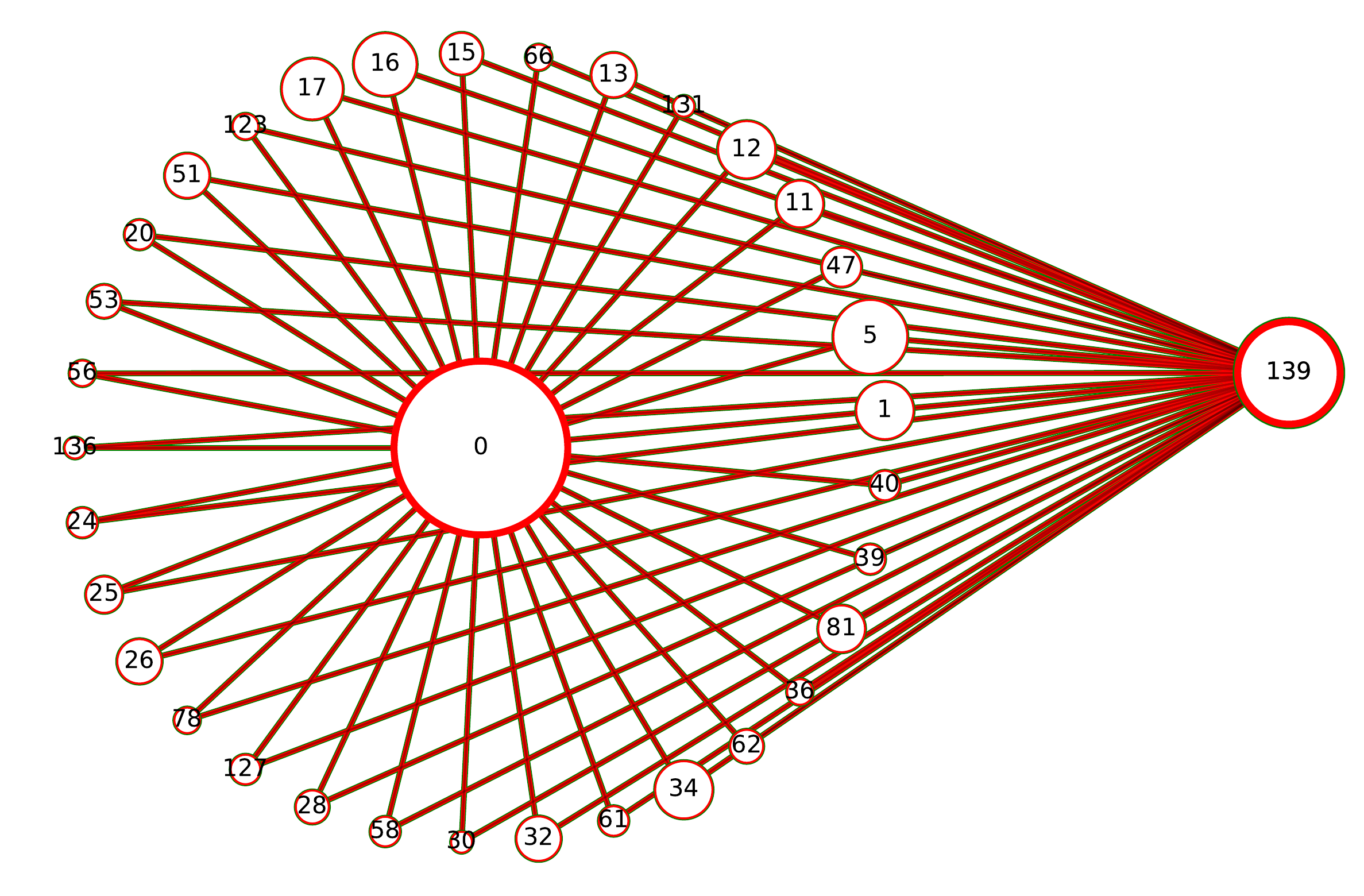}
   \end{tabular}
   &
\raisebox{-0.4\totalheight}{\includegraphics[width=0.275\linewidth]{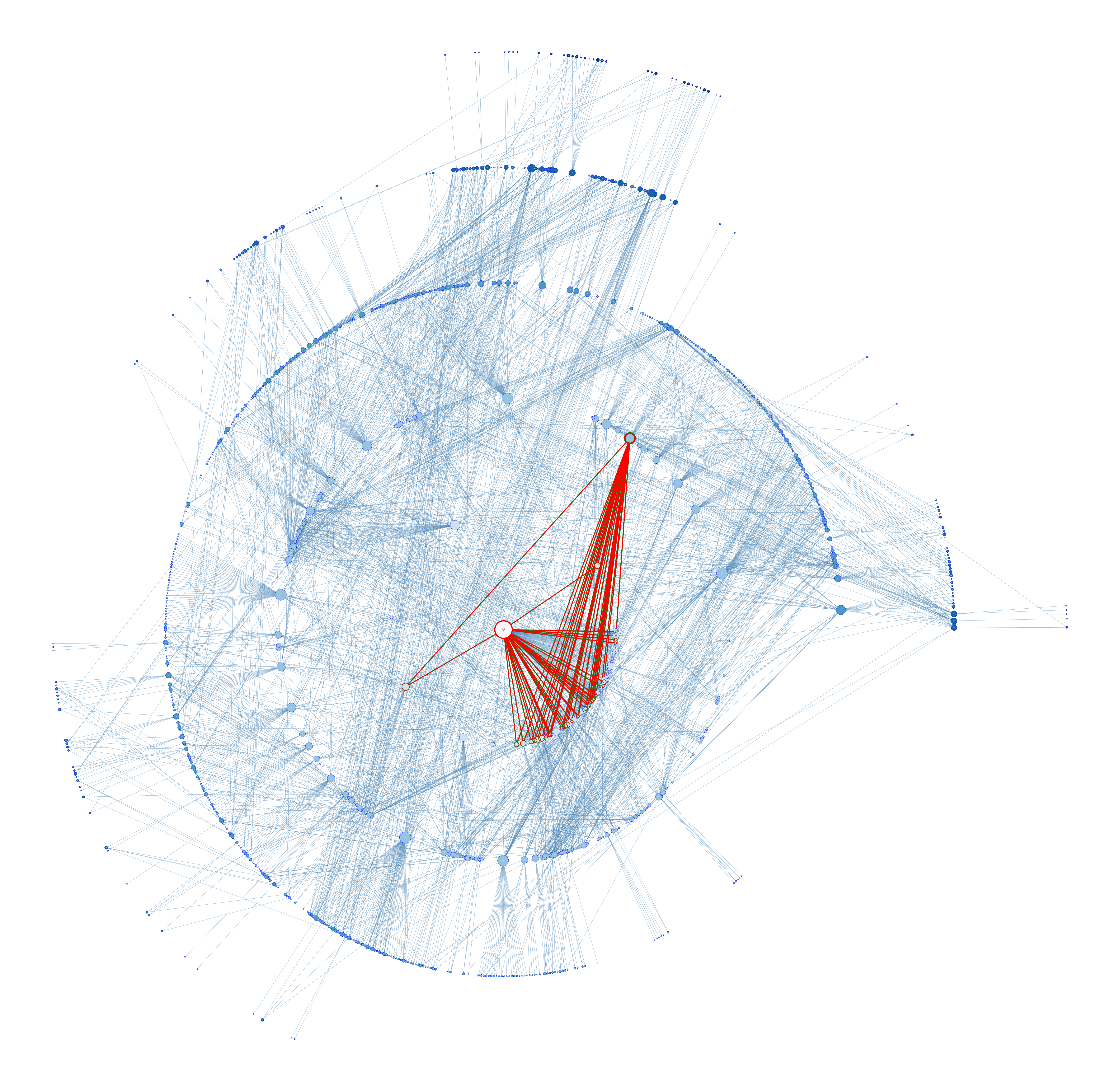}}
\end{tabular}
 \caption{\footnotesize \raggedright Logarithmic negativity produced by the three different protocols applied to each node of the  $\mathcal{G}_{PP}(N=1000,\sigma=0.4)$ network. }\label{fig:netRoutePP}
\end{figure*}
The same analysis of section \ref{sec:routingCX} was done in several networks with different sizes and topologies with very different results that we report in figures \ref{fig:netRouteER}, \ref{fig:netRouteBA}, \ref{fig:netRouteWS} and \ref{fig:netRoutePP}. A property that is not apparent in Fig.\ \ref{fig:netRoute}, is that the node with the highest enhancement of entanglement due to the multiple paths is not necessarily the one with the highest logarithmic negativity in absolute. This is the case of the ER network of Fig.\ \ref{fig:netRouteER}, in which the node with the highest entanglement, highlighted in green in the graph representation, is at distance 1 while the node with the highest difference in logarithmic negativity between the \textit{Routing} and the \textit{Shortest} protocols, highlighted in red, is at distance 3. In this case, the structure of the subgraph used throughout the \textit{Routing} is not a diamond chain and the intercorrelations among the parallel branches have limited the increase of the entanglement,as for the $\Tilde{\mathcal{D}}_N$  network in Fig.\ \ref{fig:LSD}. In any case, in this network the nodes at greater distances are the ones that are most affected by our protocol and, although in some cases many parallel paths have been disregarded, as shown by the height of the grey column, all the nodes at distance 4 received a substantial enhancement. 

The results of the simulation on the BA topology of Fig.\ \ref{fig:netRouteBA} is similar to the AS, although the first only reaches a distance of 3. The nodes with the highest absolute logarithmic negativity and the highest logarithmic negativity difference produced by the \textit{Routing} protocol coincide and are at distance 2 from Alice, whereas this time the subgraph of is a diamond with no interconnections. Also in this case distance 2 is favorable to perform quantum communications. 

The WS structure of Fig.\ \ref{fig:netRouteWS}, on the other hand, is the worst to apply the \textit{Routing protocol}. Only a few nodes, in fact, were poorly enhanced and mostly at large distances, while the logarithmic negativity averaged over all the nodes for \textit{Routing} and \textit{Shortest} is comparable. The node 44 at distance 3 is the one that received the greatest boost from our protocol, whereas node 1 (like all the other nodes at distance 1) has the highest logarithmic negativity. 

Finally, the biological network of Fig.\ \ref{fig:netRoutePP} produced the most interesting results. Once again, many nodes at distance 2 end up having more logarithmic negativity than those at distance 1, and at this distance the nodes with the same degree have the same logarithmic negativity that decreases exponentially with their degree. The nodes with highest logarithmic negativity and highest difference coincide with node 139, which is linked to Alice through 33 intermediate nodes, forming a diamond network with no interconnections.

\end{document}